\documentclass[11pt]{article}

\usepackage{amsmath}
\usepackage{graphicx}
\usepackage{authblk}
\usepackage{algorithm}
\usepackage{amsthm}
\usepackage{amsfonts}
\usepackage{amssymb}
\usepackage{algpseudocode}
\usepackage{bbm}
\usepackage{bm}
\usepackage{booktabs}
\usepackage{caption}
\usepackage{color}
\usepackage{enumerate}
\usepackage{enumitem}
\usepackage{float}
\usepackage{mathtools} 
\usepackage{hyperref}
\usepackage{natbib}
\usepackage{comment}
\usepackage{longtable}
\usepackage{rotating}
\usepackage{xcolor}
\usepackage{fullpage}

\newtheorem{theorem}{Theorem}[section]
\newtheorem{lemma}[theorem]{Lemma}

\newtheorem{proposition}{Proposition}[section]
\theoremstyle{definition}
\newtheorem{example}{Example}[section]
\newtheorem{assumption}{Assumption}[section]
\newtheorem{remark}{Remark}[section]

\algrenewcommand\algorithmicrequire{\textbf{Input:}}
\algrenewcommand\algorithmicensure{\textbf{Output:}}

\usepackage{multirow}
\newcommand{\E}{\mathbb{E}}
\newcommand{\R}{\mathbb{R}}
\newcommand{\MDD}{\mathrm{MDD}}
\newcommand{\Hcal}{\mathcal H}
\begin{document}
\title{Testing Conditional Mean Independence}
\setcounter{footnote}{1}
\author{Chung Eun Lee$^1$ and Xianyang Zhang$^{2}$ \smallskip \\
$^1$ Paul H. Chook Department of Information Systems and Statistics, Baruch College\\
$^2$ Department of Statistics, Texas A\&M University}
\date{} 

\maketitle

\begin{abstract}
This article introduces a novel nonparametric test for conditional mean independence, which formally evaluates whether a random vector $\mathbf{X}$ has an additional impact on the mean regression function of a response $Y$ when a conditioning random vector $\mathbf{Z}$ is already included. By framing the problem as classical partial effect testing, our approach goes beyond traditional parametric models to provide a highly general and flexible framework. The test statistic is constructed using martingale difference divergence (MDD) after neutralizing the effect of $\mathbf{Z}$ on $Y$ via a sieve approximation. The population residual MDD characterizes conditional mean independence under suitable moment conditions. We present a theoretical framework for the test under both a fixed number of basis functions and a growing sieve dimension. We derive the nonstandard and non-pivotal asymptotic null distributions in both regimes. To facilitate practical implementation, we propose a wild bootstrap procedure to approximate these limiting distributions and establish its asymptotic consistency. To establish the fundamental properties required for the test to be widely useful, we conduct a power analysis, demonstrating the test's consistency under fixed alternatives and establishing its asymptotic power against local alternatives. The practical utility and finite-sample performance of the proposed test are examined through simulation studies and an exploratory application to a real-world variable selection problem in regression modeling. The code for the proposed test and numerical simulations can be found at: \url{https://github.com/leece087/Conditional-Mean-Independence-Test}.
\end{abstract}

\noindent%

\section{Introduction}
Conditional independence is a fundamental concept in data analysis, with broad applications across economics, business, engineering, and biomedicine. Given its pivotal role in causal discovery and graphical modeling, it has garnered significant research interest over the past two decades, leading to the development of numerous testing procedures.

Approaches to testing conditional independence can be broadly categorized into parametric and nonparametric frameworks. Parametric tests typically assume Gaussianity or linear relationships; common examples include methods utilizing partial correlation \citep{fisher1915frequency} and tests based on structural equation models or directed acyclic graphs \citep{geiger1994learning, kano2003causal, shimizu2006linear}. Conversely, nonparametric tests leverage flexible dependence metrics to evaluate conditional independence under their respective regularity and model assumptions. Examples include methods based on Hellinger distance \citep{su2008nonparametric} and projected distance covariance \citep{fan2015projection}, the latter of which extends the framework of distance correlation \citep{szekely2007measuring}. Partial distance correlation \citep{szekely2014partial} is a related partial-dependence measure, whose zero value is not equivalent to conditional independence. Other prominent approaches involve kernel mean embeddings \citep{fukumizu2007kernel, zhang2012kernel} or regression-based methods that examine independence or covariance between estimated residuals after filtering out the influence of the conditioning variable \citep{peters2014causal, zhang2017causal, zhang2018measuring, shah2020hardness}. For a comprehensive review of nonparametric conditional independence tests, we refer readers to \citet{li2020nonparametric}.

However, when investigating the relationship between a response and its predictors, conditional mean independence often provides a more targeted assessment of the underlying dependence. Regression models, for instance, focus on the conditional mean of a response, which serves as the optimal predictor under the mean squared error criterion when the response is square-integrable. Evaluating conditional mean independence is therefore crucial for identifying the specific association between a response and its predictors and for enhancing predictive accuracy. Despite its practical importance, tests for conditional mean independence have received considerably less attention in the literature than those for general conditional independence.

This article addresses the problem of testing for conditional mean independence. Such tests are vital in regression analysis for assessing the significance of covariates \citep{fan2008sure, ShaoZhang2014JASA} and in sufficient dimension reduction for identifying the central mean subspace \citep{cook2002dimension, zhu2006fourier}. Specifically, we introduce a novel nonparametric test for the null hypothesis:
\begin{equation}\label{eq:null}
H_0: \mathbb{E}[Y \mid \mathbf{X}, \mathbf{Z}] = \mathbb{E}[Y \mid \mathbf{Z}] \quad \text{almost surely,}
\end{equation}
where $Y \in \mathbb{R}$ is the response variable, $\mathbf{X} \in \mathbb{R}^{s}$ represents the predictors of interest, and $\mathbf{Z} \in \mathbb{R}^{q}$ denotes the conditioning predictors.

Several approaches have been developed for testing conditional mean independence as defined in \eqref{eq:null}. Early work by \citet{ait2001goodness} introduced a nonparametric simplification test to determine whether specific predictors can be omitted from a regression function. However, this smoothing-based approach requires the selection of kernels for two regressions, two bandwidth parameters, and a nonnegative weight function.
Subsequently, \citet{ParkShaoYao2015EJS} developed the partial martingale difference divergence (PMDD) to measure partial mean dependence and proposed a permutation test based on this metric. Nevertheless, their metric does not fully characterize conditional mean independence (see Section 4 in \citet{ParkShaoYao2015EJS}); in particular, it can be nonzero under the null hypothesis, which can inflate the size of the resulting test, as illustrated in Example~\ref{ex:3}. Moreover, the theoretical validity of their permutation procedure was not established.
\citet{jin2018testing} proposed a test for conditional mean independence under the assumption of a linear parametric conditional mean of $Y$ given $\mathbf{Z}$. The validity of this approach relies on this restriction. Similar to \citet{ParkShaoYao2015EJS}, they employed a permutation method without providing a rigorous justification for its consistency.
More recently, several approaches have emerged that embrace machine learning methods. For instance, \citet{cai2024test} developed a test based on machine learning estimates and data splitting, while \citet{zhang2025testing} integrated deep generative neural networks with Reproducing Kernel Hilbert Space (RKHS) embeddings. Both approaches estimate nuisance functions with flexible machine-learning methods, which can be computationally demanding and sensitive to the choice of tuning parameters. Both also use sample splitting or cross-fitting to handle the effect of nuisance estimation. Our procedure instead accounts explicitly for the effect of estimating the regression function and uses a single full-sample sieve fit, which simplifies implementation.
A related line of research in time series analysis focuses on testing conditional mean independence between processes, commonly referred to as Granger causality in mean \citep{granger1969investigating, geweke1982measurement, nishiyama2011consistent, song2018measuring}. This concept assesses whether the history of one time series provides predictive information regarding the conditional mean of another, beyond the information already contained in its own past.

This paper introduces an easy-to-implement testing procedure designed to fully characterize conditional mean independence without assuming a parametric model for the regression function.
Assessing conditional mean independence is considerably more challenging than testing for its unconditional counterpart. Our test statistic is constructed in two main stages: first, we filter out the effect of $\mathbf{Z}$ on $Y$ using a sieve approximation; second, we employ the martingale difference divergence (MDD) to measure the mean dependence between the resulting residuals and $\mathbf{W}=(\mathbf{X},\mathbf{Z})$. We derive the asymptotic null distribution and the asymptotic power of our test statistic under two distinct scenarios for the number of basis functions $p$: (i) when $\mathbb{E}[Y \mid \mathbf{Z}]$ adheres to a finite-dimensional model, and (ii) when $\mathbb{E}[Y \mid \mathbf{Z}]$ is approximated by a series of finite-dimensional models with increasing complexity. The asymptotic distributions under the null and alternative hypotheses are nonstandard and non-pivotal. They retain both the oracle fluctuation and the estimation effect introduced by using residuals in place of unobserved errors. To our knowledge, these limiting distributions are new to the literature. To facilitate inference, we propose a unified wild bootstrap procedure to approximate the sampling distribution of the test statistic and establish its asymptotic consistency in both scenarios.

Existing nonparametric tests for conditional mean independence can involve several estimation steps, such as the estimation of multiple conditional mean functions (e.g., $\mathbb{E}[Y \mid \mathbf{X}, \mathbf{Z}]$, $\mathbb{E}[Y \mid \mathbf{Z}]$), variance estimations, and auxiliary refinements like bias correction or power enhancement. In contrast, our test offers a streamlined implementation that requires fewer estimates (only $\mathbb{E}[Y \mid \mathbf{Z}]$), fewer tuning parameters, and no additional refinement steps. This simplicity, rooted in the properties of the MDD metric, enables our test to utilize the entire dataset, without sample splitting (see Section~\ref{sec:numerical_simulations}).

Series (sieve) estimation of the regression function under the null has also been used to construct consistent specification tests. For example, \citet{li2003consistent} estimate additive, partially linear, and additive partially linear null models by series methods and form a Cram\'er--von Mises-type integrated conditional moment statistic from the resulting residuals. They approximate the null distribution of this statistic by a wild bootstrap and show that the test detects local alternatives of order $n^{-1/2}$. Our test belongs to this broad class of residual-based tests built on sieve estimates, but it differs from that of \citet{li2003consistent} in three respects. First, we measure the remaining dependence between the residual and $\mathbf{W}$ by the MDD. The MDD characterizes conditional mean independence and has a simple closed-form estimator based on pairwise distances. Its weighting is built into its definition, so no additional weighting function (such as the exponential weight used by \citet{li2003consistent}) or integrating measure needs to be chosen. Second, \citet{li2003consistent} let the number of basis functions grow, whereas we treat both a fixed and a growing number of basis functions. In both regimes, we derive the limiting null and local alternative distributions explicitly, through a representation that separates the oracle statistic (computed from the unobserved errors) from the estimation effect. Third, we show that a single wild bootstrap procedure is valid in both regimes, so the practitioner does not need to know which regime applies. On the other hand, the theory of \citet{li2003consistent} allows conditionally heteroskedastic errors, requiring only a bounded conditional variance, whereas our theory assumes that the regression error is independent of $\mathbf{W}$.

Both the formulation of our test statistic and its theoretical justification also differ from those of smoothing-based tests, such as those by \citet{ait2001goodness} and \citet{fan1996consistent}. While traditional smoothing methods often aim for pivotal statistics by neutralizing estimation errors through bandwidth and kernel assumptions, our approach embraces a non-pivotal statistic and accounts for the estimation effect through conditions on the basis functions of the sieve approximation. This non-pivotal nature necessitates a different theoretical treatment. We use an exact projection representation to retain both the oracle statistic and the estimation effect in the limiting distribution.

Because it does not assume a parametric model for the regression function, our test is more flexible than tests that do, such as that of \citet{jin2018testing}. It combines rigorous theoretical properties with a practical implementation characterized by minimal user intervention, no data splitting, and no need for secondary refinement. Given the current landscape of conditional mean independence testing, our method offers a powerful and computationally efficient addition to the statistical toolbox.

The remainder of this article is organized as follows. Section~\ref{sec:review} provides a brief overview of martingale difference divergence. Section~\ref{sec:CMI} introduces our conditional mean independence test, and Section~\ref{sec:bootstrap} details the bootstrap procedure and its theoretical justification for fixed and growing $p$. Section~\ref{sec:power} investigates the asymptotic power of our proposed test under fixed and local alternatives. Numerical results are presented in Section~\ref{sec:numerical_simulations}, followed by an exploratory real-data application in Section~\ref{sec:real data}. Section~\ref{sec:conclusion} concludes the paper. All mathematical proofs, supporting lemmas, and additional application results are provided in the Appendix.

\section{Preliminaries}
\subsection{Notation}\label{sec:notation}
We begin with a brief review of the notation and operations used throughout this article. Boldface symbols represent multivariate random variables, vectors, or matrices, while standard type denotes univariate random variables or scalars. For any square matrix $\mathbf{M} = [M_{ij}]_{i,j=1}^n \in \mathbb{R}^{n \times n}$, we define its $\mathcal{U}$-centered counterpart $\widetilde{\mathbf{M}} = [\widetilde{M}_{ij}]_{i,j=1}^n \in \mathbb{R}^{n \times n}$ following the methodology in \citet{szekely2014partial}, \citet{ParkShaoYao2015EJS}, and \citet{lee2020testing}:
\[
\widetilde{M}_{ij} = 
\begin{cases} 
M_{ij} - M_{i\cdot} - M_{\cdot j} + M_{\cdot \cdot} & i \neq j, \\
0 & i = j,
\end{cases}
\]
where
\[
M_{i\cdot} = \frac{1}{n-2} \sum_{l=1}^n M_{il},\qquad M_{\cdot j} = \frac{1}{n-2} \sum_{k=1}^n M_{kj},
\]
and $M_{\cdot \cdot} = \{(n-1)(n-2)\}^{-1} \sum_{k,l=1}^n M_{kl}$.

Throughout our discussion, we assume that $\E[|Y|]<\infty$. The response variable $Y$ can be decomposed as
\[
Y = \mathbb{E}[Y \mid \mathbf{Z}] + e,
\]
where $e = Y - \mathbb{E}[Y \mid \mathbf{Z}]$ is the population residual, which by construction satisfies $\mathbb{E}[e \mid \mathbf{Z}] = 0$ almost surely. Let $\boldsymbol{\phi}_p(\mathbf{z}) = (\phi_1(\mathbf{z}), \dots, \phi_p(\mathbf{z}))^\top \in \mathbb{R}^p$ denote a vector of $p$ nonconstant basis functions defined on $\mathbb{R}^q$. We define $f(\mathbf Z)=\mathbb E[Y\mid\mathbf Z]$, $\sigma^2 = \mathbb{E}[e^2]$, $\mathbf{W} = (\mathbf{X}, \mathbf{Z})$, and $\mathbf{V} = (\mathbf{W}, e)$. Let $\{G_k\}_{k=1}^{\infty}$ be a sequence of independent and identically distributed (i.i.d.) standard normal random variables. In addition, let $(\mathbf{W}', \mathbf{V}')$ be an independent copy of $(\mathbf{W}, \mathbf{V})$. We define the double-centered distance kernel $U(\mathbf{W}_1, \mathbf{W}_2)$ as:
\begin{align*}
U(\mathbf{W}_1, \mathbf{W}_2) =& \|\mathbf{W}_1 - \mathbf{W}_2\| + \mathbb{E}\|\mathbf{W} - \mathbf{W}'\| - \mathbb{E}[\|\mathbf{W} - \mathbf{W}'\| \mid \mathbf{W} = \mathbf{W}_1] 
\\ &- \mathbb{E}[\|\mathbf{W} - \mathbf{W}'\| \mid \mathbf{W}' = \mathbf{W}_2].    
\end{align*}
Using this kernel, we define
\[
K_p(\mathbf{V}_1)=-\mathbb{E}\!\left[eU(\mathbf{W},\mathbf{W}')
\breve{\boldsymbol{\phi}}_p(\mathbf{Z}')
\mid\mathbf{V}=\mathbf{V}_1\right],
\]
where $\mathbf{V}_1 = (\mathbf{W}_1, e_1)$ and $\breve{\boldsymbol{\phi}}_p(\mathbf{Z})$, the centered and standardized version of $\boldsymbol{\phi}_p(\mathbf{Z})$ with mean zero and identity covariance matrix, is defined in Condition (A2) of Assumption~\ref{assum:0} in Section~\ref{sec:CMI}. For any vector $\bm{s}$, $\|\bm{s}\|$ denotes the Euclidean norm, and for any matrix $\mathbf{M}$, $\|\mathbf{M}\|_F$ denotes the Frobenius norm.

\subsection{Review of Martingale Difference Divergence}\label{sec:review}
We provide a brief review of the martingale difference divergence (MDD), its standardized version, the martingale difference correlation (MDC), and the matrix extension known as the martingale difference divergence matrix (MDDM), as proposed by \citet{ShaoZhang2014JASA}, \citet{zhang2018conditional}, and \citet{LeeShao2018JASA}. For a random vector $\mathbf{D} \in \mathbb{R}^{r}$ and a random variable $U \in \mathbb{R}$, $U$ is said to be mean independent of $\mathbf{D}$ if
\begin{equation}
\label{eq:meanind}
\mathbb{E}[U \mid \mathbf{D}] = \mathbb{E}[U] \quad \text{almost surely.}
\end{equation}
The MDD, which measures the mean dependence of $U$ on $\mathbf{D}$, is defined as the non-negative value ${\rm MDD}(U \mid \mathbf{D})$ satisfying
\begin{equation*}
{\rm MDD}^2(U \mid \mathbf{D}) = \frac{1}{c_{r}} \int_{\mathbb{R}^{r}} \frac{|g_{U,\mathbf{D}}(\bm{s}) - g_{U}g_{\mathbf{D}}(\bm{s})|^2}{\|\bm{s}\|^{1+r}} d\bm{s},
\end{equation*}
where $g_{U,\mathbf{D}}(\bm{s}) = \mathbb{E}[U e^{\imath \bm{s}^\top \mathbf{D}}]$, $g_{U} = \mathbb{E}[U]$, and $g_{\mathbf{D}}(\bm{s}) = \mathbb{E}[e^{\imath \bm{s}^\top \mathbf{D}}]$. Here, $\imath = \sqrt{-1}$ is the imaginary unit, and $c_{r} = \pi^{(1+r)/2} / \Gamma((1+r)/2)$ is a constant.

A fundamental property of MDD is that ${\rm MDD}^2(U \mid \mathbf{D}) \geq 0$, with equality holding if and only if \eqref{eq:meanind} is satisfied. Thus, MDD fully characterizes mean independence. Under the moment conditions $\mathbb{E}[|U| + \|\mathbf{D}\|] < \infty$ and $\mathbb{E}[|U| \cdot \|\mathbf{D}\|] < \infty$, the squared MDD can be expressed as the following \citep{ShaoZhang2014JASA, lee2020testing}.
\begin{equation}\label{eq:mdd}
{\rm MDD}^2(U \mid \mathbf{D}) = -\mathbb{E}\left[(U - \mathbb{E}[U])(U' - \mathbb{E}[U']) \|\mathbf{D} - \mathbf{D}'\|\right],
\end{equation}
where $(U', \mathbf{D}')$ is an independent copy of $(U, \mathbf{D})$. Since ${\rm MDD}(U \mid \mathbf{D})$ is not scale-invariant, the MDC is defined as
\[
{\rm MDC}^2(U \mid \mathbf{D}) = \frac{{\rm MDD}^2(U \mid \mathbf{D})}{\sqrt{{\rm var}^2(U) \, {\rm dvar}^2(\mathbf{D})}},
\]
where ${\rm dvar}(\mathbf{D})$ is the distance variance of $\mathbf{D}$ defined in  \citet{szekely2007measuring}. Furthermore, to characterize the mean dependence between two random vectors $\mathbf{U} \in \mathbb{R}^t$ and $\mathbf{D}$, \citet{LeeShao2018JASA} introduced the MDDM, a real, symmetric, and positive semidefinite matrix defined as:
\[
{\rm MDDM}(\mathbf{U} \mid \mathbf{D}) = -\mathbb{E}\left[(\mathbf{U} - \mathbb{E}[\mathbf{U}])(\mathbf{U}' - \mathbb{E}[\mathbf{U}'])^\top \|\mathbf{D} - \mathbf{D}'\|\right].
\]

Given an i.i.d. sample $\{\mathbf{D}_i, U_i\}_{i=1}^{n}$, an unbiased estimator of the squared MDD can be constructed using the $\mathcal{U}$-centering approach \citep{szekely2014partial, ParkShaoYao2015EJS}. Let $\mathbf{A} = [A_{ij}]$ and $\mathbf{B} = [B_{ij}]$ be $n \times n$ matrices where $A_{ij} = \|\mathbf{D}_i - \mathbf{D}_j\|$ and $B_{ij} = \frac{1}{2}(U_i - U_j)^2$. The unbiased estimator is given by
\begin{equation*} 
\widehat{{\rm MDD}}^2(U \mid \mathbf{D}) = \frac{1}{n(n-3)} \sum_{i \neq j} \widetilde{A}_{ij} \widetilde{B}_{ij},
\end{equation*} 
where $\widetilde{\mathbf{A}}=  [\widetilde{A}_{ij}]$ and $\widetilde{\mathbf{B}}=[\widetilde{B}_{ij}]$ are the $\mathcal{U}$-centered versions of $\mathbf{A}$ and $\mathbf{B}$, respectively. The corresponding estimator for the squared MDC is
\[
\widehat{{\rm MDC}}^2(U \mid \mathbf{D}) = \frac{\widehat{{\rm MDD}}^2(U \mid \mathbf{D})}{\sqrt{\widehat{\rm var}^2(U) \widehat{{\rm dvar}}^2(\mathbf{D})}},
\]
where $\widehat{\text{var}}^2(U) = \frac{1}{n(n-3)}\sum_{i \neq j} \widetilde{B}_{ij}^2$ and $\widehat{\text{dvar}}^2(\mathbf{D}) = \frac{1}{n(n-3)}\sum_{i \neq j} \widetilde{A}_{ij}^2$.

\section{Conditional Mean Independence Test}\label{sec:CMI}
The primary objective of this paper is to test the following null hypothesis:
\begin{equation}
H_0: \mathbb{E}[Y \mid \mathbf{X}, \mathbf{Z}] = \mathbb{E}[Y \mid \mathbf{Z}] \quad \text{almost surely.} \tag{\ref{eq:null}}
\end{equation}
Under the null hypothesis, it follows that $\mathbb{E}[Y - \mathbb{E}[Y \mid \mathbf{Z}] \mid \mathbf{X}, \mathbf{Z}] = \mathbb{E}[Y \mid \mathbf{X}, \mathbf{Z}] - \mathbb{E}[Y \mid \mathbf{Z}] = 0$ almost surely. In other words, the residual $e = Y - \mathbb{E}[Y \mid \mathbf{Z}]$ is mean independent of $\mathbf{W} = (\mathbf{X}, \mathbf{Z})$, which is equivalent to 
\[
{\rm MDD}^2(e \mid \mathbf{W}) = 0.
\]
Motivated by this observation, we construct our test statistic based on a sample version of ${\rm MDD}^2(e \mid \mathbf{W})$, which provides a full characterization of conditional mean independence. We summarize the key properties of this metric in the following proposition; the proofs are omitted as they follow directly from Theorem 1 in \citet{ShaoZhang2014JASA}.

\begin{proposition}\label{prop:1} 
Suppose $\mathbb{E}[|e| + \|\mathbf{W}\|] < \infty$ and $\mathbb{E}[|e| \cdot \|\mathbf{W}\|] < \infty$. Then:
\begin{enumerate}[label=(\roman*)]
    \item ${\rm MDD}^2(e \mid \mathbf{W}) \geq 0$;
    \item ${\rm MDD}^2(e \mid \mathbf{W}) = 0$ if and only if $H_0$ is true.
\end{enumerate}
\end{proposition}

Since the error term $e$ is unobserved, it must be estimated from the data. In this work, we approximate $e$ by employing a sieve approach with a separate intercept to estimate $f(\mathbf z):=\mathbb E[Y\mid\mathbf Z=\mathbf z]$ using $p$ basis functions, where $p$ is either fixed or grows with $n$ (see the two scenarios below). Specifically, we use the vector of basis functions $\boldsymbol{\phi}_p(\mathbf{z})$ introduced in Section~\ref{sec:notation}.

Following \citet{li2007nonparametric}, and under suitable regularity conditions, the conditional mean function $f(\mathbf{z}):= \mathbb{E}[Y \mid \mathbf{Z} = \mathbf{z}]$ can be approximated by a linear combination of these basis functions and an intercept. More precisely, we assume there exists a scalar $\gamma_0$ and a vector $\bm{\beta}_0 \in \mathbb{R}^p$ such that $\sup_{\mathbf{z} \in \mathcal{S}} |f(\mathbf{z}) -\gamma_0- \bm{\beta}_0^\top \boldsymbol{\phi}_p(\mathbf{z})| \to 0$ as $p \to \infty$, where $\mathcal{S}\subseteq\mathbb{R}^q$ is a set with $\mathbb{P}(\mathbf{Z}\in\mathcal{S})=1$ (see Condition (A1) in Assumption~\ref{assum:0} and Condition (B3) in Assumption~\ref{assum:1}). Consequently, we can represent $Y$ as:
\begin{equation}\label{eq:sieve}
Y = \mathbb{E}[Y \mid \mathbf{Z}] + e = \gamma_0+{}\bm{\beta}_0^\top \boldsymbol{\phi}_p(\mathbf{Z}) + R_p + e,
\end{equation}
where $R_p = \mathbb{E}[Y \mid \mathbf{Z}] -\gamma_0- \bm{\beta}_0^\top \boldsymbol{\phi}_p(\mathbf{Z})$ denotes the approximation error, which is treated as a remainder term of smaller order; it is exactly zero in the fixed-$p$ scenario (i) below.

In practice, we observe an i.i.d. sample $\{(\mathbf{X}_i, Y_i, \mathbf{Z}_i)\}_{i=1}^n$ generated from the joint distribution of $(\mathbf{X}, Y, \mathbf{Z})$. We estimate the vector of coefficients $\bm{\beta}_0$ using the ordinary least squares (OLS) estimator obtained by regressing $Y$ on the basis functions $(1,\boldsymbol{\phi}_p(\mathbf{Z})^\top)^\top$:

\begin{align*}
\widehat{\boldsymbol{\beta}}
&= \widehat{\mathbf{Q}}^{-1}
\left(\frac{1}{n}\sum_{i=1}^{n}\{\boldsymbol{\phi}_p(\mathbf{Z}_i)
-\overline{\boldsymbol{\phi}}_p\}Y_i\right),\qquad
\widehat{\gamma}
= \overline{Y}
-\widehat{\boldsymbol{\beta}}^\top
\overline{\boldsymbol{\phi}}_p,
\end{align*}
where $\widehat{\mathbf{Q}}
= \frac{1}{n}\sum_{i=1}^{n}
(\boldsymbol{\phi}_p(\mathbf{Z}_i)
-\overline{\boldsymbol{\phi}}_p)(\boldsymbol{\phi}_p(\mathbf{Z}_i)
-\overline{\boldsymbol{\phi}}_p)^\top$, $\overline{\boldsymbol{\phi}}_p
= \frac{1}{n}\sum_{i=1}^{n}
\boldsymbol{\phi}_p(\mathbf{Z}_i)$, and $\overline{Y}
= \frac{1}{n}\sum_{i=1}^{n}Y_i$. 
Let $\widehat{e}_i = Y_i -\widehat{\gamma}- \widehat{\bm{\beta}}^\top \boldsymbol{\phi}_p(\mathbf{Z}_i)$ denote the residual for the $i$th observation. Our test statistic is defined as the sample squared MDD of the residuals given the predictors, denoted by $\widehat{\rm MDD}^2(\widehat{e} \mid \mathbf{W})$, which is computed using the pairs $\{\mathbf{W}_i, \widehat{e}_i\}_{i=1}^n$ with $\mathbf{W}_i = (\mathbf{X}_i, \mathbf{Z}_i)$. 

We note that the above expressions hold when $\widehat{\mathbf{Q}}$ is
invertible, which occurs with probability tending to one
as $n\to\infty$ under Condition (A2) in
Assumption~\ref{assum:0} and Condition (B2) in
Assumption~\ref{assum:1}, so the singular case does not affect the asymptotic results below. Our proposed test remains well defined even if $\widehat{\mathbf{Q}}$ is singular. Although the OLS coefficients themselves might lack uniqueness, the orthogonal projection onto the column space of the full design matrix ensures that the fitted values and residuals are always uniquely determined.

In the following, we establish the asymptotic null distribution of $\widehat{\rm MDD}^2(\widehat{e} \mid \mathbf{W})$ under two distinct scenarios: 
\begin{enumerate}[label=(\roman*)]
    \item The conditional mean function $f(\cdot)$ lies exactly in the space spanned by the basis functions, i.e., $f(\mathbf{z}) = \gamma_0+\bm{\beta}_0^\top \boldsymbol{\phi}_p(\mathbf{z})$ for some $\bm{\beta}_0 \in \mathbb{R}^p$ and all $\mathbf{z} \in \mathcal{S}$, where $p$ does not grow with $n.$
    \item The function $f(\cdot)$ is well-approximated by a sieve expansion where the number of basis functions $p$ diverges with the sample size $n$.
\end{enumerate}
\begin{assumption}\label{assum:0}
The following conditions hold:
\begin{itemize}
    \item[(A1)] The dimension $p$ is fixed as $n \to \infty$, and $f(\mathbf{z}) = \gamma_0+\bm{\beta}_0^\top \boldsymbol{\phi}_p(\mathbf{z})$ for some $\bm{\beta}_0 \in \mathbb{R}^p$ and all $\mathbf{z} \in \mathcal{S}$, where $\mathcal S\subseteq\mathbb R^q$ is a Borel set with $\mathbb P(\mathbf Z\in\mathcal S)=1$.
    \item[(A2)] The covariance matrix $\mathbf{Q}=\E[\{\boldsymbol\phi_p(\mathbf{Z})-\E\boldsymbol\phi_p(\mathbf{Z})\}\{\boldsymbol\phi_p(\mathbf{Z})-\E\boldsymbol\phi_p(\mathbf{Z})\}^\top]\in\mathbb{R}^{p\times p}$ is positive definite. The centered and standardized basis $\breve{\boldsymbol\phi}_p(\mathbf{Z})=\mathbf Q^{-1/2}\{\boldsymbol\phi_p(\mathbf{Z})-\E[\boldsymbol{\phi}_p(\mathbf{Z})]\}$, which has mean zero and identity covariance matrix by construction, is bounded: $\sup_{\mathbf z\in\mathcal S}\|\breve{\boldsymbol\phi}_p(\mathbf z)\|<C$ for a finite constant $C$.
    \item[(A3)] The error $e$ is independent of $\mathbf{W}$.
    \item[(A4)] $\mathbb{E}[\|\mathbf{W}\|^4 + \|\breve{\boldsymbol{\phi}}_p(\mathbf{Z})\|^4 + e^2] < C'$ for some finite constant $C'$.
\end{itemize}
\end{assumption}

\begin{theorem}\label{thm:null_fixed_p} 
Under the null hypothesis and Assumption~\ref{assum:0}, as $n \to \infty$, we have 
\begin{equation}\label{eq:null_fixed_p}
n\widehat{\rm{MDD}}^2(\widehat{e} \mid \mathbf{W}) \rightarrow^{D} \mathcal{L}_{0,p}:=\sum_{k=1}^\infty \lambda_k (G_k^2 - 1) + \bm{\mathcal{N}}^\top \breve{\mathbf{H}} \bm{\mathcal{N}} - \bm{\mathcal{T}}^\top \bm{\mathcal{N}},
\end{equation}
where $\{\lambda_k\}_{k=1}^{\infty}$ are the eigenvalues (in descending order) associated with the kernel $h(\mathbf{V}_1, \mathbf{V}_2) = -e_1 e_2 U(\mathbf{W}_1, \mathbf{W}_2)$, and the $G_k$ are mutually independent standard normals. Here, $\breve{\mathbf{H}} = \text{MDDM}(\breve{\boldsymbol{\phi}}_p(\mathbf{Z}) \mid \mathbf{W}) \in \mathbb{R}^{p \times p}$, $\bm{\mathcal{N}} \sim N(\mathbf{0}, \sigma^2 \mathbf{I}_p)$, $\sigma^2=\mathbb E e^2$, and $\bm{\mathcal{T}} \sim N(\mathbf{0}, 4 \text{var}({K}_p(\mathbf{V})))$ with $\mathbf{V} = (\mathbf{W}, e)$ as defined in Section~\ref{sec:notation}. The variables $\{G_k\}$, $\bm{\mathcal{N}}$, and $\bm{\mathcal{T}}$ are jointly Gaussian with a covariance structure detailed in Equation \eqref{eq:MCLT} of the Appendix.
\end{theorem}

Theorem~\ref{thm:null_fixed_p} provides the limiting null distribution under the first scenario, which takes a non-standard form. Notably, the first term in \eqref{eq:null_fixed_p} corresponds to the limiting distribution of $n\widehat{\rm{MDD}}^2(e \mid \mathbf{W})$ using the true unobserved errors $e$. The remaining two terms in \eqref{eq:null_fixed_p} characterize the estimation effect introduced by replacing $e$ with the residuals $\widehat{e}$ in the sample MDD.

We now turn to the asymptotic distribution of $\widehat{\rm MDD}^2(\widehat e\mid\mathbf W)$ under the second scenario, where $p$ diverges with $n$.
To derive the limiting distribution, we use a Hilbert-space representation of Euclidean distance \citep{lyons2013distance}. Let $\mathcal H$ be a real separable Hilbert
space equipped with the inner product
$\langle\cdot,\cdot\rangle_{\mathcal H}$
and the induced norm $\|\cdot\|_{\mathcal H}$. Let $\Phi:\mathbb R^{s+q}\to\mathcal H$ satisfy
\begin{equation}\label{grow:feature}
\langle\Phi(\mathbf w),\Phi(\mathbf v)\rangle_{\mathcal H}
=\|\mathbf w\|+\|\mathbf v\|-\|\mathbf w-\mathbf v\|,
\qquad \|\Phi(\mathbf w)\|_{\mathcal H}^2=2\|\mathbf w\|.
\end{equation}
Let 
$\mu_D=\mathbb E\|\mathbf W-\mathbf W'\|$ and
$m_\Phi=\mathbb E\Phi(\mathbf W)$.
In this representation, the double-centered distance kernel of Section~\ref{sec:notation} is an inner product, $U(\mathbf w,\mathbf v)=-\langle\Phi(\mathbf w)-m_\Phi,\Phi(\mathbf v)-m_\Phi\rangle_{\mathcal H}$, and, under the moment conditions of Proposition~\ref{prop:1}, \eqref{eq:mdd} gives ${\rm MDD}^2(e\mid\mathbf W)=\|\mathbb E[e\Phi(\mathbf W)]\|_{\mathcal H}^2$ because $\mathbb E[e]=0$. Thus the squared MDD is the squared norm of a mean in $\mathcal H$, and the growing-$p$ limit in Theorem~\ref{thm:null} below is the squared norm of a Gaussian element of $\mathcal H$ minus a constant.
Define the operator $\mathcal B_p:\mathbb R^p\to\mathcal H$ and residual feature by
\[
\mathcal B_p\mathbf u
=\sum_{j=1}^p u_j\beta_j^\Phi,
\qquad
h_p(\mathbf W)=\Phi(\mathbf W)-m_\Phi-\mathcal B_p\mathbf g_p(\mathbf Z),
\]
where $\beta_j^\Phi=\mathbb E[(\Phi(\mathbf W)-m_\Phi){g}_{j}(\mathbf Z)]$ and $\mathbf g_p(\mathbf z)=(g_1(\mathbf z),\ldots,g_p(\mathbf z))^\top$ is the centered orthonormal basis of the sieve space specified in Condition (B2) of Assumption~\ref{assum:1}. Because the $g_j$ are orthonormal, the terms $\beta_j^\Phi g_j(\mathbf Z)$, $j\geq1$, are mutually orthogonal in mean square, and Bessel's inequality gives $\sum_{j\geq1}\|\beta_j^\Phi\|_{\mathcal H}^2\leq\mathbb E\|\Phi(\mathbf W)-m_\Phi\|_{\mathcal H}^2<\infty$ under Condition (B1) of Assumption~\ref{assum:1}. Hence $h_p(\mathbf W)$ converges in mean square as $p\to\infty$, and we write
\begin{equation}\label{grow:limit_feature}
h_\infty=\lim_{p\to\infty}h_p,
\qquad
\boldsymbol{\Sigma}_\infty=\mathbb E[h_\infty\otimes h_\infty].
\end{equation}
Here $u\otimes v$ maps $w$ to $u\langle v,w\rangle_{\mathcal H}$ and the expression $m_\Phi+\mathcal B_p\mathbf g_p(\mathbf Z)$ represents the population projection of $\Phi(\mathbf W)$ onto the closed span of the intercept and the $\mathbf Z$-sieves. Assuming a complete basis, this yields $h_\infty(\mathbf W) = \Phi(\mathbf W) - \mathbb{E}[\Phi(\mathbf{W}) \mid \mathbf{Z}]$. 
\begin{assumption}\label{assum:1}
Suppose that $p=p_n\to\infty$ and the following conditions hold.
\begin{itemize}
\item[(B1)] The pairs $(\mathbf W_i,e_i)$ are i.i.d. with a fixed joint law, $\|\mathbf W\|\leq M<\infty$ almost surely, and $e$ is independent of $\mathbf W$, with $\mathbb E e=0$, $\sigma^2=\mathbb E e^2>0$, and $\mathbb E e^4<\infty$.
\item[(B2)] The deterministic fitted spaces admit a nested basis $g_1,g_2,\ldots$ satisfying $\mathbb E g_j(\mathbf Z)=0$ and $\mathbb E[g_j(\mathbf Z)g_k(\mathbf Z)]=\mathbf1\{j=k\}$, with
\[
\operatorname{span}\{1,\phi_1,\ldots,\phi_p\}
=\operatorname{span}\{1,g_1,\ldots,g_p\}
\quad\text{and}\quad
\sup_{\mathbf z\in\mathcal S}\|\mathbf g_p(\mathbf z)\|\leq\xi_p,
\]
where $\xi_p^2p/n\to0$. This requires $1,\phi_1(\mathbf Z),\ldots,\phi_p(\mathbf Z)$ to be linearly independent; Section~\ref{sec:numerical_simulations} explains how this is achieved for B-splines.
\item[(B3)] There are $\gamma_p$ and $\bm\beta_p$ such that
\[
\sup_{\mathbf z\in\mathcal S}|f(\mathbf z)-\gamma_p-\bm\beta_p^\top \mathbf g_p(\mathbf z)|
\leq a_p,\qquad \sqrt n\,a_p\to0.
\]
\item[(B4)] The covariance operator $\boldsymbol{\Sigma}_\infty$ satisfies that $\operatorname{tr}(\boldsymbol{\Sigma}_\infty)>0$.
\end{itemize}
\end{assumption}

Condition (B1) requires bounded covariates for the growing-sieve analysis, though this requirement is not needed for the fixed-$p$ result in Theorem \ref{thm:null_fixed_p}. Condition (B2) controls the sample regression Gram matrix, while (B3) ensures the approximation bias remains negligible relative to the feature score. Condition (B4) excludes a degenerate limiting statistic.

To illustrate conditions (B1)–(B4), we present the following example. Suppose $X$ is a Rademacher random variable independent of $Z \sim \text{Unif}(0,1)$, and consider the nested sine/cosine basis given by $g_{2k-1}(z)=\sqrt{2}\cos(2\pi kz)$ and $g_{2k}(z)=\sqrt{2}\sin(2\pi kz)$. For any even $p$, we have $\xi_p^2=p$, so setting $p \asymp n^\kappa$ with $0 < \kappa < 1/2$ satisfies Condition (B2). While a finite trigonometric regression eventually yields zero approximation error, we can consider a more general case where $f(z)=\sum_{k\geq1}k^{-a-1}\cos(2\pi kz)$ with $a>1$. In this scenario, truncation yields an error of $a_p=O(p^{-a})$, meaning any growth rate $1/(2a) < \kappa < 1/2$ satisfies Condition (B3). Finally, Condition (B4) holds because $\Vert{}h_\infty(X,Z)\Vert{}^2=1$ almost surely in this example.

\begin{theorem}\label{thm:null}
 Let $\mathcal{G}_\infty$ be a centered Gaussian random element of $\mathcal H$ with covariance operator $\sigma^2\boldsymbol{\Sigma}_\infty$, and define
\[
\mathcal L_{0,\infty}=\|\mathcal{G}_\infty\|_{\mathcal H}^2-\sigma^2\mu_D.
\]
Under the null hypothesis and Assumption~\ref{assum:1}, we have
\begin{equation}\label{eq:null_large_p}
\sup_t\left|P\{n\widehat{\rm MDD}^2(\widehat e\mid\mathbf W)\leq t\}
-P(\mathcal L_{0,\infty}\leq t)\right|\to0.
\end{equation}
\end{theorem}
Interestingly, the two limiting distributions in Theorems~\ref{thm:null_fixed_p} and \ref{thm:null} admit the same feature representation. 
\begin{proposition}\label{prop:fixed_p_feature}
Under the null hypothesis and
Assumption~\ref{assum:0} with $p$ fixed, the limiting null distribution
$\mathcal L_{0,p}$ in
Theorem~\ref{thm:null_fixed_p}
admits the equivalent representation
\begin{equation}\label{eq:fixed_p_feature_rep}
\mathcal{L}_{0,p}
\overset{D}{=}
\|\mathcal G_p\|_{\mathcal H}^2-\sigma^2\mu_D,
\end{equation}
where $\mathcal G_p$ is a centered Gaussian random element of $\mathcal H$ with covariance operator
$\sigma^2\boldsymbol{\Sigma}_p$. Here $\boldsymbol{\Sigma}_p
=\mathbb E[h_p(\mathbf W)\otimes h_p(\mathbf W)]$, and $h_p$ is defined in Section~\ref{sec:CMI} with $\mathbf g_p=\breve{\boldsymbol\phi}_p$; the limit is non-degenerate when $\sigma^2\operatorname{tr}(\boldsymbol{\Sigma}_p)>0$.
Consequently, we have
\[
n\widehat{\mathrm{MDD}}^2
(\widehat e\mid\mathbf W)
\rightarrow^{D}
\|\mathcal G_p\|_{\mathcal H}^2-\sigma^2\mu_D.
\]
\end{proposition}

\begin{remark}\label{complete:remark_tuning}\rm{
Unlike alternative procedures that require selecting multiple tuning quantities, our method depends on a single user-specified parameter: the basis functions $\boldsymbol{\phi}_p$. To select the number of basis functions $p$, several established data-driven approaches can be employed. For instance, $p$ can be chosen using generalized cross-validation \citep{craven1979smoothing}:
\[
\widehat{p} = \underset{p}{\rm argmin} \frac{\sum_{i=1}^n \{Y_i - \widehat{f}_p(\mathbf{Z}_i)\}^2}{\{1 - (p+1)/n\}^2},
\]
where $\widehat{f}_p(\mathbf{z})$ is the estimated conditional mean function using $p$ basis functions. Alternatively, $p$ can be selected using a Mallows-type information criterion \citep{mallows1973some}:
\[
\widehat{p} = \underset{p}{\rm argmin} \sum_{i=1}^n \{Y_i - \widehat{f}_p(\mathbf{Z}_i)\}^2 + 2(p+1)\widehat{\sigma}^2,
\]
where $\widehat{\sigma}^2$ is a consistent estimator of the error variance based on the residuals $\{\widehat{e}_i\}$. These criteria target prediction accuracy, whereas our theory treats $p$ as deterministic. In particular, Condition (B3) in Assumption~\ref{assum:1} requires $\sqrt n\,a_p\to0$, which may call for more basis functions than a prediction-optimal choice provides. A data-driven selection of $p$ is beyond the scope of the present work and is left for future investigation.}
\end{remark}

\begin{remark}\label{remark:basis_fct}\rm{
When $q > 1$, multivariate series functions can be utilized to approximate $f(\cdot)$. For example, one may employ tensor-product splines, formed by the product of univariate spline functions. For a bivariate case where $\mathbf{Z} = (Z_1, Z_2)$, the power series basis is $\{Z_1, Z_2, Z_1^2, Z_1 Z_2, Z_2^2, \dots\}$, with an intercept fitted separately. For splines, if the support of $\mathbf{Z}$ is a rectangle $[a,b] \times [c,d]$, tensor products are directly applicable. For non-rectangular domains, one can employ general multivariate spline basis functions as detailed in \citet{eubank1999nonparametric}. Alternatively, to mitigate the ``curse of dimensionality,'' one can adopt an additive model structure where the basis vector is defined as
\[
\boldsymbol{\phi}_p(\mathbf{Z}) := (\phi_1(Z_1), \dots, \phi_{p_0}(Z_1), \dots, \phi_1(Z_q), \dots, \phi_{p_0}(Z_q))^\top
\]
with $p = p_0 q$, where each $\{\phi_i\}_{i=1}^{p_0}$ is a set of univariate basis functions.}
\end{remark}
Building upon Theorem~\ref{thm:null_fixed_p} and Theorem~\ref{thm:null}, we define our test statistic as $T_n = n\widehat{\rm{MDD}}^2(\widehat{e} \mid \mathbf{W})$, providing a unified definition that holds regardless of the asymptotic regime. Because both formulations result in non-standard and non-pivotal limiting distributions, critical values cannot be easily obtained from standard tables. To address this, we develop a wild bootstrap procedure designed to approximate the appropriate limiting null distribution regardless of whether $p$ is fixed or diverging. This unified approach is robust because the distinct limiting distributions distinguishing the two scenarios are automatically accounted for by the bootstrap distribution, ensuring a seamless implementation in practice.

\section{Bootstrap-based Test}
\label{sec:bootstrap}
In this section, we propose a wild bootstrap procedure to estimate critical values and demonstrate its consistency. The procedure can be described as follows.
\begin{enumerate}
\item Compute the test statistic $T_n=n\widehat{\rm{MDD}}^2(\widehat{e} \mid \mathbf{W})$ based on the data $\{(\mathbf{X}_i,Y_i,\mathbf{Z}_i):i=1,2,\dots,n\}$.
\item Compute the bootstrap statistic in the following way.
\begin{enumerate}
\item Generate i.i.d.\ standard normal random variables $\eta_1^*,\ldots,\eta_n^*$, independent of the data.
\item Generate the bootstrapped response variables $\{Y^{*}_i\}_{i=1}^n$ by
\[
Y^{*}_i = \widehat\gamma+\widehat{\bm{\beta}}^\top \boldsymbol{\phi}_p(\mathbf{Z}_i) + \widehat{e}_{i} \eta_i^*,\qquad i=1,2,\dots,n.
\]
\item Compute the bootstrap statistic
$T_n^*=n \widehat{\text{MDD}}^{2} (\widehat{e}^{*} \mid \bm{W})$ based on the residuals $\{\widehat{e}^{*}_i\}_{i=1}^n$ from the least squares regression of $Y_i^*$ on $(1,\boldsymbol{\phi}_p(\mathbf{Z}_i)^\top)^\top$, $i=1,\ldots,n$.
\end{enumerate}
\item Repeat Step 2 $N_{\rm boot}$ times and collect the bootstrap statistics $T_{n,b}^*$ for $b=1,2,\dots,N_{\rm boot}$.
\item Compute the empirical $(1-\alpha)$ quantile $Q_{1-\alpha}$ of $T_{n,1}^*,\ldots,T_{n,N_{\rm boot}}^*$.
\item Reject the null if $T_n >Q_{1-\alpha}$.
\end{enumerate}

\begin{remark}\label{remark:last}\rm{
We highlight several key distinctions between our proposed test and existing methodologies. The tests developed by \citet{ParkShaoYao2015EJS} and \citet{jin2018testing} rely on permutation procedures to estimate critical values, whereas our approach approximates the limiting null distribution using a wild bootstrap procedure. Importantly, while the aforementioned permutation-based tests lack rigorous theoretical justification for their consistency, the validity of our wild bootstrap is established in Theorems~\ref{thm:null_fixed_p_bootstrap} and \ref{thm:bootstrap}. Furthermore, existing metrics often fail to fully characterize conditional mean independence. For instance, the PMDD in \citet{ParkShaoYao2015EJS} does not possess the property described in Proposition~\ref{prop:1}, whereas the test in \citet{jin2018testing} is restricted by its reliance on a linear regression framework. In contrast, our test does not assume a parametric model for the regression function, and the population quantity on which it is based characterizes conditional mean independence; see Proposition~\ref{prop:1}.
}
\end{remark}

\begin{remark}\label{complete:remark_metric}\rm{
It is noteworthy that our testing framework can be extended by replacing the Euclidean distance with a general metric $d(\cdot, \cdot)$ in a space of strong negative type \citep{lyons2013distance}, leading to the metric $-\mathbb{E}[\{e-\mathbb{E}[e]\}\{e'-\mathbb{E}[e']\} d(\mathbf{W}, \mathbf{W}')]$. As shown in \citet{zhou2021modified} and \citet{lee2024conditional}, this generalized metric retains the desirable characterization property in Proposition~\ref{prop:1}. Furthermore, we expect the theoretical results in Sections~\ref{sec:CMI}, \ref{sec:bootstrap}, and \ref{sec:power} to remain valid under similar or identical conditions to those in Assumptions~\ref{assum:0}, \ref{assum:1}, \ref{assum:2}, and \ref{assum:3}, although we do not pursue the details here. Common examples of such metrics include kernel-induced semimetrics (see \citet{sejdinovic2013equivalence}):
\[
d(\mathbf{w}_1, \mathbf{w}_2) = \mathcal{K}(\mathbf{w}_1, \mathbf{w}_1) + \mathcal{K}(\mathbf{w}_2, \mathbf{w}_2) - 2\mathcal{K}(\mathbf{w}_1, \mathbf{w}_2),
\]
where $\mathcal{K}$ is a characteristic kernel, such as the Gaussian or Laplacian kernel. Additional examples of suitable distances can be found in \citet{chakraborty2021new}.
}
\end{remark}

The following theorems establish the asymptotic validity of our bootstrap-based tests under the null hypothesis in both scenarios. We adopt the definition of bootstrap consistency from \citet{li2003consistent}. Specifically, for a bootstrap statistic $J_n^*$ depending on the sample $\{R_i\}_{i=1}^n$, we say $J_n^*$ converges to $J$ in distribution in probability if, for any subsequence $J_n^{*'}$, there exists a further subsequence $J_n^{*''}$ such that the conditional distribution of $J_n^{*''}$ given $\{R_i\}$ converges to the distribution of $J$ for almost every sequence $\{R_i\}_{i=1}^\infty$. We denote this convergence as $J_n^*\rightarrow^{D^*}J$. 

\begin{assumption}\label{assum:2}
For fixed $p$, assume the following conditions hold:
\begin{itemize}
\item[(i)] $\mathbb{E}[\|\mathbf{W}\|^8 + \| \breve{\boldsymbol{\phi}}_p(\mathbf{Z})\|^8 + e^8] < C$ for some finite constant $C > 0$;
\item[(ii)] The bootstrap weights $\{\eta_i^*\}$ are i.i.d. standard normal random variables and independent of the observed sample.
\end{itemize}
\end{assumption}

\begin{theorem}\label{thm:null_fixed_p_bootstrap} 
Under the null hypothesis and Assumptions~\ref{assum:0} and \ref{assum:2}, as $n \to \infty$, we have
\begin{equation*}
T_n^* \rightarrow^{D^*} \sum_{k=1}^\infty \lambda_k (G_k^2 - 1) + \bm{\mathcal{N}}^\top \breve{\mathbf{H}} \bm{\mathcal{N}} - \bm{\mathcal{T}}^\top \bm{\mathcal{N}},
\end{equation*}
where $\{G_k\}$, $\bm{\mathcal{N}}$, $\bm{\mathcal{T}}$, and $\breve{\mathbf{H}}$ are defined as in Theorem~\ref{thm:null_fixed_p}.
\end{theorem}
\begin{remark} \rm{While Assumption~\ref{assum:2} specifies that the bootstrap weights $\{\eta_i^*\}$ are i.i.d. standard normal, Theorem~\ref{thm:null_fixed_p_bootstrap} remains valid for any i.i.d. sequence that is independent of the observed sample and has zero mean, unit variance, and a finite fourth moment, i.e., $\mathbb{E}[(\eta_i^*)^4] < \infty$, such as the Rademacher or Mammen's two-point distribution. The proof uses Gaussianity in two places: the conditional Gaussian limit of the multiplier sums, and the value $\operatorname{var}(\eta_i^{*2})=2$ in the conditional variance of the diagonal terms. For such weights, the conditional Lindeberg--Feller theorem replaces the first, the constant $2$ becomes $\mathbb{E}[(\eta_i^*)^4]-1$, and the remaining bounds continue to hold. We nevertheless maintain the standard normal assumption to ensure a unified treatment across both asymptotic scenarios, because the proof of Theorem~\ref{thm:bootstrap} for growing $p$ uses Gaussian weights.
}
\end{remark}

Under the conditions of Theorem~\ref{thm:null_fixed_p_bootstrap}, if in addition $\sigma^2\operatorname{tr}(\boldsymbol{\Sigma}_p)>0$, with $\boldsymbol{\Sigma}_p$ defined in Proposition~\ref{prop:fixed_p_feature}, then $\mathcal L_{0,p}$ is non-degenerate with a continuous distribution function, and Theorems~\ref{thm:null_fixed_p} and \ref{thm:null_fixed_p_bootstrap} imply that the test rejecting $H_0$ when $T_n$ exceeds the $(1-\alpha)$ quantile of the bootstrap distribution of $T_n^*$ has asymptotic rejection probability $\alpha$ under $H_0$. In the growing-$p$ regime, the corresponding conditions $\sigma^2>0$ and $\operatorname{tr}(\boldsymbol{\Sigma}_\infty)>0$ are imposed by Conditions (B1) and (B4) in Assumption~\ref{assum:1}, so Theorems~\ref{thm:null} and \ref{thm:bootstrap} imply the same conclusion.

\begin{assumption}\label{assum:3}
The bootstrap weights $\{\eta_i^*\}$ are i.i.d. standard normal variables and independent of the observed sample.
\end{assumption}
\begin{theorem}\label{thm:bootstrap}
Under the null hypothesis and Assumptions~\ref{assum:1} and \ref{assum:3}, we have
\[
\sup_t\left|P^*\left(T_n^*\leq t\right)-P(\mathcal L_{0,\infty}\leq t)\right|\rightarrow^{p}0,
\]
where $P^*$ denotes the conditional probability given the observed sample.
\end{theorem}

\section{Asymptotic Power under Alternatives}
\label{sec:power}
A useful conditional mean independence test should detect departures from the null hypothesis. We therefore study the asymptotic power of the proposed test, which we refer to as the CMDD test, under fixed and local alternatives in both the fixed-$p$ and growing-$p$ regimes. Since conditional mean independence testing asks whether $\mathbf{X}$ contributes to the mean regression function after adjusting for $\mathbf{Z}$, in the spirit of classical partial-effect testing, we consider the following sequence of alternatives:
\begin{equation}\label{eq:local_alt}
H_{1,n}: Y = f(\mathbf{Z}) + c_n \Delta(\mathbf{X}, \mathbf{Z}) + \varepsilon,
\end{equation}
where $\Delta(\mathbf{X}, \mathbf{Z})$ is a non-trivial measurable function satisfying $\mathbb{E}[\Delta(\mathbf{X}, \mathbf{Z}) \mid \mathbf{Z}] = 0$ almost surely, and the error $\varepsilon$ satisfies $\mathbb{E}[\varepsilon]=0$ and is independent of $(\mathbf{X},\mathbf{Z})$. Thus, $\mathbb{E}[\Delta(\mathbf{X}, \mathbf{Z}) \boldsymbol{\phi}_p(\mathbf{Z})] = {\bf 0}$, so that the sieve projection captures only $f(\mathbf{Z})$. The sequence $c_n$ determines the strength of the alternative: a constant value, such as $c_n = 1$, gives a fixed alternative, whereas $c_n \to 0$ gives a local alternative.
With independent copies denoted by primes, define
\[
C(\mathbf V)=-\varepsilon\mathbb E[U(\mathbf W,\mathbf W')\Delta(\mathbf W')\mid\mathbf W],\quad
\boldsymbol\nu_\Delta=2\mathbb E[U(\mathbf W,\mathbf W')\Delta(\mathbf W)\breve{\boldsymbol\phi}_p(\mathbf Z')],
\]
\[
\begin{aligned}
m_\Delta&=\mathbb E[\Delta(\mathbf W)\Phi(\mathbf W)],\\
\mu_\Delta&=\|m_\Delta\|_{\mathcal H}^2={\rm MDD}^2(\Delta(\mathbf W)\mid\mathbf W)>0,
\end{aligned}
\]
where $\mathbf V=(\mathbf W,\varepsilon)$; that is, throughout this section $\varepsilon$ plays the role of the error $e$ in Section~\ref{sec:notation}. The kernel $h$, the function $K_p$, and the null-reference limit $\mathcal L_{0,p}$, $\mathcal L_{0,\infty}$ in Theorems~\ref{thm:power_local_fixed_p}, \ref{thm:power_local_growing_p}, \ref{thm:power_local_boot}, \ref{thm:power_local_boot_growing_p} are formed using $\varepsilon$.

For fixed $p$, with $h_p$ defined as in Proposition~\ref{prop:fixed_p_feature}, we have $m_\Delta=\mathbb E[\Delta(\mathbf W)h_p(\mathbf W)]$ because $\mathbb E[\Delta(\mathbf W)\mid\mathbf Z]=0$. In the growing-$p$ setting, $m_\Delta=\mathbb E[\Delta(\mathbf W)h_\infty(\mathbf W)]$ follows from the same orthogonality and the $L^2$ convergence of $h_p$ to $h_\infty$, together with $\mathbb E[\Delta(\mathbf W)^2]<\infty$ as assumed in Theorem~\ref{thm:power_local_growing_p}.

\begin{theorem}\label{thm:power_local_fixed_p}
Under the alternative hypothesis in \eqref{eq:local_alt}, Assumption~\ref{assum:0} with $\varepsilon$ in place of $e$, and the additional assumptions that
\[
\E[\Delta(\mathbf{W})^4]<\infty
\quad\text{and}\quad
\mu_\Delta>0,
\]
the following statements hold:
\begin{enumerate}
\item[(i)] If $\sqrt{n}|c_n|\rightarrow\infty$ and $\sup_n|c_n|<\infty$, then, as $n \to \infty$,
\begin{equation*}
n\widehat{\rm{MDD}}^2(\widehat{e} \mid \mathbf{W}) \xrightarrow{p} \infty.
\end{equation*}

\item[(ii)] If $\sqrt n\,c_n\to\tau\in\mathbb R$, then, as $n \to \infty$,
\begin{equation*}
n\widehat{\rm{MDD}}^2(\widehat{e} \mid \mathbf{W})
\rightarrow^{D} 
\mathcal{L}_{p,\tau}:=\mathcal L_{0,p}
+\tau\mathcal C_\Delta+\tau\mathcal D_\Delta+\tau^2\mu_\Delta.
\end{equation*}

\item[(iii)] If $\sqrt{n}c_n\rightarrow 0$, then, as $n \to \infty$,
\begin{equation*}
n\widehat{\rm{MDD}}^2(\widehat{e} \mid \mathbf{W}) \rightarrow^{D} \mathcal{L}_{0,p}.
\end{equation*}
\end{enumerate}
Here
\[
\mathcal{C}_\Delta \sim N\{0, 4 \text{var}({C}(\mathbf{V}))\},
\qquad
\mathcal{D}_\Delta \sim N(0, \sigma^2\|\boldsymbol{\nu}_\Delta\|^2),
\]
with $C(\mathbf{V})$ and $\boldsymbol\nu_\Delta$ defined above. The variables $\{G_k\}$, $\bm{\mathcal{N}}$, $\bm{\mathcal{T}}$, $\mathcal{C}_\Delta$, and $\mathcal{D}_\Delta$ are jointly Gaussian with covariance defined in \eqref{eq:MCLT_alternative} of the Appendix.
\end{theorem}

\begin{theorem}\label{thm:power_local_growing_p}
Under the alternative hypothesis in \eqref{eq:local_alt}, Assumption~\ref{assum:1} with $\varepsilon$ replacing $e$, and $\mathbb{E}[\Delta(\mathbf{W})^2]<\infty$, the following statements hold:
\begin{enumerate}
\item[(i)] If $\sqrt n\,|c_n|\to\infty$ and $\sup_n|c_n|<\infty$, then, as $n \to \infty$,
\[
n\widehat{\rm{MDD}}^2(\widehat{e} \mid \mathbf{W})/(nc_n^2)\rightarrow^{p}\mu_\Delta .
\]
\item[(ii)] If $\sqrt n\,c_n\to\tau\in\mathbb R$, then, as $n \to \infty$,
\[
n\widehat{\rm{MDD}}^2(\widehat{e} \mid \mathbf{W})\rightarrow^{D}\mathcal L_{\infty,\tau}
:=\|\mathcal G_\infty+\tau m_\Delta\|_{\mathcal H}^2-\sigma^2\mu_D .
\]
\item[(iii)] If $\sqrt n\,c_n\to0$, then, as $n \to \infty$,
\[n\widehat{\rm{MDD}}^2(\widehat{e} \mid \mathbf{W})\rightarrow^{D}\mathcal L_{0,\infty}.\]
\end{enumerate}
\end{theorem}
Theorem~\ref{thm:power_local_growing_p} shows that the test retains non-trivial power against local alternatives of order $n^{-1/2}$ even when the number of basis functions diverges. This property is shared by other non-smoothing (global) tests, such as the empirical-process-based significance test of \citet{delgado2001significance} and the series-based specification tests of \citet{li2003consistent}. In contrast, smoothing-based tests, such as those of \citet{fan1996consistent} and \citet{ait2001goodness}, cannot detect alternatives at this rate, although they can be more powerful against high-frequency alternatives. This reflects the usual trade-off between global tests and local smoothing tests.

It is also worth noting that our test does not require the sample splitting or cross-fitting used by the machine learning-based tests in \citet{cai2024test} and \citet{zhang2025testing}: the full sample is used both to estimate $\mathbb{E}[Y\mid\mathbf{Z}]$ and to compute the test statistic, which keeps the implementation simple.

\begin{theorem}\label{thm:power_local_boot}
Assume that the alternative hypothesis in \eqref{eq:local_alt}, the conditions in Theorem~\ref{thm:power_local_fixed_p}, and Assumption~\ref{assum:2} with $\varepsilon$ in place of $e$ continue to hold. Additionally assume $\mathbb E[\Delta(\mathbf W)^8]<\infty$ and $\sigma^2=\mathbb E[\varepsilon^2]>0$. Let $Q_{(1-\alpha), \mathcal L_{0,p}}$ and $Q_{(1-\alpha),n}^*$ be the $(1-\alpha)$ quantiles of $\mathcal L_{0,p}$ and of the conditional distribution of $T_n^*$ given the sample, respectively. Then, the following statements hold:
\begin{enumerate}
\item[(i)] If $\sqrt{n}|c_n|\rightarrow\infty$ and $\sup_n|c_n|<\infty$, then, as $n \to \infty$,
\begin{equation*}
P\left(n\widehat{\rm{MDD}}^2(\widehat{e} \mid \mathbf{W}) > Q_{(1-\alpha),n}^* \mid H_{1,n}\right) \rightarrow 1.
\end{equation*}

\item[(ii)] If $\sqrt n\,c_n\to\tau\in\mathbb R$, then, as $n \to \infty$,
\begin{equation*}
P\left(n\widehat{\rm{MDD}}^2(\widehat{e} \mid \mathbf{W}) > Q_{(1-\alpha),n}^* \mid H_{1,n}\right)
\rightarrow
P\left(\mathcal{L}_{p,\tau} > Q_{(1-\alpha),\mathcal{L}_{0,p}}\right).
\end{equation*}

\item[(iii)] If $\sqrt{n}c_n\rightarrow 0$, then, as $n \to \infty$,
\begin{equation*}
P\left(n\widehat{\rm{MDD}}^2(\widehat{e} \mid \mathbf{W}) > Q_{(1-\alpha),n}^* \mid H_{1,n}\right)
\rightarrow \alpha.
\end{equation*}
\end{enumerate}
\end{theorem}
The conditions $\sigma^2>0$ and $\mu_\Delta>0$ ensure that $\mathcal L_{0,p}$ is non-degenerate, because $\mu_\Delta=\|\mathbb E[\Delta(\mathbf W)h_p(\mathbf W)]\|_{\mathcal H}^2\leq\mathbb E[\Delta(\mathbf W)^2]\operatorname{tr}(\boldsymbol{\Sigma}_p)$ by the Cauchy--Schwarz inequality, so that $\sigma^2\operatorname{tr}(\boldsymbol{\Sigma}_p)>0$; its distribution function is then continuous and strictly increasing at $Q_{(1-\alpha),\mathcal L_{0,p}}$. When $c_n\to0$, as in parts (ii) and (iii), the proof shows that $T_n^*\rightarrow^{D^*}\mathcal L_{0,p}$, which together with these two properties gives $Q_{(1-\alpha),n}^*\rightarrow^{p}Q_{(1-\alpha),\mathcal L_{0,p}}$; the proof of part (i), which includes fixed alternatives, needs only $Q_{(1-\alpha),n}^*=O_p(1)$. In the growing-$p$ regime, the corresponding nondegeneracy requirements, $\sigma^2>0$ and $\operatorname{tr}(\boldsymbol{\Sigma}_\infty)>0$, are imposed by Conditions (B1) and (B4) of Assumption~\ref{assum:1}.

\begin{theorem}\label{thm:power_local_boot_growing_p}
Assume that the alternative hypothesis in \eqref{eq:local_alt}, the conditions of Theorem~\ref{thm:power_local_growing_p}, and Assumption~\ref{assum:3} continue to hold. Let $Q_{(1-\alpha), \mathcal L_{0,\infty}}$ and $Q_{(1-\alpha),n}^*$ be the $(1-\alpha)$ quantiles of $\mathcal L_{0,\infty}$ and of the conditional distribution of $T_n^*$ given the sample, respectively. Then, the following statements hold:
\begin{enumerate}
\item[(i)] If $\sqrt n\,|c_n|\to\infty$ and $\sup_n|c_n|<\infty$, then, as $n\to \infty$,
\[
P\left(n\widehat{\rm{MDD}}^2(\widehat{e} \mid \mathbf{W})>Q_{(1-\alpha),n}^*\mid H_{1,n}\right)\to1 .
\]
\item[(ii)] If $\sqrt n\,c_n\to\tau\in\mathbb R$, then, as $n\to \infty$,
\[
P\left(n\widehat{\rm{MDD}}^2(\widehat{e} \mid \mathbf{W})>Q_{(1-\alpha),n}^*\mid H_{1,n}\right)\to
P(\mathcal L_{\infty,\tau}>Q_{(1-\alpha),\mathcal L_{0,\infty}}).
\]
\item[(iii)] If $\sqrt n\,c_n\to0$, then, as $n\to \infty$, \[P\left(n\widehat{\rm{MDD}}^2(\widehat{e} \mid \mathbf{W}) > Q_{(1-\alpha),n}^* \mid H_{1,n}\right)
\rightarrow \alpha.
\]
\end{enumerate}
\end{theorem}

\section{Numerical Simulations}
\label{sec:numerical_simulations}
In this section, we evaluate the finite-sample performance of the proposed test (CMDD) in comparison with four existing methods: the LMDD test of \citet{jin2018testing}, the PMDD test of \citet{ParkShaoYao2015EJS}, the smoothing-based goodness-of-fit test (GOFT) of \citet{ait2001goodness}, and the partial mean independence test (PMIT) of \citet{cai2024test}. For the series-based estimation of the conditional mean $f(\cdot)$, we employ an additive basis
\[
\boldsymbol{\phi}_p(\mathbf{z}) = (\phi_1(z_1), \dots, \phi_{p_0}(z_1), \dots, \phi_1(z_q), \dots, \phi_{p_0}(z_q))^\top,
\]
where $\mathbf{z} = (z_1, \dots, z_q)$ and $\{\phi_i\}_{i=1}^{p_0}$ is a set of univariate B-spline basis functions, so that $p_0$ is the number of basis functions for each component of $\mathbf{z}$ and $p=p_0q$ is the total number of basis functions; see Remark~\ref{remark:basis_fct}. Each univariate basis consists of the $p_0$ columns returned by the function \texttt{bs} in the R package \texttt{splines} with its default option \texttt{intercept = FALSE}, and a separate intercept is included in the regression. Since the $p_0+1$ functions of the full B-spline basis sum to one, their columns add up to the intercept column; the option \texttt{intercept = FALSE} drops the first of these functions and thus removes this exact collinearity. In Examples~\ref{ex:1}--\ref{ex:3}, where $q=1$ and hence $p=p_0$, we use cubic B-splines (\texttt{degree = 3}) with \texttt{df} set to $p_0$ and with boundary knots fixed at $\pm1$ in Example~\ref{ex:1} and at $\pm3.5$ in Examples~\ref{ex:2} and~\ref{ex:3}. With no interior knots supplied, \texttt{bs} places the $p_0-3$ interior knots at the sample quantiles of orders $j/(p_0-2)$, $j=1,\dots,p_0-3$, of the observations of $Z$ that lie within the boundary knots. These interior knots depend on the data, whereas our theory treats the basis functions as deterministic (Assumptions~\ref{assum:0} and~\ref{assum:1}), so this implementation is not formally covered by our theoretical results.

In our wild bootstrap procedure, the weights $\{\eta^*_i\}$ are generated from a standard normal distribution, and we set the number of bootstrap replicates to $N_{\rm boot} = 499$. For the permutation-based tests of \citet{jin2018testing} and \citet{ParkShaoYao2015EJS}, we similarly utilize 499 permutations. The GOFT procedure by \citet{ait2001goodness} requires several user-specified parameters, including two bandwidths $(h_1, h_2)$, two kernel functions $K_1(\cdot)$ and $K_2(\cdot)$ for estimating $\mathbb{E}[Y \mid \mathbf{Z}]$ and $\mathbb{E}[Y \mid \mathbf{X}, \mathbf{Z}]$, and a weight function $\mathcal{W}$. We tested several configurations, including those suggested in the original study, and report the best results. Specifically, we considered $(h_1, h_2) = (b_1 n^{-1/4.75}, b_2 n^{-1/4.25})$ with $(b_1, b_2) \in \{(0.75, 0.79), (0.65, 0.5)\}$, Gaussian kernels for $K_1$ and $K_2$, and a weight function $\mathcal{W}(Y, \mathbf{Z}) = \mathbf{1}\{\sqrt{Y^2 + \|\mathbf{Z}\|^2} \leq a\}$ for $a \in \{0.75, 1, 1.25\}$. For the PMIT method \citep{cai2024test}, we report the rejection rates obtained using their power enhancement procedure. The conditional mean functions for PMIT are estimated via extreme gradient boosting \citep{chen2016xgboost} using the \texttt{xgboost} R package, with a data-splitting ratio of 0.8 and the power enhancement parameter set to 1, as suggested by the authors.

Empirical size and power are calculated based on 500 simulation runs for each example at nominal significance levels of $\alpha \in \{10\%, 5\%, 1\%\}$.

\begin{example}\label{ex:1}\rm{
In our first example, we consider a linear data-generating process:
\[
Y = Z + r X + 0.1\varepsilon,
\]
where $Z$ and $X$ are independent random variables that each follow a uniform distribution on $[-1, 1]$, $\varepsilon \sim N(0, 1)$ is the error term, and $r \in \{0, 0.1, 0.25, 0.5\}$ controls the degree of dependence. When $r = 0$, the null hypothesis holds ($Y$ is conditionally mean independent of $X$ given $Z$), allowing us to assess the empirical size of the tests. The results are summarized in Table~\ref{table:size_ex1}.
}
\end{example}

\begin{table}[!htbp]
\centering
\caption{Empirical sizes for Example \ref{ex:1}, where the results are obtained based on 500 replicates.}\label{table:size_ex1}
\footnotesize\setlength{\tabcolsep}{3.5pt}%
\begin{tabular}{@{}ccccccccc@{}}
\toprule
\multirow{2}{*}{\textbf{Sample Size}} & \multirow{2}{*}{\boldmath$\alpha$} & \multicolumn{3}{c}{\textbf{CMDD}} & \textbf{LMDD} & \textbf{PMDD} & \textbf{GOFT} & \textbf{PMIT} \\
\cmidrule(lr){3-5}
& & $p_0=10$ & $p_0=25$ & $p_0=50$ & & & & \\
\midrule
\multirow{3}{*}{$n=100$} 
& 10\% & 9.0  & 9.4  & 11.6 & 10.2 & 10.2 & 21.2 & 11.2 \\
& 5\%  & 5.0  & 6.4  & 7.2  & 5.4  & 5.0  & 7.0  & 6.8  \\
& 1\%  & 2.2  & 1.6  & 2.8  & 1.0  & 1.0  & 1.4  & 1.4  \\
\midrule
\multirow{3}{*}{$n=200$} 
& 10\% & 10.2 & 11.2 & 11.2 & 8.6  & 10.4 & 12.8 & 12.6 \\
& 5\%  & 5.4  & 5.6  & 5.8  & 5.2  & 7.0  & 2.8  & 7.2  \\
& 1\%  & 1.6  & 1.6  & 1.6  & 0.8  & 1.6  & 0.4  & 2.4  \\
\bottomrule
\end{tabular}

\end{table}

\begin{table}[!htbp]
\centering
\caption{Empirical powers for Example \ref{ex:1}, where the results are obtained based on 500 replicates.}
\label{table:power_ex1}
\footnotesize\setlength{\tabcolsep}{3.5pt}%
\begin{tabular}{@{}cccccccccc@{}}
\toprule
\textbf{Sample Size} & \textbf{$r$} & \boldmath$\alpha$ & \multicolumn{3}{c}{\textbf{CMDD}} & \textbf{LMDD} & \textbf{PMDD} & \textbf{GOFT} & \textbf{PMIT} \\
\cmidrule(lr){4-6}
& & & $p_0=10$ & $p_0=25$ & $p_0=50$ & & & & \\
\midrule
\multirow{9}{*}{$n=100$} 
& \multirow{3}{*}{0.1} 
  & 10\% & 100.0 & 99.8 & 99.8 & 100.0 & 52.8 & 100.0 & 29.6 \\
& & 5\%  & 100.0 & 99.8 & 97.6 & 100.0 & 49.0 & 100.0 & 12.0 \\
& & 1\%  & 99.4  & 98.6 & 92.8 & 99.6  & 38.8 & 98.0  & 2.0  \\
\cmidrule(lr){2-10}
& \multirow{3}{*}{0.25} 
  & 10\% & 100.0 & 100.0 & 100.0 & 100.0 & 86.6 & 100.0 & 89.2 \\
& & 5\%  & 100.0 & 100.0 & 100.0 & 100.0 & 85.4 & 100.0 & 70.6 \\
& & 1\%  & 100.0 & 100.0 & 100.0 & 100.0 & 81.6 & 100.0 & 23.4 \\
\cmidrule(lr){2-10}
& \multirow{3}{*}{0.5} 
  & 10\% & 100.0 & 100.0 & 100.0 & 100.0 & 99.8 & 100.0 & 99.8 \\
& & 5\%  & 100.0 & 100.0 & 100.0 & 100.0 & 99.8 & 100.0 & 97.2 \\
& & 1\%  & 100.0 & 100.0 & 100.0 & 100.0 & 99.8 & 100.0 & 70.4 \\
\midrule
\multirow{9}{*}{$n=200$} 
& \multirow{3}{*}{0.1} 
  & 10\% & 100.0 & 100.0 & 100.0 & 100.0 & 64.4 & 100.0 & 92.0 \\
& & 5\%  & 100.0 & 100.0 & 100.0 & 100.0 & 59.8 & 100.0 & 69.4 \\
& & 1\%  & 100.0 & 100.0 & 100.0 & 100.0 & 48.8 & 100.0 & 20.8 \\
\cmidrule(lr){2-10}
& \multirow{3}{*}{0.25} 
  & 10\% & 100.0 & 100.0 & 100.0 & 100.0 & 96.8 & 100.0 & 100.0 \\
& & 5\%  & 100.0 & 100.0 & 100.0 & 100.0 & 96.4 & 100.0 & 100.0 \\
& & 1\%  & 100.0 & 100.0 & 100.0 & 100.0 & 95.0 & 100.0 & 99.2 \\
\cmidrule(lr){2-10}
& \multirow{3}{*}{0.5} 
  & 10\% & 100.0 & 100.0 & 100.0 & 100.0 & 100.0 & 100.0 & 100.0 \\
& & 5\%  & 100.0 & 100.0 & 100.0 & 100.0 & 100.0 & 100.0 & 100.0 \\
& & 1\%  & 100.0 & 100.0 & 100.0 & 100.0 & 100.0 & 100.0 & 100.0 \\
\bottomrule
\end{tabular}
\end{table}
Note that $\mathbb E[Y\mid Z]$ is linear in this example, which aligns with the structural assumptions in \citet{jin2018testing}. Tables~\ref{table:size_ex1}--\ref{table:power_ex1} summarize the empirical sizes and powers for Example~\ref{ex:1}. We observe that the proposed CMDD test, along with the LMDD and PMDD tests, generally has smaller size distortions than GOFT in these settings. In contrast, GOFT over-rejects at the $\alpha = 10\%$ level, and PMIT tends to over-reject mildly, particularly at the $\alpha = 10\%$ and $5\%$ significance levels. In terms of power, CMDD, GOFT, and LMDD demonstrate comparably high empirical power. As $n$ increases from 100 to 200, the powers of PMDD and PMIT increase, whereas CMDD, LMDD, and GOFT already have power of at least 92.8\% at $n=100$.

\begin{example}\label{ex:2}\rm{
In this example, we consider a case where $Y$ depends on $Z$ nonlinearly, while $Z$ and $X$ are linearly related. Specifically, the data are generated according to:
\[
Y = Z^2 - 1 + r X + 0.1\varepsilon,
\]
where $(X, Z)$ follows a bivariate normal distribution with mean zero, unit variances, and a correlation coefficient of 0.25. The error term $\varepsilon$ is drawn from a standard normal distribution, and $r \in \{0, 0.1, 0.25, 0.5\}$.
}\end{example}

\begin{table}[!htbp]
\centering
\caption{Empirical sizes for Example \ref{ex:2}, where the results are obtained based on 500 replicates.}
\label{table:size_ex2}
\footnotesize\setlength{\tabcolsep}{3.5pt}%
\begin{tabular}{@{}ccccccccc@{}}
\toprule
\multirow{2}{*}{\textbf{Sample Size}} & \multirow{2}{*}{\boldmath$\alpha$} & \multicolumn{3}{c}{\textbf{CMDD}} & \textbf{LMDD} & \textbf{PMDD} & \textbf{GOFT} & \textbf{PMIT} \\
\cmidrule(lr){3-5}
& & $p_0=10$ & $p_0=25$ & $p_0=50$ & & & & \\
\midrule
\multirow{3}{*}{$n=100$} 
& 10\% & 8.8  & 9.2  & 8.6  & 26.4 & 11.2 & 0.0  & 16.4 \\
& 5\%  & 4.6  & 5.4  & 5.6  & 12.8 & 6.8  & 0.0  & 11.6 \\
& 1\%  & 1.0  & 1.2  & 2.4  & 2.8  & 1.8  & 0.0  & 5.6  \\
\midrule
\multirow{3}{*}{$n=200$} 
& 10\% & 10.4 & 11.6 & 11.0 & 41.2 & 11.0 & 0.2  & 15.0 \\
& 5\%  & 6.2  & 6.4  & 7.2  & 26.4 & 7.0  & 0.0  & 10.2 \\
& 1\%  & 1.8  & 2.4  & 2.4  & 7.6  & 2.8  & 0.0  & 5.6  \\
\bottomrule
\end{tabular}
\end{table}

\begin{table}[!htbp]
\centering
\caption{Empirical powers for Example \ref{ex:2}, where the results are obtained based on 500 replicates.}
\label{table:power_ex2}
\footnotesize\setlength{\tabcolsep}{3.5pt}%
\begin{tabular}{@{}cccccccccc@{}}
\toprule
\textbf{Sample Size} & \textbf{$r$} & \boldmath$\alpha$ & \multicolumn{3}{c}{\textbf{CMDD}} & \textbf{LMDD} & \textbf{PMDD} & \textbf{GOFT} & \textbf{PMIT} \\
\cmidrule(lr){4-6}
& & & $p_0=10$ & $p_0=25$ & $p_0=50$ & & & & \\
\midrule
\multirow{9}{*}{$n=100$} 
& \multirow{3}{*}{0.1} 
  & 10\% & 100.0 & 100.0 & 100.0 & 34.2 & 18.2 & 3.6  & 27.6 \\
& & 5\%  & 100.0 & 100.0 & 100.0 & 19.8 & 12.0 & 1.0  & 15.0 \\
& & 1\%  & 100.0 & 100.0 & 100.0 & 7.4  & 5.8  & 0.2  & 5.4  \\
\cmidrule(lr){2-10}
& \multirow{3}{*}{0.25} 
  & 10\% & 100.0 & 100.0 & 100.0 & 69.8 & 58.2 & 90.8 & 66.0 \\
& & 5\%  & 100.0 & 100.0 & 100.0 & 58.8 & 48.2 & 83.4 & 51.0 \\
& & 1\%  & 100.0 & 100.0 & 100.0 & 35.2 & 31.0 & 66.0 & 17.0 \\
\cmidrule(lr){2-10}
& \multirow{3}{*}{0.5} 
  & 10\% & 100.0 & 100.0 & 100.0 & 99.0 & 95.8 & 100.0 & 90.2 \\
& & 5\%  & 100.0 & 100.0 & 100.0 & 97.0 & 93.6 & 100.0 & 84.4 \\
& & 1\%  & 100.0 & 100.0 & 100.0 & 91.6 & 89.4 & 100.0 & 59.4 \\
\midrule
\multirow{9}{*}{$n=200$} 
& \multirow{3}{*}{0.1} 
  & 10\% & 100.0 & 100.0 & 100.0 & 93.2 & 62.8 & 21.6 & 60.2 \\
& & 5\%  & 100.0 & 100.0 & 100.0 & 86.0 & 48.0 & 10.6 & 46.8 \\
& & 1\%  & 100.0 & 100.0 & 100.0 & 86.0 & 19.6 & 1.8  & 21.2 \\
\cmidrule(lr){2-10}
& \multirow{3}{*}{0.25} 
  & 10\% & 100.0 & 100.0 & 100.0 & 100.0 & 99.0 & 100.0 & 92.2 \\
& & 5\%  & 100.0 & 100.0 & 100.0 & 99.8 & 98.2 & 100.0 & 90.0 \\
& & 1\%  & 100.0 & 100.0 & 100.0 & 99.6 & 91.6 & 100.0 & 76.8 \\
\cmidrule(lr){2-10}
& \multirow{3}{*}{0.5} 
  & 10\% & 100.0 & 100.0 & 100.0 & 100.0 & 100.0 & 100.0 & 99.8 \\
& & 5\%  & 100.0 & 100.0 & 100.0 & 100.0 & 100.0 & 100.0 & 99.2 \\
& & 1\%  & 100.0 & 100.0 & 100.0 & 100.0 & 100.0 & 100.0 & 96.4 \\
\bottomrule
\end{tabular}
\end{table}
Tables~\ref{table:size_ex2}--\ref{table:power_ex2} present the simulation results for Example~\ref{ex:2}. We observe that CMDD and PMDD control the empirical size reasonably well at the $10\%$ and $5\%$ levels, with some over-rejection at the $1\%$ level. In contrast, LMDD and PMIT exhibit noticeable over-rejection of the null hypothesis. For LMDD, this distortion likely stems from the nonlinear dependence between $Y$ and $Z$, which violates its underlying model assumptions. Conversely, GOFT is strongly conservative, with empirical sizes of at most 0.2\%. Regarding empirical power, our proposed CMDD test rejects in every replicate in all settings, so it matches or exceeds the power of every competing test. This example highlights the robustness and superior power of our approach in the presence of nonlinear conditional mean structures.

\begin{example}\label{ex:3}\rm{
In this example, we consider a setting where the predictor of interest and the conditioning variable are dependent. The data are generated as follows:
\[
Y = Z + r X + U_1, \quad X = Z + U_2,
\]
where $Z, U_1$, and $U_2$ are independent standard normal random variables, and we set $r \in \{0, 0.1, 0.25, 0.5\}$.
}\end{example}

\begin{table}[!htbp]
\centering
\caption{Empirical sizes for Example \ref{ex:3}, where the results are obtained based on 500 replicates.}\label{table:size_ex3}
\footnotesize\setlength{\tabcolsep}{3.5pt}%
\begin{tabular}{@{}ccccccccc@{}}
\toprule
\multirow{2}{*}{\textbf{Sample Size}} & \multirow{2}{*}{\boldmath$\alpha$} & \multicolumn{3}{c}{\textbf{CMDD}} & \textbf{LMDD} & \textbf{PMDD} & \textbf{GOFT} & \textbf{PMIT} \\
\cmidrule(lr){3-5}
& & $p_0=10$ & $p_0=25$ & $p_0=50$ & & & & \\
\midrule
\multirow{3}{*}{$n=100$} 
& 10\% & 8.0  & 7.2  & 7.2  & 2.2  & 35.4 & 13.0 & 31.6 \\
& 5\%  & 3.6  & 5.2  & 5.2  & 0.4  & 27.8 & 8.4  & 16.0 \\
& 1\%  & 1.0  & 1.6  & 2.2  & 0.0  & 16.4 & 2.8  & 4.8  \\
\midrule
\multirow{3}{*}{$n=200$} 
& 10\% & 10.4 & 10.4 & 11.6 & 3.6  & 44.4 & 13.2 & 36.2 \\
& 5\%  & 5.6  & 6.2  & 8.4  & 1.6  & 41.4 & 7.8  & 20.2 \\
& 1\%  & 2.0  & 1.6  & 2.8  & 0.0  & 31.2 & 2.2  & 3.8  \\
\bottomrule
\end{tabular}
\end{table}

\begin{table}[!htbp]
\centering
\caption{Empirical powers for Example \ref{ex:3}, where the results are obtained based on 500 replicates.}
\label{table:power_ex3}
\footnotesize\setlength{\tabcolsep}{3.5pt}%
\begin{tabular}{@{}cccccccccc@{}}
\toprule
\multirow{2}{*}{\textbf{Sample Size}} & \multirow{2}{*}{\textbf{$r$}} & \multirow{2}{*}{\boldmath$\alpha$} & \multicolumn{3}{c}{\textbf{CMDD}} & \textbf{LMDD} & \textbf{PMDD} & \textbf{GOFT} & \textbf{PMIT} \\
\cmidrule(lr){4-6}
& & & $p_0=10$ & $p_0=25$ & $p_0=50$ & & & & \\
\midrule
\multirow{9}{*}{$n=100$} 
& \multirow{3}{*}{0.1} 
& 10\% & 20.2 & 19.0 & 14.0 & 11.0 & 79.0 & 31.6 & 35.4 \\
& & 5\%  & 13.6 & 12.8 & 9.6  & 4.4  & 73.4 & 20.8 & 20.0 \\
& & 1\%  & 5.4  & 4.0  & 4.2  & 1.0  & 57.0 & 10.4 & 6.8 \\
\cmidrule{2-10}
& \multirow{3}{*}{0.25} 
& 10\% & 69.2 & 60.6 & 49.2 & 56.4 & 98.6 & 53.6 & 47.0\\
& & 5\%  & 58.0 & 53.6 & 41.4 & 41.0 & 97.4 & 40.6 & 29.0 \\
& & 1\%  & 36.6 & 31.8 & 24.4 & 14.2 & 93.2 & 25.2 & 10.2\\
\cmidrule{2-10}
& \multirow{3}{*}{0.5} 
& 10\% & 99.6 & 98.8 & 90.8 & 98.8 & 100.0 & 87.4 & 79.2 \\
& & 5\%  & 99.0 & 98.4 & 88.0 & 97.2 & 100.0 & 81.2 & 61.0 \\
& & 1\%  & 96.2 & 90.8 & 78.4 & 84.2 & 100.0 & 68.0 & 30.8 \\
\midrule
\multirow{9}{*}{$n=200$} 
& \multirow{3}{*}{0.1} 
& 10\% & 38.6 & 35.6 & 32.6 & 26.4 & 89.8 & 35.8 & 49.0 \\
& & 5\%  & 30.8 & 26.8 & 25.6 & 14.0 & 86.8 & 26.0 & 30.6\\
& & 1\%  & 12.8 & 14.6 & 14.4 & 4.0  & 81.4 & 12.6 & 8.0 \\
\cmidrule{2-10}
& \multirow{3}{*}{0.25} 
& 10\% & 94.6 & 92.6 & 87.6 & 87.4 & 99.6 & 69.8 & 79.0 \\
& & 5\%  & 89.8 & 87.4 & 83.4 & 78.8 & 99.6 & 58.0 & 60.4 \\
& & 1\%  & 77.2 & 74.8 & 70.4 & 50.4 & 99.6 & 39.8 & 24.4\\
\cmidrule{2-10}
& \multirow{3}{*}{0.5} 
& 10\% & 100.0 & 100.0 & 100.0 & 100.0 & 100.0 & 98.2 & 98.6 \\
& & 5\%  & 100.0 & 100.0 & 100.0 & 100.0 & 100.0 & 96.4 & 95.2 \\
& & 1\%  & 100.0 & 100.0 & 100.0 & 99.6  & 100.0 & 92.2 & 76.2 \\
\bottomrule
\end{tabular}
\end{table}

This example, adapted from \citet{ParkShaoYao2015EJS}, illustrates a scenario where the PMDD test fails to correctly assess the null hypothesis. As noted in Section 4.1 of \citet{ParkShaoYao2015EJS}, their numerical results report a positive value of the population measure under the null hypothesis. Consequently, the empirical sizes of the PMDD test are expected to be inflated. According to the empirical sizes reported in Table~\ref{table:size_ex3}, the PMDD and PMIT tests substantially over-reject the null hypothesis and GOFT over-rejects mildly, whereas the LMDD test tends to be conservative and under-rejects. In contrast, our proposed CMDD test generally has smaller size distortions, although it also over-rejects in some settings. As shown in Table~\ref{table:power_ex3}, PMDD has high alternative rejection rates, but these cannot be interpreted as size-adjusted power comparisons. Conversely, the CMDD test generally achieves higher power than LMDD and, for the larger values of $r$, GOFT.

\begin{example}\label{ex:4}\rm{
In this example, the predictors of interest, $\mathbf{X} = (X_1, \dots, X_5)^\top$, and the conditioning variables, $\mathbf{Z} \in \mathbb{R}^5$, are both multivariate. The response variable $Y$ is generated according to:
\[ 
Y = \|\mathbf{Z}\|^2 + r(X_1 + X_2 + X_3) + 0.1\varepsilon, 
\]
where $r \in \{0, 0.1, 0.25, 0.5\}$, and the components of $\mathbf{Z}$ and $\mathbf{X}$, as well as the error term $\varepsilon$, are independently drawn from a $t$-distribution with 10 degrees of freedom.
}\end{example}

\begin{table}[!htbp]
\centering
\caption{Empirical sizes for Example \ref{ex:4}, where the results are obtained based on 500 replicates.}
\label{table:size_ex4}
\footnotesize\setlength{\tabcolsep}{3.5pt}%
\begin{tabular}{@{}ccccccccc@{}}
\toprule
\multirow{2}{*}{\textbf{Sample Size}} & \multirow{2}{*}{\boldmath$\alpha$} & \multicolumn{3}{c}{\textbf{CMDD}} & \textbf{LMDD} & \textbf{PMDD} & \textbf{GOFT} & \textbf{PMIT} \\
\cmidrule(lr){3-5}
& & $p_0=5$ & $p_0=10$ & $p_0=15$ & & & & \\
\midrule
\multirow{3}{*}{$n=100$}
& 10\% & 10.2 & 9.4 & 8.8 & 32.8 & 22.2 & 19.2 & 13.8 \\
& 5\%  & 6.2  & 7.2 & 6.8 & 20.8 & 12.2 & 0.8  & 8.8  \\
& 1\%  & 2.0  & 3.0 & 3.6 & 6.2  & 4.8  & 0.2  & 4.4  \\
\midrule
\multirow{3}{*}{$n=200$}
& 10\% & 11.2 & 9.0 & 11.2 & 35.4 & 24.4 & 21.0 & 16.0 \\
& 5\%  & 5.8  & 6.4 & 6.4  & 24.6 & 15.6 & 0.0  & 9.8  \\
& 1\%  & 2.0  & 2.2 & 2.2  & 8.4  & 5.0  & 0.0  & 3.2  \\
\bottomrule
\end{tabular}
\end{table}

\begin{table}[!htbp]
\centering
\caption{Empirical powers for Example \ref{ex:4}, where the results are obtained based on 500 replicates.}
\label{table:power_ex4}
\footnotesize\setlength{\tabcolsep}{3.5pt}%
\begin{tabular}{@{}cccccccccc@{}}
\toprule
\multirow{2}{*}{\textbf{Sample Size}} & \multirow{2}{*}{$r$} & \multirow{2}{*}{\boldmath$\alpha$} & \multicolumn{3}{c}{\textbf{CMDD}} & \textbf{LMDD} & \textbf{PMDD} & \textbf{GOFT} & \textbf{PMIT} \\
\cmidrule(lr){4-6}
& & & $p_0=5$ & $p_0=10$ & $p_0=15$ & & & & \\
\midrule
\multirow{9}{*}{$n=100$} 
& \multirow{3}{*}{$0.1$} 
& 10\% & 100.0 & 100.0 & 98.6 & 100.0 & 100.0 & 100.0 & 39.8 \\
& & 5\%  & 100.0 & 100.0 & 98.6 & 100.0 & 100.0 & 100.0 & 23.6 \\
& & 1\%  & 100.0 & 100.0 & 96.8 & 100.0 & 100.0 & 100.0 & 3.8 \\
\cmidrule(lr){2-10}
& \multirow{3}{*}{$0.25$} 
& 10\% & 100.0 & 100.0 & 100.0 & 100.0 & 100.0 & 100.0 & 66.0 \\
& & 5\%  & 100.0 & 100.0 & 100.0 & 100.0 & 100.0 & 100.0 & 40.6 \\
& & 1\%  & 100.0 & 100.0 & 100.0 & 100.0 & 100.0 & 100.0 & 12.8 \\
\cmidrule(lr){2-10}
& \multirow{3}{*}{$0.5$} 
& 10\% & 100.0 & 100.0 & 100.0 & 100.0 & 100.0 & 100.0 & 88.8 \\
& & 5\%  & 100.0 & 100.0 & 100.0 & 100.0 & 100.0 & 100.0 & 76.2 \\
& & 1\%  & 100.0 & 100.0 & 99.8 & 100.0 & 100.0 & 100.0 & 39.0 \\
\midrule
\multirow{9}{*}{$n=200$}
& \multirow{3}{*}{$0.1$} 
& 10\% & 100.0 & 100.0 & 100.0 & 100.0 & 100.0 & 100.0 & 98.2 \\
& & 5\%  & 100.0 & 100.0 & 100.0 & 100.0 & 100.0 & 100.0 & 96.8 \\
& & 1\%  & 100.0 & 100.0 & 100.0 & 100.0 & 100.0 & 100.0 & 77.0 \\
\cmidrule(lr){2-10}
& \multirow{3}{*}{$0.25$}
& 10\% & 100.0 & 100.0 & 100.0 & 100.0 & 100.0 & 100.0 & 99.8 \\
& & 5\%  & 100.0 & 100.0 & 100.0 & 100.0 & 100.0 & 100.0 & 99.8 \\
& & 1\%  & 100.0 & 100.0 & 100.0 & 100.0 & 100.0 & 100.0 & 97.4 \\
\cmidrule(lr){2-10}
& \multirow{3}{*}{$0.5$}
& 10\% & 100.0 & 100.0 & 100.0 & 100.0 & 100.0 & 100.0 & 100.0 \\
& & 5\%  & 100.0 & 100.0 & 100.0 & 100.0 & 100.0 & 100.0 & 100.0 \\
& & 1\%  & 100.0 & 100.0 & 100.0 & 100.0 & 100.0 & 100.0 & 100.0 \\
\bottomrule
\end{tabular}
\end{table}

As shown in Table~\ref{table:size_ex4}, the empirical sizes of the proposed CMDD test are close to the nominal level at 10\%, somewhat above it at 5\% (between 5.8\% and 7.2\%), and more clearly above it at 1\% (between 2.0\% and 3.6\%). In contrast, LMDD, PMDD, and PMIT tend to over-reject the null hypothesis, exhibiting significant size inflation. The empirical power results are summarized in Table~\ref{table:power_ex4}. We observe that the PMIT test shows lower alternative rejection rates compared to the other methods. While the CMDD, LMDD, PMDD, and GOFT tests all achieve powers near unity for every positive value of $r$ considered, the results for LMDD and PMDD are less practically meaningful given their highly inflated empirical sizes. 

Overall, the simulation results indicate that the CMDD test has reasonable size control and good power in many of the scenarios considered, including cases where $Y$ depends on ${\bm{Z}}$ either linearly or nonlinearly, settings with univariate or multivariate predictors, and instances where $\mathbf{X}$ and $\mathbf{Z}$ are dependent. Some over-rejection occurs at the 1\% level, at the 5\% level in Example~\ref{ex:4}, and when the number of basis functions is large relative to the sample size, for example, $p_0=50$ in Example~\ref{ex:1} with $n=100$ and in Example~\ref{ex:3} with $n=200$.

\section{Data Illustration}\label{sec:real data}
In this section, we utilize the World Health Organization (WHO) life expectancy dataset to illustrate the practical utility of our testing methodology for variable selection in regression models. After removing observations with missing values, our dataset consists of life expectancy records for 130 countries from the year 2015. Life expectancy serves as the response variable, while 16 other factors, such as adult mortality and infant deaths, are treated as potential predictors. The dataset is available on Kaggle at \url{https://www.kaggle.com/datasets/kumarajarshi/life-expectancy-who}.

Following the approach of \citet{szekely2014partial} and \citet{ParkShaoYao2015EJS}, we first standardize all variables. We then employ a forward selection procedure using the MDC \citep{ShaoZhang2014JASA} to identify a set of predictors. We assume a generalized additive regression model of the form:
\[
Y = \widetilde{\gamma}_0 + \sum_{i \in S_0} f_i(X_i) + \varepsilon,
\]
where $S_0 \subseteq \{1, \dots, 16\}$ represents the active set of predictors and $\mathbb{E}[f_i(X_i)] = 0$ for all $i \in S_0$. 

We compare our method with the tests proposed by \citet{ParkShaoYao2015EJS}, \citet{jin2018testing}, and \citet{cai2024test} to assess whether an additional predictor provides significant incremental value to the model. Specifically, let $\{\widetilde{X}_1, \dots, \widetilde{X}_{j-1}\}$ be the predictors already selected, and let $\widetilde{X}_j$ be a candidate for inclusion. Our objective is to determine whether $\widetilde{X}_j$ significantly influences the conditional mean of the response. This is equivalent to testing whether the residual $e_{j-1} = Y - \mathbb{E}[Y \mid \widetilde{X}_1, \dots, \widetilde{X}_{j-1}]$ is mean independent of the expanded set $\{\widetilde{X}_1, \dots, \widetilde{X}_j\}$. Formally, we test the null hypothesis:
\[
H_0^{(j)}: {\rm MDD}^2(e_{j-1} \mid \widetilde{X}_1, \dots, \widetilde{X}_j) = 0, \quad j \geq 1,
\]
where $e_0 = Y - \mathbb{E}[Y]$. For $j \geq 2$, we estimate $e_{j-1}$ by the residual $\widehat{e}_{j-1}$ from the sieve regression described in the forward selection procedure below.

We exclude the test by \citet{ait2001goodness} from this comparison as it does not provide a straightforward mechanism for variable selection. In contrast, the methodologies in \citet{ParkShaoYao2015EJS}, \citet{jin2018testing}, and \citet{cai2024test}, alongside our proposed approach, offer standardized dependence measures such as partial MDC and linear MDC. These metrics allow for a direct comparison of the relative importance of different variables. Additionally, as a benchmark for variable selection performance, we apply Group Lasso regularization to a generalized additive model (GAM) containing all variables. This is implemented using five cubic splines per predictor, with the regularization parameter selected via cross-validation.

To evaluate predictive performance, we split the data into a training set ($n=65$) and a testing set. The training set is used to select predictors and fit the model following the configuration described in Section~\ref{sec:numerical_simulations}. For our sieve approximation, we set $p=10$ based on the training sample size; here $p$ is the number of B-spline basis functions for each selected predictor, that is, $p_0$ in the notation of Section~\ref{sec:numerical_simulations}. Sensitivity analysis with $p \in [5, 15]$ yielded consistent results, as detailed in Appendix~\ref{sec:who}.

Subsequently, we apply the fitted regression model to the $n_{\text{test}} = 65$ observations in the testing set to evaluate its predictive performance. The prediction accuracy is measured using the Prediction Mean Squared Error (PMSE), defined as:
\[ 
\text{PMSE} = \frac{1}{n_{\text{test}}} \sum_{i=1}^{n_{\text{test}}} (Y_i - \widehat{Y}_i)^2, 
\]
where $Y_i$ is the observed life expectancy and $\widehat{Y}_i$ is the value predicted by the model trained on the training set. To identify the set of predictors for the additive regression model, we implement the following forward selection procedure:
\begin{enumerate}
    \item \textbf{Initial Selection:} Identify the first predictor, $\widetilde{X}_1$, that exhibits the largest MDC with the response in the training set:
    \[
    \widetilde{X}_1 = \underset{X_i \in \mathcal{X}}{\text{argmax}} \, \widehat{\text{MDC}}^2(Y \mid X_i),
    \]
    where $\mathcal{X} = \{X_1, \dots, X_{16}\}$. Let $S=\varnothing$. We perform the mean independence test of \citet{lee2020testing} with $N_{\rm boot}=499$ bootstrap replicates and standard normal weights $\{\eta_i^*\}$ to test $H_0^{(1)}: \mathbb{E}[Y \mid \widetilde{X}_1] = \mathbb{E}[Y]$ almost surely. If the resulting $p$-value is less than $0.05$, we reject the null hypothesis, update $S=\{\widetilde X_1\}$, and proceed to select additional predictors; otherwise, the selection process terminates.  
    \item \textbf{Candidate Identification:} To select the $j$-th predictor (for $j \geq 2$), we define $\widetilde{X}_j$ as:
    \[
    \widetilde{X}_j = \underset{X_i \in \mathcal{X} \setminus S}{\text{argmax}} \, \widehat{\text{MDC}}^2(\widehat{e}_{j-1} \mid \widetilde{X}_1, \dots, \widetilde{X}_{j-1}, X_i).
    \]
    Here, $\widehat{e}_{j-1}$ represents the residuals from the sieve approximation detailed in Section~\ref{sec:numerical_simulations} using the previously selected predictors. For $j=2$, $\widehat{e}_1 = Y - \widehat\gamma_{1}- \widehat{\bm{\beta}}_1^\top \boldsymbol{\phi}_p(\widetilde{X}_1)$ with $p=10$.

    \item \textbf{Hypothesis Testing:} We apply the bootstrap-based test described in Section~\ref{sec:bootstrap}, with $N_{\rm boot}=499$ and standard normal weights, to compute the $p$-value for $H_0^{(j)}$. The $p$-value is calculated as $\frac{1}{N_{\rm boot}}\sum_{b=1}^{N_{\rm boot}} \mathbf{1}(T_{n, b}^* \geq T_n)$. If the $p$-value is below $0.05$, we update the active set $S = S \cup \{\widetilde{X}_j\}$ and return to Step 2.

    \item \textbf{Iteration:} This process is repeated for $H_0^{(j)}$ ($j \geq 3$), where the residual $\widehat{e}_{j-1} = \widehat{e}_{j-2} -\widehat{\gamma}_{j-1}-\widehat{\bm{\beta}}_{j-1}^\top \boldsymbol{\phi}_p(\widetilde{X}_{j-1})$ is obtained by filtering the previous residual $\widehat{e}_{j-2}$ through a sieve approximation using $\widetilde{X}_{j-1}$.

    \item \textbf{Termination:} The procedure continues until the test fails to reject the null hypothesis $H_0^{(k)}$ at the $5\%$ level. The final model then consists of the $k-1$ previously selected predictors.
\end{enumerate}
Because each candidate is chosen to maximize an estimated dependence measure before it is tested, the $5\%$ thresholds in this procedure act as a stopping rule rather than as selection-adjusted error control, and the procedure should be regarded as exploratory.

The selected sets of predictors identified by each method are summarized in Table~\ref{table:variable selection}.

\begin{table}[!htbp]
\centering
\caption{Selected predictors by different methods.}
\label{table:variable selection}
\begin{tabular}{@{}lp{0.78\textwidth}@{}}
\toprule
\textbf{Method} & \textbf{Selected Predictors} \\
\midrule
\textbf{CMDD} & Income composition of resources, Adult mortality, BMI, Hepatitis B \\
\textbf{PMDD} & Income composition of resources, Adult mortality, Polio, BMI, HIV/AIDS, Schooling, Thinness 1–19 years \\
\textbf{LMDD} & Income composition of resources, Adult mortality, Hepatitis B, Polio \\
\textbf{PMIT} & Income composition of resources \\
\textbf{Group Lasso} & Income composition of resources, Adult mortality, BMI, Under five deaths, Polio, Diphtheria, Thinness 1–19 years, Thinness 5–9 years, Schooling \\
\bottomrule
\end{tabular}
\end{table}
All four testing approaches agree that \textit{Income Composition of Resources} is the primary predictor of life expectancy. However, the methods diverge regarding the total number of predictors selected for the regression model. Both the proposed CMDD test and the LMDD test identify four predictors, though they differ on one specific variable. In contrast, the PMDD approach selects seven predictors, the Group Lasso method selects nine, and the PMIT test identifies only a single predictor.

After determining the selected set of predictors for each method, we fit a GAM using the training data. Following the implementation in \citet{ParkShaoYao2015EJS}, we utilize the \texttt{gam} function from the \texttt{mgcv} R package, employing five cubic splines for each selected predictor. 

Table~\ref{table:prediction} summarizes the PMSE results for each method. For this training--testing split, the model selected by CMDD attains the smallest PMSE, while using fewer predictors than the PMDD and Group Lasso models. Because these results are based on a single split of a moderately sized dataset, they should be regarded as illustrative rather than as a definitive comparison of predictive performance.
\begin{table}[!htbp]
\centering
\caption{Prediction accuracy for five methods.}
\label{table:prediction}
\begin{tabular}{@{}lccccc@{}}
\toprule
& \textbf{CMDD} & \textbf{PMDD} & \textbf{LMDD} & \textbf{PMIT} & \textbf{Group Lasso} \\
\midrule
PMSE & 0.0861 & 0.1162 & 0.1229 & 0.1696 & 0.1337\\
\bottomrule
\end{tabular}
\end{table}

\section{Discussion and Conclusions}
\label{sec:conclusion}
In this article, we have proposed a novel nonparametric test for conditional mean independence between a response $Y$ and a set of predictors $\mathbf{X}$ given the conditioning variables $\mathbf{Z}$. Our approach utilizes a sieve approximation to filter out the effect of $\mathbf{Z}$ from $Y$, followed by the application of MDD to measure the remaining mean dependence on $(\mathbf X,\mathbf Z)$. We derived non-standard limiting distributions under both the null and alternative hypotheses when the conditional mean $\mathbb{E}[Y \mid \mathbf{Z}]$ follows a finite-dimensional model with fixed $p$, or is approximated by a sieve expansion of increasing complexity with $p\to\infty$.
To facilitate practical inference, we introduced a wild bootstrap procedure and established its asymptotic consistency in both regimes. In our numerical studies, the proposed CMDD test performed comparably to or better than existing methods in terms of empirical size and power in the settings considered. Compared to the smoothing-based test of \citet{ait2001goodness} and the machine-learning-based test of \citet{cai2024test}, our procedure is simpler to implement: it requires fewer tuning parameters and, unlike the latter, no data splitting. Furthermore, unlike the permutation-based approaches of \citet{ParkShaoYao2015EJS} and \citet{jin2018testing}, our bootstrap test is theoretically justified, and the population MDD on which it is based fully characterizes conditional mean independence without parametric assumptions on the regression function. Our theory does, however, assume that the error $e$ is independent of $\mathbf{W}$, and extending it to conditionally heteroskedastic errors is an interesting topic for future research.

We conclude by suggesting several promising directions for future research. First, extending our framework to evaluate conditional mean independence for a multivariate response $\mathbf{Y}$ would be highly valuable. This would necessitate the development of multivariate sieve models to effectively remove the influence of $\mathbf{Z}$ from $\mathbf{Y}$ before assessing dependence. Second, another important extension involves testing for conditional quantile independence, defined as $Q_\theta(Y \mid \mathbf{X}, \mathbf{Z}) = Q_\theta(Y \mid \mathbf{Z})$ almost surely, where $Q_\theta$ denotes the $\theta$-th conditional quantile. Developing a robust test for this hypothesis would provide a more comprehensive understanding of the conditional relationship across different parts of the response distribution. Finally, establishing a rigorous data-driven method for choosing $p$ that maintains the asymptotic guarantees of our test is an important direction for future work.

\bibliographystyle{agsm}
\bibliography{CMI_ref_revise}

\newpage
\appendix
\setcounter{equation}{0}
\renewcommand{\theequation}{\thesection.\arabic{equation}}
\renewcommand{\theHequation}{\thesection.\arabic{equation}}
\setcounter{table}{0}
\renewcommand{\thetable}{\thesection.\arabic{table}}
\renewcommand{\theHtable}{\thesection.\arabic{table}}
\section{Technical Proofs}
This appendix contains the proofs of Lemmas~\ref{lemma:sim1}--\ref{lemma:boostrap_remainder_alt} and Theorems~\ref{thm:null_fixed_p}--\ref{thm:power_local_boot_growing_p}; additional results for the real data illustration are given in Appendix~\ref{sec:who}. Throughout, $n\widehat{\rm MDD}^{2*}(\widehat e\mid\mathbf W)$ denotes the bootstrap statistic $T_n^*=n\widehat{\rm MDD}^{2}(\widehat e^*\mid\mathbf W)$ defined in Section~\ref{sec:bootstrap} of the main text.

As in the main text, $P^*$ and $\E^*$ denote the probability and expectation conditional on the observed sample. For a random quantity $X_n^*$ that depends on the sample and on the bootstrap weights $\{\eta_i^*\}_{i=1}^n$, and for positive constants $a_n$, we write $X_n^*=o_p^*(a_n)$ if $P^*(|X_n^*|>\delta a_n)\rightarrow^{p}0$ for every $\delta>0$, and $X_n^*=O_p^*(a_n)$ if for every $\delta>0$ there exist $K_\delta<\infty$ and $n_\delta$ such that $P\{P^*(|X_n^*|>K_\delta a_n)>\delta\}<\delta$ for all $n\ge n_\delta$; when $X_n^*$ is a vector, a matrix, or an element of $\mathcal H$, $|\cdot|$ is replaced by the relevant norm. 

Consequently, a quantity that depends only on the sample and is $O_p(a_n)$ (respectively, $o_p(a_n)$) is also $O_p^*(a_n)$ (respectively, $o_p^*(a_n)$), and, by Markov's inequality, $\E^*|X_n^*|^2=O_p(a_n^2)$ (respectively, $o_p(a_n^2)$) implies $X_n^*=O_p^*(a_n)$ (respectively, $o_p^*(a_n)$). For a constant $c$, we write $X_n^*\rightarrow^{p^*}c$ if $X_n^*-c=o_p^*(1)$.

For fixed \(p\), following the argument in \citet[Chapter 15.6]{li2007nonparametric}, we work with centered and orthonormalized basis functions without loss of generality. The transformed basis in Condition (A2) of Assumption~\ref{assum:0}
\[
\breve{\boldsymbol{\phi}}_p(\mathbf Z)
=
\mathbf Q^{-1/2}
\{\boldsymbol{\phi}_p(\mathbf Z)-\E[\boldsymbol{\phi}_p(\mathbf Z)]\}
\]
satisfies
\[
\mathbb \E[\breve{\boldsymbol{\phi}}_p(\mathbf Z)]
=\mathbf 0,
\qquad
\mathbb \E[
\breve{\boldsymbol{\phi}}_p(\mathbf Z)
\breve{\boldsymbol{\phi}}_p(\mathbf Z)^\top
]
=\mathbf I_p.
\]
Since \(\mathbf Q^{-1/2}\) is nonsingular, this transformation preserves the linear span of the basis functions after the intercept is included. Consequently, the corresponding population and empirical least-squares projection spaces remain unchanged. Therefore, without loss of generality, we assume \(\mathbb E[\boldsymbol{\phi}_p(\mathbf Z)]=\mathbf 0\) and \(\mathbf Q=\mathbf I_p\), so that \(\breve{\boldsymbol{\phi}}_p=\boldsymbol{\phi}_p\), in Assumptions~\ref{assum:0} throughout the proofs.

For the feature-space arguments in both regimes, recall that $\Phi:\mathbb R^{s+q}\to\mathcal H$ satisfies
\begin{equation}\label{grow:feature_supp}
\langle\Phi(\mathbf w),\Phi(\mathbf v)\rangle_{\mathcal H}
=\|\mathbf w\|+\|\mathbf v\|-\|\mathbf w-\mathbf v\|,
\qquad \|\Phi(\mathbf w)\|_{\mathcal H}^2=2\|\mathbf w\|.
\end{equation}
Write $m_\Phi=\E\Phi(\mathbf W)$ and $\mu_D=\E\|\mathbf W-\mathbf W'\|$, where $\mathbf W'$ is an independent copy of $\mathbf W$. For growing $p$, let $\mathbf g_p=(g_1,\ldots,g_p)^\top$ be the centered orthonormal basis in Condition (B2); for fixed $p$, set $\mathbf g_p=\breve{\boldsymbol\phi}_p$. The operator $\mathcal B_p:\mathbb R^p\to\mathcal H$ and the residual feature are defined as
\[
\mathcal B_p\mathbf u
= \sum_{j=1}^p u_j\beta_j^\Phi,
\qquad
h_p(\mathbf W)=\Phi(\mathbf W)-m_\Phi-\mathcal B_p\mathbf{g}_{p}(\mathbf Z),
\]
where $\beta_j^\Phi=\mathbb E[(\Phi(\mathbf W)-m_\Phi){g}_{j}(\mathbf Z)]$.

\subsection{Proofs for limiting null distributions}
We first collect four lemmas used below. We then prove the fixed-$p$ and growing-$p$ null limits, establish the feature representation of the fixed-$p$ limit, and prove bootstrap validity in the same two regimes. The proofs follow the order of the results in the main text.

\noindent\textit{Notation for the fixed-$p$ proofs.} We write $\boldsymbol\phi_{p,k}=\boldsymbol\phi_p(\mathbf Z_k)$, $\overline{\boldsymbol\phi}_p=n^{-1}\sum_{k=1}^n\boldsymbol\phi_{p,k}$, and $\widehat{\mathbf Q}=n^{-1}\sum_{k=1}^n(\boldsymbol\phi_{p,k}-\overline{\boldsymbol\phi}_p)(\boldsymbol\phi_{p,k}-\overline{\boldsymbol\phi}_p)^\top$ for the sample covariance matrix of the basis. For a sequence $a=(a_1,\ldots,a_n)$, which may involve the bootstrap weights, let
\[
\begin{aligned}
\boldsymbol\xi(a)&=\widehat{\mathbf Q}^{-1}\frac{1}{\sqrt n}\sum_{k=1}^n(\boldsymbol\phi_{p,k}-\overline{\boldsymbol\phi}_p)a_k,\\
\boldsymbol\xi^\circ(a)&=\frac{1}{\sqrt n}\sum_{k=1}^n\boldsymbol\phi_{p,k}a_k,
\end{aligned}
\]
where products of sequences are taken entrywise; for example, $\boldsymbol\xi(e\eta^*)$ is $\boldsymbol\xi(a)$ with $a_k=e_k\eta_k^*$. With $A_{ij}=\|\mathbf W_i-\mathbf W_j\|$, $R_i=\sum_jA_{ij}$, $R_{++}=\sum_{i,j}A_{ij}$, $\mathbf S_{ij}=\frac12(\boldsymbol\phi_{p,i}-\boldsymbol\phi_{p,j})(\boldsymbol\phi_{p,i}-\boldsymbol\phi_{p,j})^\top$, and a tilde denoting the $\mathcal U$-centered version defined in Section~\ref{sec:notation} of the main text (applied entrywise to vector- and matrix-valued arrays), let
\[
\mathbf S_n=\frac{1}{n(n-3)}\sum_{i\neq j}\widetilde A_{ij}\widetilde{\mathbf S}_{ij},
\qquad
\mathbf w_{\mathbf F}=\frac{1}{\sqrt n(n-3)}\sum_{i\neq j}\widetilde A_{ij}\widetilde{\mathbf F}_{ij}
\]
for any array of $p$-vectors $\{\mathbf F_{ij}\}$, such as $\mathbf R_{ij}=(e_i-e_j)(\boldsymbol\phi_{p,i}-\boldsymbol\phi_{p,j})$ or $\mathbf R_{ij}^*=(e_i\eta_i^*-e_j\eta_j^*)(\boldsymbol\phi_{p,i}-\boldsymbol\phi_{p,j})$.

\begin{lemma}\label{lemma:sim1}
Consider the case when $p$ is fixed, and let $\boldsymbol\xi$, $\boldsymbol\xi^\circ$, $\mathbf S_n$, $\mathbf w_{\mathbf R}$, and $\mathbf w_{\mathbf R^*}$ be as defined above.
\begin{enumerate}
\item Under the null and Assumption~\ref{assum:0}, we have
\begin{align*}
\boldsymbol\xi(e)^\top\mathbf S_n\boldsymbol\xi(e)
&=\boldsymbol\xi^\circ(e)^\top\mathbf S_n\boldsymbol\xi^\circ(e)+o_p(1),\\
\mathbf w_{\mathbf R}^\top\boldsymbol\xi(e)
&=\mathbf w_{\mathbf R}^\top\boldsymbol\xi^\circ(e)+o_p(1).
\end{align*}
\item Under the null, Assumption~\ref{assum:0}, and Assumption~\ref{assum:2}, we have
\begin{align*}
\boldsymbol\xi(e\eta^*)^\top\mathbf S_n\boldsymbol\xi(e\eta^*)
&=\boldsymbol\xi^\circ(e\eta^*)^\top\mathbf S_n\boldsymbol\xi^\circ(e\eta^*)+o_p^*(1),\\
\mathbf w_{\mathbf R^*}^\top\boldsymbol\xi(e\eta^*)
&=\mathbf w_{\mathbf R^*}^\top\boldsymbol\xi^\circ(e\eta^*)+o_p^*(1).
\end{align*}
\end{enumerate}

\end{lemma}
\begin{proof}[Proof of Lemma~\ref{lemma:sim1}]
For each assertion, we first remove the centering of the basis by $\overline{\boldsymbol\phi}_p$ and then the factor $\widehat{\mathbf Q}^{-1}$.

\textit{Preliminary bounds.} Since $\E[\boldsymbol\phi_p(\mathbf Z)]=\mathbf 0$ and $\E[\boldsymbol\phi_p(\mathbf Z)\boldsymbol\phi_p(\mathbf Z)^\top]=\mathbf I_p$, the multivariate central limit theorem and the law of large numbers give
\begin{align}
\overline{\boldsymbol\phi}_p&=O_p(n^{-1/2}),\label{eq:phi_bar}\\
\widehat{\mathbf Q}&=\frac1n\sum_{k=1}^n\boldsymbol\phi_{p,k}\boldsymbol\phi_{p,k}^\top-\overline{\boldsymbol\phi}_p\overline{\boldsymbol\phi}_p^\top\rightarrow^p\mathbf I_p,\label{eq:Q}\\
\|\widehat{\mathbf Q}-\mathbf I_p\|_2&=o_p(1),\label{eq:Q-I}
\end{align}
where $\|\cdot\|_2$ denotes the spectral norm of a matrix. Next, $\widetilde{\mathbf A}$ is symmetric, has zero diagonal, and, by \eqref{eq:row_col_sum} in the proof of Lemma~\ref{lemma:test_statistic}, has zero row and column sums. Hence $\sum_{i\neq j}\widetilde A_{ij}\widetilde{\mathbf F}_{ij}=\sum_{i\neq j}\widetilde A_{ij}\mathbf{F}_{ij}$ for any array $\{\mathbf{F}_{ij}\}$, and terms of $\mathbf{F}_{ij}$ that depend on a single index do not contribute; in particular,
\[
\sum_{i\neq j}\widetilde A_{ij}\widetilde{\mathbf S}_{ij}=\sum_{i\neq j}\widetilde A_{ij}\mathbf S_{ij}=-\sum_{i\neq j}\widetilde A_{ij}\boldsymbol\phi_{p,i}\boldsymbol\phi_{p,j}^\top.
\]%
Moreover, $\E\sum_{i\neq j}\widetilde A_{ij}^2=O(n^2)$, because, by \eqref{eq:ucenter}, $\widetilde A_{ij}$ is a signed sum of $A_{ij}$, $R_i/(n-2)$, $R_j/(n-2)$, and $R_{++}/\{(n-1)(n-2)\}$, whose second moments are bounded uniformly in $n$ when $\E\|\mathbf W\|^2<\infty$. Using these facts and the Cauchy--Schwarz inequality, we obtain
\begin{equation}\label{eq:sim1_II}
\begin{split}
\|\mathbf S_n\|_2
&\le\Big\|\frac{1}{n(n-3)}\sum_{i\neq j}\widetilde A_{ij}\boldsymbol\phi_{p,i}\boldsymbol\phi_{p,j}^\top\Big\|_2\\
&\le\frac{1}{n(n-3)}\Big(\sum_{i\neq j}\widetilde A_{ij}^2\Big)^{1/2}\Big(\sum_{i\neq j}\|\boldsymbol\phi_{p,i}\boldsymbol\phi_{p,j}^\top\|_2^2\Big)^{1/2}=O_p(1).
\end{split}
\end{equation}

\textit{First assertion.} Write $\boldsymbol\xi$ and $\boldsymbol\xi^\circ$ for $\boldsymbol\xi(e)$ and $\boldsymbol\xi^\circ(e)$. Since the observations are independent, $e$ is centered and independent of $\mathbf W$, and the basis has identity covariance and mean zero,
\[
\begin{aligned}
\E\|\boldsymbol\xi^\circ\|^2
&=\frac1n\sum_{k=1}^n\E[e_k^2\|\boldsymbol\phi_{p,k}\|^2]\\
&=\sigma^2\E\|\boldsymbol\phi_p(\mathbf Z)\|^2=\sigma^2p.
\end{aligned}
\]
Thus $\|\boldsymbol\xi^\circ\|=O_p(1)$ by Markov's inequality, because $p$ is fixed. By \eqref{eq:phi_bar} and the facts that $\|\widehat{\mathbf Q}^{-1}\|_2=O_p(1)$ and $n^{-1/2}\sum_{k=1}^n e_k=O_p(1)$, we obtain
\[
\boldsymbol\xi=\widehat{\mathbf Q}^{-1}\boldsymbol\xi^\circ-\widehat{\mathbf Q}^{-1}\overline{\boldsymbol\phi}_p\,\frac{1}{\sqrt n}\sum_{k=1}^ne_k=\widehat{\mathbf Q}^{-1}\boldsymbol\xi^\circ+o_p(1).
\]
Because $\|\mathbf S_n\|_2=O_p(1)$ by \eqref{eq:sim1_II} and $\|\boldsymbol\xi^\circ\|=O_p(1)$ by the moment bound above, this implies the first line of
\begin{align*}
\boldsymbol\xi^\top\mathbf S_n\boldsymbol\xi
&=(\widehat{\mathbf Q}^{-1}\boldsymbol\xi^\circ)^\top\mathbf S_n(\widehat{\mathbf Q}^{-1}\boldsymbol\xi^\circ)+o_p(1),\\
\mathbf w_{\mathbf R}^\top\boldsymbol\xi
&=\mathbf w_{\mathbf R}^\top\widehat{\mathbf Q}^{-1}\boldsymbol\xi^\circ+o_p(1),
\end{align*}
and the second line follows by the same arguments, because $\|\mathbf w_{\mathbf R}\|=O_p(1)$. Indeed, let $\mathbf v_i=\sum_{j\neq i}\widetilde A_{ij}\boldsymbol\phi_{p,j}$. Since $\widetilde{\mathbf A}$ has zero diagonal and zero row sums, $\sum_{i\neq j}\widetilde A_{ij}\widetilde{\mathbf R}_{ij}=-2\sum_ie_i\mathbf v_i$; since, moreover, $e$ is independent of $\mathbf W$ and $\|\mathbf v_i\|^2\le C^2n\sum_j\widetilde A_{ij}^2$,
\[
\E\|\mathbf w_{\mathbf R}\|^2=\frac{4\sigma^2}{n(n-3)^2}\sum_{i=1}^n\E\|\mathbf v_i\|^2\le\frac{4\sigma^2C^2}{(n-3)^2}\E\sum_{i\neq j}\widetilde A_{ij}^2=O(1).
\]

It remains to replace $\widehat{\mathbf Q}^{-1}\boldsymbol\xi^\circ$ by $\boldsymbol\xi^\circ$. Let $\mathbf d=(\widehat{\mathbf Q}^{-1}-\mathbf I_p)\boldsymbol\xi^\circ$, so that $\widehat{\mathbf Q}^{-1}\boldsymbol\xi^\circ=\boldsymbol\xi^\circ+\mathbf d$. Since $\mathbf S_n$ is symmetric,
\begin{align*}
(\widehat{\mathbf Q}^{-1}\boldsymbol\xi^\circ)^\top\mathbf S_n(\widehat{\mathbf Q}^{-1}\boldsymbol\xi^\circ)
&=\boldsymbol\xi^{\circ\top}\mathbf S_n\boldsymbol\xi^\circ+2\mathbf d^\top\mathbf S_n\widehat{\mathbf Q}^{-1}\boldsymbol\xi^\circ-\mathbf d^\top\mathbf S_n\mathbf d,\\
\mathbf w_{\mathbf R}^\top\widehat{\mathbf Q}^{-1}\boldsymbol\xi^\circ
&=\mathbf w_{\mathbf R}^\top\boldsymbol\xi^\circ+\mathbf w_{\mathbf R}^\top\mathbf d.
\end{align*}
By \eqref{eq:Q}, \eqref{eq:sim1_II}, and the moment bound above,
\begin{align*}
|\mathbf d^\top\mathbf S_n\widehat{\mathbf Q}^{-1}\boldsymbol\xi^\circ|
&\le\|\mathbf d\|\,\|\mathbf S_n\|_2\,\|\widehat{\mathbf Q}^{-1}\boldsymbol\xi^\circ\|\\
&\le\|\widehat{\mathbf Q}^{-1}-\mathbf I_p\|_2\,\|\widehat{\mathbf Q}^{-1}\|_2\,\|\boldsymbol\xi^\circ\|^2\,\|\mathbf S_n\|_2=o_p(1).
\end{align*}
Using similar arguments, we deduce that $\mathbf d^\top\mathbf S_n\mathbf d\rightarrow^p0$ and $\mathbf w_{\mathbf R}^\top\mathbf d\rightarrow^p0$, which proves the first assertion.

\textit{Second assertion.} Write $\boldsymbol\xi^*$ and $\boldsymbol\xi^{\circ*}$ for $\boldsymbol\xi(e\eta^*)$ and $\boldsymbol\xi^\circ(e\eta^*)$. Besides the algebraic identities displayed there, the proof of the first assertion uses only the bounds $\|\widehat{\mathbf Q}^{-1}\|_2=O_p(1)$, $\|\widehat{\mathbf Q}^{-1}-\mathbf I_p\|_2=o_p(1)$, $\|\mathbf S_n\|_2=O_p(1)$, and $\overline{\boldsymbol\phi}_p=O_p(n^{-1/2})$, which involve only the sample and therefore also hold with $O_p^*$ and $o_p^*$, together with the fact that $n^{-1/2}\sum_{k=1}^ne_k$, $\|\boldsymbol\xi^\circ\|$, and $\|\mathbf w_{\mathbf R}\|$ are $O_p(1)$. With $e_k$ replaced by $e_k\eta_k^*$, these three quantities are $O_p^*(1)$. Indeed, $\sum_{i\neq j}\widetilde A_{ij}\widetilde{\mathbf R}_{ij}^*=-2\sum_ie_i\eta_i^*\mathbf v_i$ in the same way, and, since $\{\eta_k^*\}$ are i.i.d.\ with mean zero and unit variance and independent of the sample, their conditional second moments are
\[
\frac1n\sum_{k=1}^ne_k^2,\qquad
\frac1n\sum_{k=1}^n\|\boldsymbol\phi_{p,k}\|^2e_k^2,\qquad
\frac{4}{n(n-3)^2}\sum_{i=1}^ne_i^2\|\mathbf v_i\|^2.
\]
The first two are $O_p(1)$ by the law of large numbers, and the expectation of the third equals $\E\|\mathbf w_{\mathbf R}\|^2=O(1)$. Since products of $O_p^*(1)$ and $o_p^*(1)$ terms are $o_p^*(1)$, the proof of the first assertion applies verbatim with $o_p(1)$ replaced by $o_p^*(1)$, which proves the second assertion. For later use, we note that $\boldsymbol\xi^*=\widehat{\mathbf Q}^{-1}\boldsymbol\xi^{\circ*}-\widehat{\mathbf Q}^{-1}\overline{\boldsymbol\phi}_p\,n^{-1/2}\sum_{k=1}^ne_k\eta_k^*$. Applying $\|a+b\|^2\le2\|a\|^2+2\|b\|^2$ and the first two conditional second moments displayed above gives
\begin{equation}\label{eq:phi_e}
\E^*\|\boldsymbol\xi^*\|^2\le2\|\widehat{\mathbf Q}^{-1}\|_2^2\Big(\frac1n\sum_{k=1}^n\|\boldsymbol\phi_{p,k}\|^2e_k^2+\|\overline{\boldsymbol\phi}_p\|^2\,\frac1n\sum_{k=1}^ne_k^2\Big)=O_p(1).
\end{equation}

\end{proof}

\begin{lemma}\label{lemma:test_statistic}
Let $\mathbf P_n$ be the orthogonal projection onto the column space of the full design matrix, including the intercept, and let $\mathbf\Pi_n=\mathbf I_n-\mathbf P_n$. For any $\mathbf v\in\R^n$, put $\widehat e_v=\mathbf\Pi_n\mathbf v$. Then
\begin{equation*}
\begin{aligned}
n\widehat{\MDD}^{\,2}(\widehat{e}_v\mid \mathbf W)
&=\frac{n}{n-3}\left\|\frac1{\sqrt n}\sum_{i=1}^n(\mathbf \Pi_n\mathbf v)_i\Phi(\mathbf W_i)\right\|_{\Hcal}^2
-\frac{\mathbf v^\top \mathbf \Pi_n \mathbf D_d\mathbf \Pi_n \mathbf v}{n-3}\\
&=\frac{n}{n-3}\left\|\frac1{\sqrt n}\sum_{i=1}^n(\mathbf \Pi_n\mathbf v)_i\Phi(\mathbf W_i)\right\|_{\Hcal}^2
-\frac{1}{n-3}\sum_i d_i(\mathbf \Pi_n \mathbf v)_i^2,
\end{aligned}
\end{equation*}
where $\mathbf D_d=\text{diag}(d_1, \cdots, d_n)$, $d_i=2R_i/(n-2)
-R_{++}/\{(n-1)(n-2)\}$, $R_i=\sum_j A_{ij}
$, and $R_{++}=\sum_{i,j}A_{ij}$, $A_{ij}=\|\mathbf W_i-\mathbf W_j\|$.
\end{lemma}
The proof below uses only that $\mathbf\Pi_n$ is symmetric and $\mathbf\Pi_n\mathbf1=\mathbf0$, so the identity applies to both the fixed-$p$ and growing-$p$ fits; the explicit growing-$p$ formula for $\mathbf P_n$ is given in \eqref{eq:P_n, Pi_n} below.
\begin{proof}[Proof of Lemma~\ref{lemma:test_statistic}]
Let $\mathbf A=(A_{ij})$, and let $\widetilde{\mathbf A}=(\widetilde A_{ij})$ be its $\mathcal U$-centered version (see Section~\ref{sec:notation} of the main text), which is symmetric with zero diagonal:
\begin{equation}\label{eq:ucenter}
\widetilde A_{ij}=\begin{cases}
A_{ij}-\dfrac{R_i+R_j}{n-2}+\dfrac{R_{++}}{(n-1)(n-2)},&i\neq j,\\[6pt]
0,&i=j.
\end{cases}
\end{equation}
This matrix has zero row sums. Indeed, for each fixed $i$,
\begin{align*}
\sum_{j}\widetilde A_{ij}
&=R_i-\frac{(n-1)R_i+\sum_{j\neq i}R_j}{n-2}
+\frac{(n-1)R_{++}}{(n-1)(n-2)}\\
&=R_i-\frac{(n-2)R_i+R_{++}}{n-2}
+\frac{R_{++}}{n-2}=0.
\end{align*}
Hence 
\begin{equation}\label{eq:row_col_sum}
\widetilde {\mathbf A} \mathbf1={\bf 0}, \qquad \mathbf1^\top\widetilde{\mathbf A}={\bf 0}.
\end{equation}
Write $\mathbf u=(u_1,\ldots,u_n)^\top=\mathbf\Pi_n\mathbf v$, so that $\widehat e_v=\mathbf u$. Since $\mathbf\Pi_n$ is symmetric and $\mathbf\Pi_n\mathbf 1=\mathbf 0$ by \eqref{eq:zero_design}, we have $\sum_{i=1}^n u_i=\mathbf 1^\top\mathbf\Pi_n\mathbf v=0$. Recall from Section~\ref{sec:review} of the main text that
\[
n\widehat{\MDD}^{\,2}(\widehat{e}_v\mid \mathbf W)=\frac{1}{n-3}\sum_{i\neq j}\widetilde A_{ij}\widetilde V_{ij},
\]
where $\widetilde V_{ij}$ is the $\mathcal{U}$-centered version of $V_{ij}=\frac{1}{2}\{(\widehat{e}_v)_i-(\widehat{e}_v)_j\}^2=\frac{1}{2}(u_i-u_j)^2$. Since $\widetilde{\mathbf A}$ has zero row and column sums, we have
\begin{align*}
\sum_{i\neq j}\widetilde A_{ij}\widetilde V_{ij}
&=\sum_{i\neq j}\widetilde{\widetilde A}_{ij}V_{ij}=\sum_{i\neq j}\widetilde A_{ij}V_{ij}
=\frac12\sum_{i,j=1}^n\widetilde A_{ij}(u_i-u_j)^2\\
&=-\sum_{i,j}\widetilde A_{ij}u_iu_j
=-\mathbf u^\top\widetilde{\mathbf A}\mathbf u.
\end{align*}
Thus,
\begin{equation}\label{eq:base}
n\widehat{\MDD}^{\,2}(\widehat{e}_v\mid \mathbf W)
=-\frac{\mathbf u^\top\widetilde{\mathbf A}\mathbf u}{n-3}.
\end{equation}

Next, let $c_i=R_i/(n-2)-R_{++}/\{2(n-1)(n-2)\}$, so that $d_i=2c_i$. By \eqref{eq:ucenter}, $\widetilde A_{ij}=A_{ij}-c_i-c_j$ for $i\neq j$, whereas $\widetilde A_{ii}=0=A_{ii}-2c_i+d_i$ because $A_{ii}=0$. Hence $\widetilde{\mathbf A}=\mathbf A-\mathbf c\mathbf 1^\top-\mathbf 1\mathbf c^\top+\mathbf D_d$ with $\mathbf c=(c_1,\ldots,c_n)^\top$, and, since $\mathbf 1^\top\mathbf u=0$,
\begin{equation}\label{eq:keycenter}
\mathbf u^\top\widetilde{\mathbf A}\mathbf u
=\mathbf u^\top\mathbf A\mathbf u+\mathbf u^\top\mathbf D_d\mathbf u
=\mathbf u^\top\mathbf A\mathbf u+\sum_{i=1}^n d_iu_i^2,
\end{equation}
where the second term accounts for the zero diagonal of $\widetilde{\mathbf A}$. By the definition of $\Phi$ in \eqref{grow:feature_supp}, $A_{ij}=\|\mathbf W_i\|+\|\mathbf W_j\|-\langle\Phi(\mathbf W_i),\Phi(\mathbf W_j)\rangle_{\Hcal}$. Since $\sum_iu_i=0$, the first two terms do not contribute to $\mathbf u^\top\mathbf A\mathbf u$, so that $\mathbf u^\top\mathbf A\mathbf u=-\|\sum_{i=1}^n u_i\Phi(\mathbf W_i)\|_{\Hcal}^2$. Substituting this into \eqref{eq:keycenter} and then into \eqref{eq:base} gives
\[
n\widehat{\MDD}^{\,2}(\widehat{e}_v\mid \mathbf W)
=\frac{n}{n-3}\left\|\frac1{\sqrt n}\sum_{i=1}^n u_i\Phi(\mathbf W_i)\right\|_{\Hcal}^2
-\frac{\mathbf u^\top\mathbf D_d\mathbf u}{n-3}.
\]
Since $u_i=(\mathbf\Pi_n\mathbf v)_i$ and $\mathbf u^\top\mathbf D_d\mathbf u=\mathbf v^\top\mathbf\Pi_n\mathbf D_d\mathbf\Pi_n\mathbf v=\sum_i d_iu_i^2$, this is the claimed identity, which completes the proof.
\end{proof}

\begin{lemma}\label{lemma:fixed_p_bootstrap_remainder} We consider the asymptotic expansions of the bootstrap statistic under the fixed $p$ scenario. Under the null, Assumptions~\ref{assum:0} and \ref{assum:2}, we have
\[
\begin{aligned}
n\widehat{\rm{MDD}}^{2*}(\widehat{e}\mid\mathbf{W})
&=\frac{1}{n-3}\sum_{i\neq j}\widetilde{A}_{ij}\widetilde{B}_{ij}^{**}
+\boldsymbol\xi(e\eta^*)^\top\mathbf S_n\,\boldsymbol\xi(e\eta^*)\\
&\quad-\mathbf w_{\mathbf R^*}^\top\boldsymbol\xi(e\eta^*)+\mathcal R_{n,1}^*,
\end{aligned}
\]%
where $\widetilde{B}_{ij}^{**}$ and $\widetilde{\mathbf{R}}_{ij}^*$ are the $\mathcal{U}$-centered versions of $B_{ij}^{**}=\frac{1}{2}(e_i\eta_i^*-e_j\eta_j^*)^2$ and $\mathbf{R}_{ij}^*=(e_i\eta_i^*-e_j\eta_j^*)(\boldsymbol\phi_{p,i}-\boldsymbol\phi_{p,j})$; $\boldsymbol\xi(e\eta^*)$, $\mathbf S_n$, and $\mathbf w_{\mathbf R^*}$ are defined in the notation for the fixed-$p$ proofs at the beginning of this section; and ${\mathcal{R}}^*_{n,1}=o_p^*(1)$.
\end{lemma}
\begin{proof}[Proof of Lemma~\ref{lemma:fixed_p_bootstrap_remainder}]

Throughout the proof, we write
\[
a_n=\widehat\gamma-\gamma_0,\qquad
\mathbf b_n=\widehat{\boldsymbol\beta}-\boldsymbol\beta_0,\qquad
\varrho_k=a_n+\boldsymbol\phi_{p,k}^\top\mathbf b_n.
\]
Under the null, $a_n=O_p(n^{-1/2})$ and $\|\mathbf b_n\|=O_p(n^{-1/2})$. The estimated residuals are $\widehat e_k=e_k+\boldsymbol\phi_{p,k}^\top(\boldsymbol\beta_0-\widehat{\boldsymbol\beta})+(\gamma_0-\widehat\gamma)=e_k-\varrho_k$. Let $\mathbf v_i=\sum_{j\neq i}\widetilde A_{ij}\boldsymbol\phi_{p,j}$ for $i=1,\ldots,n$. We repeatedly use the following facts, which hold conditionally on the sample because $\{\eta_i^*\}$ are i.i.d.\ with mean zero and unit variance and independent of the sample:
\begin{enumerate}
\item[(F1)] for any array $\{c_{ij}\}$ that depends only on the sample, we have
\[
\E^*\Big[\Big(\sum_{i\neq j}c_{ij}\eta_i^*\eta_j^*\Big)^2\Big]=\sum_{i\neq j}c_{ij}(c_{ij}+c_{ji})\le 2\sum_{i\neq j}c_{ij}^2;
\]
\item[(F2)] for any vectors $\{\mathbf u_i\}$ that depend only on the sample, $\E^*\|\sum_{i}\eta_i^*\mathbf u_i\|^2=\sum_{i}\|\mathbf u_i\|^2$;
\item[(F3)] since $\sup_{\mathbf z}\|\boldsymbol\phi_p(\mathbf z)\|\le C$, the Cauchy--Schwarz inequality gives
\[
\|\mathbf v_i\|^2\le C^2n\sum_{j}\widetilde A_{ij}^2;
\]
under Assumption~\ref{assum:2}, $\sum_{i\neq j}\widetilde A_{ij}^2(1+e_i^2)=O_p(n^2)$, so that $\sum_{i}(1+e_i^2)\|\mathbf v_i\|^2=O_p(n^3)$; and $\|\mathbf M_n\|_2=O_p(1)$ for $\mathbf M_n=n^{-1}(n-3)^{-1}\sum_{i\neq j}\widetilde A_{ij}\boldsymbol\phi_{p,i}\boldsymbol\phi_{p,j}^\top$ by \eqref{eq:sim1_II}.
\end{enumerate}

In (F1), expanding the square leaves only index pairs $(i,j)=(k,l)$ or $(i,j)=(l,k)$, because every other product contains a multiplier to the first power and has conditional mean zero. This explains both the equality and the factor $2$ in the bound.

\smallskip\noindent\textit{Step 1: the bootstrap coefficient.} Since $Y_k^*=\widehat\gamma+\boldsymbol\phi_{p,k}^\top\widehat{\boldsymbol\beta}+\widehat e_k\eta_k^*$, the least-squares fit of $Y_k^*$ on $\boldsymbol\phi_{p,k}$ with an intercept gives
\begin{equation}\label{eq:beta^*-beta_hat_lemma}
\widehat{\boldsymbol\beta}^*-\widehat{\boldsymbol\beta}
=\frac{1}{\sqrt n}\,\boldsymbol\xi(\widehat e\eta^*)
=\mathbf s_n^*+\mathbf r_n^*,
\end{equation}
where, since $\widehat e_k=e_k-\varrho_k$,
\[
\mathbf s_n^*=\frac{1}{\sqrt n}\,\boldsymbol\xi(e\eta^*),
\qquad
\mathbf r_n^*=-\frac{1}{\sqrt n}\,\boldsymbol\xi(\varrho\eta^*).
\]
By \eqref{eq:phi_e}, $\|\mathbf s_n^*\|=O_p^*(n^{-1/2})$. By (F2) and $\varrho_k^2\le 2a_n^2+2\|\mathbf b_n\|^2\|\boldsymbol\phi_{p,k}\|^2$,
\begin{align*}
\E^*\|\mathbf r_n^*\|^2
&=\frac{1}{n^2}\sum_{k=1}^n\|\widehat{\mathbf Q}^{-1}(\boldsymbol\phi_{p,k}-\overline{\boldsymbol\phi}_p)\|^2\varrho_k^2\\
&\le\frac{2\|\widehat{\mathbf Q}^{-1}\|_2^2}{n^2}\sum_{k=1}^n\|\boldsymbol\phi_{p,k}-\overline{\boldsymbol\phi}_p\|^2\big(a_n^2+\|\mathbf b_n\|^2\|\boldsymbol\phi_{p,k}\|^2\big)=O_p(n^{-2}),
\end{align*}
because $a_n^2+\|\mathbf b_n\|^2=O_p(n^{-1})$, $\|\widehat{\mathbf Q}^{-1}\|_2=O_p(1)$, and both $n^{-1}\sum_{k}\|\boldsymbol\phi_{p,k}-\overline{\boldsymbol\phi}_p\|^2$ and $n^{-1}\sum_{k}\|\boldsymbol\phi_{p,k}-\overline{\boldsymbol\phi}_p\|^2\|\boldsymbol\phi_{p,k}\|^2$ are $O_p(1)$. Hence $\|\mathbf r_n^*\|=O_p^*(n^{-1})$.

\smallskip\noindent\textit{Step 2: the bootstrap residuals.} The bootstrap residuals are
\[
\widehat e_i^*=Y_i^*-\widehat\gamma^*-\boldsymbol\phi_{p,i}^\top\widehat{\boldsymbol\beta}^*
=\widehat e_i\eta_i^*+(\widehat\gamma-\widehat\gamma^*)-\boldsymbol\phi_{p,i}^\top(\widehat{\boldsymbol\beta}^*-\widehat{\boldsymbol\beta}).
\]
By $\widehat e_i=e_i-\varrho_i$ and \eqref{eq:beta^*-beta_hat_lemma}, $\widehat e_i^*=(\widehat\gamma-\widehat\gamma^*)+f_i^*-g_i^*$, where
\[
f_i^*=e_i\eta_i^*-\boldsymbol\phi_{p,i}^\top\mathbf s_n^*,
\qquad
g_i^*=\varrho_i\eta_i^*+\boldsymbol\phi_{p,i}^\top\mathbf r_n^*.
\]
Since the intercept term $\widehat\gamma-\widehat\gamma^*$ does not depend on $i$, we have $\widehat e_i^*-\widehat e_j^*=(f_i^*-f_j^*)-(g_i^*-g_j^*)$; here $f_i^*$ carries the leading terms, and $g_i^*$ collects all terms that involve $a_n$ or $\mathbf b_n$.

\smallskip\noindent\textit{Step 3: expansion of the statistic.} By definition,
\[
n\widehat{\rm MDD}^{2*}(\widehat e\mid\mathbf W)=\frac{1}{n-3}\sum_{i\neq j}\widetilde A_{ij}\widetilde V_{ij}^*,
\]
where $\widetilde V_{ij}^*$ is the $\mathcal U$-centered version of $V_{ij}^*=\frac12(\widehat e_i^*-\widehat e_j^*)^2$. As in the proof of Lemma~\ref{lemma:test_statistic}, $\sum_{i\neq j}\widetilde A_{ij}\widetilde F_{ij}=\sum_{i\neq j}\widetilde A_{ij}F_{ij}$ for any array $\{F_{ij}\}$, so the tilde on $V_{ij}^*$ can be dropped. Moreover, $\widetilde{\mathbf A}$ is symmetric, has zero diagonal, and, by \eqref{eq:row_col_sum}, has zero row and column sums, so that terms depending on a single index vanish; hence, for any numbers $f_1,\ldots,f_n$ and $g_1,\ldots,g_n$,
\begin{equation*}
\sum_{i\neq j}\widetilde A_{ij}(f_i-f_j)(g_i-g_j)=-2\sum_{i\neq j}\widetilde A_{ij}f_ig_j.\tag{$\star$}
\end{equation*}
By Step 2, $V_{ij}^*=\frac12(f_i^*-f_j^*)^2-(f_i^*-f_j^*)(g_i^*-g_j^*)+\frac12(g_i^*-g_j^*)^2$. The first part equals $B_{ij}^{**}-\mathbf R_{ij}^{*\top}\mathbf s_n^*+\mathbf s_n^{*\top}\mathbf S_{ij}\mathbf s_n^*$. Summing it against $\widetilde A_{ij}/(n-3)$, restoring the tildes, and using $\sqrt n\,\mathbf s_n^*=\boldsymbol\xi(e\eta^*)$ yields the three leading terms in
\begin{equation}\label{eq:nMDD_boot_lemma}
\begin{aligned}
n\widehat{\rm MDD}^{2*}(\widehat e\mid\mathbf W)
&=\frac{1}{n-3}\sum_{i\neq j}\widetilde A_{ij}\widetilde B_{ij}^{**}
+\boldsymbol\xi(e\eta^*)^\top\mathbf S_n\,\boldsymbol\xi(e\eta^*)\\
&\quad-\mathbf w_{\mathbf R^*}^\top\boldsymbol\xi(e\eta^*)+\mathcal R_{n,1}^*,
\end{aligned}
\end{equation}
and $\mathcal R_{n,1}^*$ is the contribution of the other two parts. Applying $(\star)$ to them gives
\[
\mathcal R_{n,1}^*=\frac{1}{n-3}\sum_{i\neq j}\widetilde A_{ij}\,(2f_i^*-g_i^*)\,g_j^*.
\]
By Step 2, $2f_i^*-g_i^*=(2e_i-\varrho_i)\eta_i^*-\boldsymbol\phi_{p,i}^\top(2\mathbf s_n^*+\mathbf r_n^*)$ and $g_j^*=\varrho_j\eta_j^*+\boldsymbol\phi_{p,j}^\top\mathbf r_n^*$. We expand the product and use the symmetry of $\widetilde A_{ij}$, $\sum_{j\neq i}\widetilde A_{ij}\boldsymbol\phi_{p,j}=\mathbf v_i$, and $\sum_{i\neq j}\widetilde A_{ij}\boldsymbol\phi_{p,i}\boldsymbol\phi_{p,j}^\top=n(n-3)\mathbf M_n$. This gives $\mathcal R_{n,1}^*=\sum_{m=1}^4\mathcal R_{n,1,m}^*$, where
\begin{align*}
\mathcal R_{n,1,1}^*&=\frac{1}{n-3}\sum_{i\neq j}\widetilde A_{ij}(2e_i-\varrho_i)\varrho_j\eta_i^*\eta_j^*,\\
\mathcal R_{n,1,2}^*&=\frac{1}{n-3}\Big\{\sum_{i}(2e_i-\varrho_i)\eta_i^*\mathbf v_i\Big\}^\top\mathbf r_n^*,\\
\mathcal R_{n,1,3}^*&=-\frac{1}{n-3}\Big(\sum_{i}\varrho_i\eta_i^*\mathbf v_i\Big)^\top(2\mathbf s_n^*+\mathbf r_n^*),\\
\mathcal R_{n,1,4}^*&=-n\,(2\mathbf s_n^*+\mathbf r_n^*)^\top\mathbf M_n\mathbf r_n^*.
\end{align*}

\smallskip\noindent\textit{Step 4: the remainder is negligible.} Let $\varrho_{\max}^2=\max_k\varrho_k^2\le2(a_n^2+C^2\|\mathbf b_n\|^2)=O_p(n^{-1})$. By (F1), (F3), and $(2e_i-\varrho_i)^2\le8e_i^2+2\varrho_{\max}^2$,
\[
\E^*[(\mathcal{R}_{n,1,1}^*)^2]\le\frac{2\varrho_{\max}^2}{(n-3)^2}\sum_{i\neq j}\widetilde A_{ij}^2(8e_i^2+2\varrho_{\max}^2)=O_p(n^{-1}).
\]
By (F2) and (F3), $\E^*\|\sum_i(2e_i-\varrho_i)\eta_i^*\mathbf v_i\|^2\le\sum_i(8e_i^2+2\varrho_{\max}^2)\|\mathbf v_i\|^2=O_p(n^3)$ and $\E^*\|\sum_i\varrho_i\eta_i^*\mathbf v_i\|^2\le\varrho_{\max}^2\sum_i\|\mathbf v_i\|^2=O_p(n^2)$. Since $\|\mathbf r_n^*\|=O_p^*(n^{-1})$ and $\|2\mathbf s_n^*+\mathbf r_n^*\|=O_p^*(n^{-1/2})$ by Step 1, and $\|\mathbf M_n\|_2=O_p(1)$ by (F3), the Cauchy--Schwarz inequality shows that $\mathcal R_{n,1,2}^*$, $\mathcal R_{n,1,3}^*$, and $\mathcal R_{n,1,4}^*$ are all $O_p^*(n^{-1/2})$. Combining these bounds, we conclude that $\mathcal{R}_{n,1}^*=o_p^*(1)$.

\end{proof}

\begin{lemma}\label{lemma:t_rate} Let $\boldsymbol\Phi=(\Phi(\mathbf W_1),\ldots,\Phi(\mathbf W_n))^\top$, and let $t_i=(\mathbf\Pi_n\boldsymbol\Phi)_i$ for $i=1,\ldots,n$, where $\mathbf\Pi_n$ is the residual projection matrix defined in Lemma~\ref{lemma:test_statistic}. Under Assumption~\ref{assum:1}, we have
\begin{equation}\label{eq:t_rate}\max_i
\|t_i\|_{\mathcal H}^2
=
O_p(\xi_p^2).\end{equation}
\end{lemma}
\begin{proof}[Proof of Lemma~\ref{lemma:t_rate}]

\noindent Here $\mathbf P_n$ is the matrix of the orthogonal projection onto the column space of the full design matrix (the intercept and the $p$ basis functions), and $\mathbf\Pi_n=\mathbf I_n-\mathbf P_n$. Write $\mathbf b_{p,i}=\mathbf g_p(\mathbf Z_i)-\overline{\mathbf g}_p$, where $\overline{\mathbf g}_p=n^{-1}\sum_i\mathbf g_p(\mathbf Z_i)$, and let $\widehat{\mathbf Q}_b=n^{-1}\sum_i\mathbf b_{p,i}\mathbf b_{p,i}^\top$ be the centered sample Gram matrix. We also write $\Phi_j=\Phi(\mathbf W_j)$, so that $\|\Phi_j\|_{\mathcal H}^2=2\|\mathbf W_j\|\le2M$ by (B1). Let $\ell_i:=(\mathbf P_n)_{ii}=n^{-1}+n^{-1}\mathbf b_{p,i}^\top\widehat{\mathbf Q}_b^{-1}\mathbf b_{p,i}$. As shown in the proof of Theorem~\ref{thm:null}, the event $\|\widehat{\mathbf Q}_b^{-1}\|_2\le2$ has probability tending to one; on this event, $\|\mathbf b_{p,i}\|\le\|\mathbf g_p(\mathbf Z_i)\|+\|\overline{\mathbf g}_p\|\le2\xi_p$ gives $\ell_i\le n^{-1}(1+8\xi_p^2)$ for every $i$. Since $\xi_p^2\ge\E\|\mathbf g_p(\mathbf Z)\|^2=p\ge1$, it follows that
\begin{equation}\label{eq:ell_i}
\max_i\ell_i
=
O_p\left(\frac{\xi_p^2}{n}\right).
\end{equation}
Because $\mathbf P_n$ is symmetric and idempotent, $\sum_j(\mathbf P_n)_{ij}^2=(\mathbf P_n^2)_{ii}=\ell_i$, so the Cauchy--Schwarz inequality gives $\|(\mathbf P_n\boldsymbol\Phi)_i\|_{\mathcal H}^2\le\ell_i\sum_j\|\Phi_j\|_{\mathcal H}^2\le2Mn\ell_i$. Since $t_i=\Phi_i-(\mathbf P_n\boldsymbol\Phi)_i$ and $\|t_i\|_{\mathcal H}^2\le2\|\Phi_i\|_{\mathcal H}^2+2\|(\mathbf P_n\boldsymbol\Phi)_i\|_{\mathcal H}^2$, we obtain
\[
\max_i\|t_i\|_{\mathcal H}^2\le4M\Big(1+n\max_i\ell_i\Big)=O_p(\xi_p^2).
\]

\end{proof}

\begin{proof}[Proof of Theorem~\ref{thm:null_fixed_p}]

We use the notation introduced at the beginning of this section. In the fixed-$p$ setting, $\widehat{\bm{\beta}}-\bm{\beta}_0=\widehat{\mathbf Q}^{-1}n^{-1}\sum_{k=1}^n(\boldsymbol\phi_{p,k}-\overline{\boldsymbol\phi}_p)e_k=n^{-1/2}\boldsymbol\xi(e)$, so the difference between the estimated residuals can be expanded as
\begin{align*}
\widehat{e}_i-\widehat{e}_j
&=e_i-e_j+(\boldsymbol\phi_{p,i}-\boldsymbol\phi_{p,j})^\top(\bm{\beta}_0-\widehat{\bm{\beta}})\\
&=e_i-e_j-n^{-1/2}(\boldsymbol\phi_{p,i}-\boldsymbol\phi_{p,j})^\top\boldsymbol\xi(e).
\end{align*}
With $B_{ij}=(e_i-e_j)^2/2$ and $\mathbf R_{ij}=(e_i-e_j)(\boldsymbol\phi_{p,i}-\boldsymbol\phi_{p,j})$, it follows that
\[
\frac{(\widehat{e}_i-\widehat{e}_j)^2}{2}
=B_{ij}-\frac{\mathbf R_{ij}^\top\boldsymbol\xi(e)}{\sqrt n}
+\frac{\boldsymbol\xi(e)^\top\mathbf S_{ij}\boldsymbol\xi(e)}{n}.
\]
Since $\mathcal U$-centering is linear, this leads to the following decomposition of the test statistic:
\[
n\widehat{\rm{MDD}}^2(\widehat{e}\mid\mathbf{W})
=\frac{1}{n-3}\sum_{i\neq j}\widetilde{A}_{ij}\widetilde{B}_{ij}
-\mathbf w_{\mathbf R}^\top\boldsymbol\xi(e)
+\boldsymbol\xi(e)^\top\mathbf S_n\boldsymbol\xi(e),
\]
where $\widetilde{B}_{ij}$ is the $\mathcal{U}$-centered version of $B_{ij}$; recall that $\mathbf S_n$ and $\mathbf w_{\mathbf R}$ are built from the $\mathcal U$-centered versions of $A_{ij}=\|\mathbf{W}_i-\mathbf{W}_j\|$, $\mathbf S_{ij}$ and $\mathbf R_{ij}$.

By Lemma~\ref{lemma:sim1} and the facts that $\widehat{\mathbf{Q}}\rightarrow^{p}\mathbf{I}_p$ and $\overline{\boldsymbol{\phi}}_p\rightarrow^{p}0$ as $n\to\infty$, we have
\begin{equation}\label{eq:nmdd_fixed_p_stat}
n\widehat{\rm{MDD}}^2(\widehat{e}\mid\mathbf{W})
=\frac{1}{n-3}\sum_{i\neq j}\widetilde{A}_{ij}\widetilde{B}_{ij}
-\mathbf w_{\mathbf R}^\top\boldsymbol\xi^\circ(e)
+\boldsymbol\xi^\circ(e)^\top\mathbf S_n\boldsymbol\xi^\circ(e)
+\widetilde{\mathcal{R}}_n,
\end{equation}
where $\widetilde{\mathcal{R}}_n$ is asymptotically negligible.

Note that $n^{-1/2}\mathbf w_{\mathbf R}$, $\{n(n-3)\}^{-1}\sum_{i\neq j}\widetilde{A}_{ij}\widetilde{B}_{ij}$ and $\mathbf S_n$ all have the form $\{n(n-3)\}^{-1}\sum_{i\neq j}\widetilde{A}_{ij}\widetilde{F}_{ij}$, with $F=\mathbf R$, $B$ and $\mathbf S$, respectively, and can therefore be expressed as fourth-order $\mathcal{U}$-statistics:
\[
\frac{1}{n(n-3)}\sum_{i\neq j}\widetilde{A}_{ij}\widetilde{F}_{ij}
=\binom{n}{4}^{-1}\sum_{i<j<q<r}h_F(\mathbf{V}_i,\mathbf{V}_j,\mathbf{V}_q,\mathbf{V}_r),
\]
where $\mathbf{V}_i=(\mathbf{W}_i,e_i)$ and the symmetric kernel $h_F$ is
\[
h_F(\mathbf{V}_i,\mathbf{V}_j,\mathbf{V}_q,\mathbf{V}_r)
=\frac{1}{4!}\sum_{(s,t,u,v)}^{(i,j,q,r)}A_{st}\left(F_{uv}+F_{st}-F_{su}-F_{tv}\right),
\]
with $\sum_{(s,t,u,v)}^{(i,j,q,r)}$ denoting the summation over all permutations of the indices $(i,j,q,r)$. We write $h_1=h_{\mathbf R}\in\mathbb{R}^p$, $h_2=h_B\in\mathbb{R}$ and $h_3=h_{\mathbf S}\in\mathbb{R}^{p\times p}$; since $\mathbf S_{ij}$ involves only $\mathbf W_i$ and $\mathbf W_j$, we write $h_3(\mathbf{W}_i,\mathbf{W}_j,\mathbf{W}_q,\mathbf{W}_r)$.

By the strong law of large numbers for $\mathcal{U}$-statistics \citep{hoeffding1961strong, lee1990u}, it follows that
\begin{equation}\label{eq:converges_p_refined}
\mathbf S_n\rightarrow^{p}\breve{\mathbf{H}}:=-\E\left[\boldsymbol{\phi}_p(\mathbf{Z})\boldsymbol{\phi}_p(\mathbf{Z}')^\top\|\mathbf{W}-\mathbf{W}'\|\right],
\end{equation}
because the entries of $h_3$ are integrable and $\E[\mathbf S_n]=\E[h_3]=\breve{\mathbf H}$; the latter follows by taking expectations term by term in $h_3$ and using $\E[\boldsymbol\phi_p(\mathbf Z)]=\mathbf 0$ and $\E[\boldsymbol\phi_p(\mathbf Z)\boldsymbol\phi_p(\mathbf Z)^\top]=\mathbf I_p$. For integrability, by Condition (A2), every entry of $\mathbf S_{uv}$ is bounded in absolute value by $\frac12\|\boldsymbol\phi_{p,u}-\boldsymbol\phi_{p,v}\|^2\le2C^2$, so that $\E|\{h_3(\mathbf W_i,\mathbf W_j,\mathbf W_q,\mathbf W_r)\}_{k,l}|\le8C^2\E\|\mathbf W-\mathbf W'\|<\infty$ for all $k,l$. Similarly, $\E\|h_1(\mathbf{V}_i,\mathbf{V}_j,\mathbf{V}_q,\mathbf{V}_r)\|^2<\infty$ under Condition (A4), because $\mathbf R_{ij}$ is linear in the errors and $e$ is independent of $\mathbf W$.

We show below that
\[
\frac{1}{n-3}\sum_{i\neq j}\widetilde{A}_{ij}\widetilde{B}_{ij}=\mathcal{U}_{n,1}+\mathcal{R}_{n,1},
\qquad
\mathbf w_{\mathbf R}^\top\boldsymbol\xi^\circ(e)=\mathcal{U}_{n,3}+\mathcal{R}_{n,2},
\]
where
\[
\mathcal{U}_{n,1}=\frac{6}{n}\sum_{i\neq j}h_{2,2}(\mathbf{V}_i,\mathbf{V}_j),
\qquad
\mathcal{U}_{n,3}=\frac{4}{\sqrt{n}}\sum_{i=1}^n h_{1,1}(\mathbf{V}_i)^\top\boldsymbol\xi^\circ(e),
\]
the projection kernels are
\begin{align*}
h_{2,2}(\mathbf{V}_1,\mathbf{V}_2)&=\frac{1}{6}U(\mathbf{W}_1,\mathbf{W}_2)(-e_1e_2)\in\mathbb{R},\\
h_{1,1}(\mathbf{V}_1)&=\frac{1}{2}\E\left[U(\mathbf{W},\mathbf{W}')\{-e\boldsymbol{\phi}_p(\mathbf{Z}')\}\mid\mathbf{V}=\mathbf{V}_1\right]\in\mathbb{R}^p,
\end{align*}
and the remainders $\mathcal{R}_{n,1},\mathcal{R}_{n,2}$ are asymptotically negligible.

The Hoeffding decomposition (see \citet{serfling2009approximation}), applied to the square-integrable kernel $h_1$, gives
\[
\mathbf w_{\mathbf R}=\frac4{\sqrt n}\sum_{i=1}^n h_{1,1}(\mathbf V_i)+O_p(n^{-1/2}).
\]
Here the factor $4$ is the order of the kernel, and the sum of higher-order Hoeffding terms has variance $O(n^{-1})$ after multiplication by $\sqrt n$. Multiplying the remainder by $\boldsymbol\xi^\circ(e)=O_p(1)$ gives $\mathcal R_{n,2}=o_p(1)$. For the first term, we use the following exact centering identity, which leads to an asymptotically negligible remainder under the second-moment condition $\E e^2<\infty$ in Condition (A4). Since $\widetilde{\mathbf A}$ has zero diagonal and, by \eqref{eq:row_col_sum}, zero row and column sums, the terms of $B_{ij}=\frac12e_i^2+\frac12e_j^2-e_ie_j$ that depend on a single index do not contribute, and $\sum_{i\neq j}\widetilde A_{ij}\widetilde B_{ij}=-\sum_{i\neq j}\widetilde A_{ij}e_ie_j$. Since $6h_{2,2}(\mathbf V_i,\mathbf V_j)=-U(\mathbf W_i,\mathbf W_j)e_ie_j$ by the definition of $h_{2,2}$, it follows that
\[
\mathcal R_{n,1}=\frac{3}{n-3}\,\mathcal U_{n,1}+\mathcal E_n,\qquad
\mathcal E_n=-\frac{1}{n-3}\sum_{i\neq j}\{\widetilde A_{ij}-U(\mathbf W_i,\mathbf W_j)\}e_ie_j.
\]
Because the $e_i$ are i.i.d.\ with mean zero and independent of $\mathbf W_1,\ldots,\mathbf W_n$, only the products with $\{k,l\}=\{i,j\}$ contribute to $\E[\mathcal E_n^2]$, so that
\[
\E[\mathcal E_n^2]=\frac{2\sigma^4}{(n-3)^2}\sum_{i\neq j}\E\{\widetilde A_{ij}-U(\mathbf W_i,\mathbf W_j)\}^2=O(n^{-1}).
\]
Here we used that, by \eqref{eq:ucenter} and the definition of $U$, $\widetilde A_{ij}-U(\mathbf W_i,\mathbf W_j)$ is a signed sum of the deviations of $R_i/(n-2)$, $R_j/(n-2)$ and $R_{++}/\{(n-1)(n-2)\}$ from $\E[\|\mathbf W_i-\mathbf W'\|\mid\mathbf W_i]$, $\E[\|\mathbf W_j-\mathbf W'\|\mid\mathbf W_j]$ and $\E\|\mathbf W-\mathbf W'\|$, respectively, and that each deviation has second moment $O(n^{-1})$ when $\E\|\mathbf W\|^2<\infty$. Finally, $h_{2,2}$ is degenerate and $\E[h_{2,2}(\mathbf V_1,\mathbf V_2)^2]=\sigma^4\E[U(\mathbf W_1,\mathbf W_2)^2]/36<\infty$, so that $\E[\mathcal U_{n,1}^2]=O(1)$. Therefore $\mathcal R_{n,1}=o_p(1)$.

\noindent Together with \eqref{eq:nmdd_fixed_p_stat}, this decomposition leads to the representation
\begin{equation}\label{eq:nMDD_sim1_fixed_p_final}
n\widehat{\rm{MDD}}^2(\widehat{e}\mid\mathbf{W})=\mathcal{U}_{n,1}+\mathcal{U}_{n,2}-\mathcal{U}_{n,3}+\mathcal{R}_n,
\end{equation}
where $\mathcal{U}_{n,2}=\boldsymbol\xi^\circ(e)^\top\mathbf S_n\boldsymbol\xi^\circ(e)$ and $\mathcal{R}_n$ is asymptotically negligible.

We next derive the spectral decomposition of the kernel $6h_{2,2}$ from the feature map $\Phi$ in \eqref{grow:feature_supp}. Let $m_\Phi=\E\Phi(\mathbf W)$, $\mu_D=\E\|\mathbf W-\mathbf W'\|$, and
\[
\boldsymbol\chi(\mathbf V)=e\{\Phi(\mathbf W)-m_\Phi\}\in\mathcal H.
\]
Expanding the inner product using \eqref{grow:feature_supp} and comparing with the definition of $U$ in Section~\ref{sec:notation} of the main text gives $\langle\Phi(\mathbf w)-m_\Phi,\Phi(\mathbf v)-m_\Phi\rangle_{\mathcal H}=-U(\mathbf w,\mathbf v)$, so that
\[
6h_{2,2}(\mathbf V_1,\mathbf V_2)=-e_1e_2U(\mathbf W_1,\mathbf W_2)=\langle\boldsymbol\chi(\mathbf V_1),\boldsymbol\chi(\mathbf V_2)\rangle_{\mathcal H}.
\]
Taking $\mathbf w=\mathbf v$ gives $\|\Phi(\mathbf w)-m_\Phi\|_{\mathcal H}^2=-U(\mathbf w,\mathbf w)=2\E\|\mathbf w-\mathbf W'\|-\mu_D$. Since $e$ is independent of $\mathbf W$ by Condition (A3), it follows that
\[
\E\|\boldsymbol\chi(\mathbf V)\|_{\mathcal H}^2=\sigma^2\E\|\Phi(\mathbf W)-m_\Phi\|_{\mathcal H}^2=\sigma^2\mu_D<\infty.
\]
Because $\|\Phi(\mathbf w)-\Phi(\mathbf v)\|_{\mathcal H}^2=2\|\mathbf w-\mathbf v\|$ by \eqref{grow:feature_supp}, $\Phi$ is continuous, so we may take $\mathcal H$ to be the closed linear span of $\Phi(\mathbb R^{s+q})$, which is separable. The covariance operator $\boldsymbol\Sigma_{\boldsymbol\chi}=\E[\boldsymbol\chi(\mathbf V)\otimes\boldsymbol\chi(\mathbf V)]$ is then positive, self-adjoint and trace class. By the spectral theorem for compact self-adjoint operators \citep{schwartz1963linear}, its positive eigenvalues, repeated according to multiplicity, are $\lambda_1\ge\lambda_2\ge\cdots>0$, with orthonormal eigenvectors $u_1,u_2,\ldots\in\mathcal H$, and its null space is the orthogonal complement of $\{u_k\}$. Let
\[
\psi_k(\mathbf V)=\lambda_k^{-1/2}\langle\boldsymbol\chi(\mathbf V),u_k\rangle_{\mathcal H}.
\]
Then $\E[\psi_k(\mathbf V)\psi_l(\mathbf V)]=(\lambda_k\lambda_l)^{-1/2}\langle\boldsymbol\Sigma_{\boldsymbol\chi}u_k,u_l\rangle_{\mathcal H}=\mathbf 1(k=l)$, because the $u_k$ are orthonormal eigenvectors. Moreover, for every $u$ in the null space of $\boldsymbol\Sigma_{\boldsymbol\chi}$, we have $\E\langle\boldsymbol\chi(\mathbf V),u\rangle_{\mathcal H}^2=\langle\boldsymbol\Sigma_{\boldsymbol\chi}u,u\rangle_{\mathcal H}=0$; applying this to a countable dense subset of the null space shows that $\boldsymbol\chi(\mathbf V)=\sum_k\lambda_k^{1/2}\psi_k(\mathbf V)u_k$ almost surely. Parseval's identity then yields, almost surely,
\begin{equation}\label{eq:kernel_decom_final}
\begin{aligned}
6h_{2,2}(\mathbf{V}_1,\mathbf{V}_2)&=\sum_{k=1}^{\infty}\lambda_k\psi_k(\mathbf{V}_1)\psi_k(\mathbf{V}_2),\\
\|\boldsymbol\chi(\mathbf V)\|_{\mathcal H}^2&=\sum_{k=1}^{\infty}\lambda_k\psi_k(\mathbf V)^2,
\end{aligned}
\end{equation}
where the sums are finite if $\boldsymbol\Sigma_{\boldsymbol\chi}$ has finitely many positive eigenvalues. Finally, note that $\E[6h_{2,2}(\mathbf v,\mathbf V)\psi_k(\mathbf V)]=\lambda_k^{-1/2}\langle\boldsymbol\chi(\mathbf v),\boldsymbol\Sigma_{\boldsymbol\chi}u_k\rangle_{\mathcal H}=\lambda_k\psi_k(\mathbf v)$, so $\{\lambda_k,\psi_k\}_{k\ge1}$ are the eigenvalues and orthonormal eigenfunctions associated with the kernel $h=6h_{2,2}$ in Theorem~\ref{thm:null_fixed_p}; by \eqref{eq:kernel_decom_final}, this kernel has no other nonzero eigenvalues.

Moreover, $\E[\psi_k(\mathbf V)]=\lambda_k^{-1/2}\langle\E\boldsymbol\chi(\mathbf V),u_k\rangle_{\mathcal H}=0$ for all $k$, because $\E\boldsymbol\chi(\mathbf V)=\E[e]\,\E[\Phi(\mathbf W)-m_\Phi]=0$ by Condition (A3).

To derive the limiting distribution, we use the truncation method. Let $\boldsymbol{\psi}^K(\mathbf{V}_i) = (\psi_1(\mathbf{V}_i), \dots, \psi_K(\mathbf{V}_i))^\top \in \mathbb{R}^K$. By the multivariate Central Limit Theorem (CLT), for any fixed positive integer $K$, we have:
\begin{equation}\label{eq:MCLT}
    \begin{pmatrix}
        \frac{1}{\sqrt{n}} \sum_{i=1}^n \boldsymbol{\psi}^K(\mathbf{V}_i) \\
        \boldsymbol\xi^\circ(e) \\
        \frac{1}{\sqrt{n}} \sum_{i=1}^n 4h_{1,1}(\mathbf{V}_i)
    \end{pmatrix} \rightarrow^{D} N(\mathbf{0}, \widetilde{\boldsymbol{\Sigma}}),
\end{equation}%
where the joint covariance matrix $\widetilde{\boldsymbol{\Sigma}}$ is given by:
\[
\widetilde{\boldsymbol{\Sigma}} = \begin{pmatrix}
    \mathbf{I}_K & \mathbb{E}[\boldsymbol{\psi}^K(\mathbf{V}) \boldsymbol{\phi}_p(\mathbf{Z})^\top e] & \mathbb{E}[\boldsymbol{\psi}^K(\mathbf{V}) 4h_{1,1}(\mathbf{V})^\top] \\
    \mathbb{E}[e \boldsymbol{\phi}_p(\mathbf{Z}) \boldsymbol{\psi}^K(\mathbf{V})^\top] & \sigma^2 \mathbf{I}_p & \mathbb{E}[e \boldsymbol{\phi}_p(\mathbf{Z}) 4h_{1,1}(\mathbf{V})^\top] \\
    \mathbb{E}[4h_{1,1}(\mathbf{V}) \boldsymbol{\psi}^K(\mathbf{V})^\top] & \mathbb{E}[4h_{1,1}(\mathbf{V}) e \boldsymbol{\phi}_p(\mathbf{Z})^\top] & \text{var}\{4h_{1,1}(\mathbf{V})\}
\end{pmatrix}.
\]
For fixed $K$, define the truncated statistic
\[
\mathcal U_{n,1}^K=\frac1n\sum_{i\ne j}\sum_{k=1}^K\lambda_k\psi_k(\mathbf V_i)\psi_k(\mathbf V_j).
\]
By \eqref{eq:MCLT}, the law of large numbers for $n^{-1}\sum_i\psi_k(\mathbf V_i)^2$, and the continuous mapping theorem,
\begin{align}
\mathcal{U}_{n,1}^K
&=\sum_{k=1}^K\lambda_k\left[\left\{\frac{1}{\sqrt{n}}\sum_{i=1}^n\psi_k(\mathbf{V}_i)\right\}^2-\frac{1}{n}\sum_{i=1}^n\psi_k(\mathbf{V}_i)^2\right],\nonumber\\
\mathcal{U}_{n,1}^K+\mathcal{U}_{n,2}-\mathcal{U}_{n,3}
&\rightarrow^{D}\sum_{k=1}^K\lambda_k(G_k^2-1)+\boldsymbol{\mathcal{N}}^\top\breve{\mathbf{H}}\boldsymbol{\mathcal{N}}-\boldsymbol{\mathcal{T}}^\top\boldsymbol{\mathcal{N}},\label{eq:fixed_K_dist}
\end{align}%
where $\{G_k\}$ are independent standard normal random variables, $\boldsymbol{\mathcal{N}} \sim N(\mathbf{0}, \sigma^2 \mathbf{I}_p)$, and $\boldsymbol{\mathcal{T}} \sim N(\mathbf{0}, \text{var}(4h_{1,1}(\mathbf{V})))$. Here $(G_1,\ldots,G_K)^\top$, $\boldsymbol{\mathcal N}$, and $\boldsymbol{\mathcal T}$ are jointly distributed as the limit $N(\mathbf 0,\widetilde{\boldsymbol\Sigma})$ in \eqref{eq:MCLT}; since $4h_{1,1}=2K_p$, $\text{var}(4h_{1,1}(\mathbf V))=4\,\text{var}(K_p(\mathbf V))$, as in Theorem~\ref{thm:null_fixed_p}. 

Given that $\mathbb{E}[6^2 h_{2,2}(\mathbf{V}, \mathbf{V}')^2] = \mathbb{E}[U(\mathbf{W}, \mathbf{W}')^2 e^2 (e')^2] = \sum_{k=1}^\infty \lambda_k^2 < \infty$ under the moment conditions $\mathbb{E}[\|\mathbf{W}\|^2 + e^2] < \infty$ and $\mathbb{E}[\|\mathbf{W}\|^2 e^2] < \infty$, it follows that:
\[
\mathbb{E}\left[ (\mathcal{U}_{n,1} - \mathcal{U}_{n,1}^K)^2 \right] = \frac{2(n-1)}{n}\sum_{k=K+1}^\infty \lambda_k^2 \to 0 \quad \text{as } K \to \infty.
\]
By Markov's inequality, for any $\delta > 0$:
\begin{equation}\label{eq:limsup}
    \lim_{K \to \infty} \limsup_{n \to \infty} P\left( \left| (\mathcal{U}_{n,1} + \mathcal{U}_{n,2} - \mathcal{U}_{n,3}) - (\mathcal{U}_{n,1}^K + \mathcal{U}_{n,2} - \mathcal{U}_{n,3}) \right| \geq \delta \right) = 0.
\end{equation}
Applying Theorem 2 in \citet{dehling2009new} alongside \eqref{eq:nMDD_sim1_fixed_p_final}, \eqref{eq:fixed_K_dist}, and \eqref{eq:limsup}, we conclude:
\begin{equation*}
    n\widehat{\rm{MDD}}^2(\widehat{e} \mid \mathbf{W}) \rightarrow^{D} \sum_{k=1}^\infty \lambda_k (G_k^2 - 1) + \boldsymbol{\mathcal{N}}^\top \breve{\mathbf{H}} \boldsymbol{\mathcal{N}} - \boldsymbol{\mathcal{T}}^\top \boldsymbol{\mathcal{N}}.
\end{equation*}%
\end{proof}

\begin{proof}[Proof of Theorem~\ref{thm:null}]
For simplicity of notation, let
\begin{gather*}
\mathbf{g}_p(\mathbf{Z}) = (g_1(\mathbf{Z}), \cdots, g_p(\mathbf{Z}))^\top,\qquad
\mathbf{g}_{p,i}=\mathbf{g}_p(\mathbf{Z}_i),\\
\bar{\mathbf{g}}_p=n^{-1}\sum_i \mathbf{g}_{p,i},\qquad
\mathbf{b}_{p,i}=\mathbf{g}_{p,i}-\bar{ \mathbf{g}}_p
\end{gather*}%
and
\[
\mathbf{B}_n=\begin{pmatrix}\mathbf{b}_{p,1}^\top \\ \vdots\\ \mathbf{b}_{p,n}^\top\end{pmatrix} \in \R^{n\times p},
\qquad
\widehat{\mathbf{Q}}_b=\frac{1}{n}\mathbf{B}_n^\top \mathbf{B}_n\in \R^{p \times p}, \qquad \mathbf{G}=\begin{pmatrix}\mathbf{g}_{p,1}^\top\\ \vdots\\ \mathbf{g}_{p,n}^\top\end{pmatrix}\in \R^{n \times p}.
\]
By Condition (B2), the full OLS design matrix, including the intercept, has the same column space as $[{\bf 1}, \mathbf{G}]\in \R^{n\times (p+1)}$, so the OLS fitted values and residuals are those of the regression on $[{\bf 1},\mathbf G]$.
The fitted and residual projections are
\begin{equation}\label{eq:P_n, Pi_n}
\mathbf{P}_n=\frac{\mathbf1\mathbf1^\top}{n}
+\frac1n\mathbf{B}_n\widehat{\mathbf{Q}}_b^{-1}\mathbf{B}_n^\top,\qquad \mathbf{\Pi}_n=\mathbf{I}_n-\mathbf{P}_n,
\end{equation}
which provide us with
\begin{equation}\label{eq:zero_design}
\mathbf{\Pi}_n\mathbf1=\bf 0,\quad \mathbf{\Pi}_n\mathbf{G}=\bf 0.
\end{equation}
Let $\widehat{\mathbf{Q}}_{g}=n^{-1}\sum_i \mathbf{g}_{p,i}\mathbf{g}_{p,i}^\top$. Since
$\E[\mathbf{g}_p\mathbf{g}_p^\top]=\mathbf{I}_p$, we have $1\le p=\E\|\mathbf g_p\|^2\le\xi_p^2$, so that $p/n\le\xi_p^2p/n\to0$ and $\xi_p^2/n\le\xi_p^2p/n\to0$ by Condition (B2); moreover, $\E\|\bar{\mathbf{g}}_p\|^2=p/n\to0$ and 
\[
\E\|\widehat{\mathbf{Q}}_{g}-\mathbf{I}_p\|_F^2
=\frac1n\{\E\|\mathbf{g}_p\|^4-p\}
\le\frac{\xi_p^2p}{n}\to0, 
\]
which implies that
\[
\|\widehat{\mathbf{Q}}_b-\mathbf{I}_p\|_{2}
\le\|\widehat{\mathbf{Q}}_{g}-\mathbf{I}_p\|_F+\|\bar{\mathbf{g}}_p\|^2=o_p(1).
\]
Hence, with probability tending to one,
\[
\lambda_{\min}(\widehat{\mathbf{Q}}_b)\ge\frac12,\qquad
\|\widehat{\mathbf{Q}}_b^{-1}\|_{2}\le2.
\]
This guarantees that $\widehat{\mathbf{Q}}_b^{-1}$ involved in $\mathbf{P}_n$ remains stable
when $p\to\infty$.

 Under the null hypothesis and by \eqref{eq:zero_design}, the residual becomes:
\begin{equation}\label{eq:e_hat}
\widehat e=\mathbf{\Pi}_nY
=\mathbf{\Pi}_n(\gamma_p\mathbf 1+\mathbf{G}\boldsymbol{\beta}_p+R_p+e)
=\mathbf{\Pi}_nR_p+\mathbf{\Pi}_ne,
\end{equation}%
where $Y=(Y_1,\ldots,Y_n)^\top$, $e=(e_1,\ldots,e_n)^\top$, and $R_p=(R_{p,1}, \cdots, R_{p,n})^\top$ with
\[
R_{p,i}
=
f(\mathbf{Z}_i)-\gamma_p-\boldsymbol{\beta}_p^\top \mathbf{g}_p(\mathbf{Z}_i),
\qquad
|R_{p,i}|\le a_p\quad\text{by Condition (B3)}.
\]

By the definitions of $h_p(\mathbf W)$ and $h_\infty(\mathbf W)$, we have
\[
\E [h_p(\mathbf W)]=0,\quad \E[g_j(\mathbf{Z})h_p(\mathbf W)]=0\ (j\le p),\quad
\E\|h_p(\mathbf W)-h_\infty(\mathbf W)\|_{\Hcal}^2\to0;
\]
for the last property, the orthonormality of the $g_j$ and Bessel's inequality give
\[
\sum_{j\ge1}\|\beta_j^\Phi\|_{\Hcal}^2\le\E\|\Phi(\mathbf W)-m_\Phi\|_{\Hcal}^2=\mu_D,
\]
so that $h_p$ converges in $L^2$ and $\E\|h_p(\mathbf W)-h_\infty(\mathbf W)\|_{\Hcal}^2=\sum_{j>p}\|\beta_j^\Phi\|_{\Hcal}^2\to0$.
We denote 
\[\boldsymbol\Phi:=(\Phi(\mathbf W_1),\cdots,\Phi(\mathbf W_n))^\top \in \mathcal{H}^n,\quad \mathbf{h}_p:=(h_p(\mathbf W_1),\cdots,h_p(\mathbf W_n))^\top\in \mathcal{H}^n.\]

By Lemma~\ref{lemma:test_statistic}, our test statistic becomes
\[
T_n
:=n\widehat{\MDD}^{\,2}(\widehat e\mid \mathbf W)=
\frac{n}{n-3}
\left\|
\frac1{\sqrt n}\sum_{i=1}^n \widehat{e}_i\Phi(\mathbf W_i)
\right\|_{\Hcal}^2
-
\frac{1}{n-3}\sum_i d_i \widehat{e}_i^2,
\]
where $d_i=2R_i/(n-2)
-R_{++}/\{(n-1)(n-2)\}$, $R_i=\sum_j A_{ij}
$, and $R_{++}=\sum_{i,j}A_{ij}$, $A_{ij}=\|\mathbf W_i-\mathbf W_j\|$.

Under the null and Assumption~\ref{assum:1}, we shall show that
\[
T_n=n\widehat{\MDD}^{\,2}(\widehat e\mid \mathbf W)
\rightarrow^D
\mathcal{L}_{0,\infty}:=\|\mathcal{G}_\infty\|_{\Hcal}^2-\sigma^2\mu_D,
\]
where $\mathcal{G}_\infty$ is centered Gaussian in $\Hcal$ with covariance
$\sigma^2\boldsymbol{\Sigma}_\infty$, $\boldsymbol{\Sigma}_\infty=\E[h_\infty\otimes h_\infty]$.

We treat the feature-score term and the diagonal correction separately. For the first term, we make three successive replacements and then apply a Hilbert-space central limit theorem. Each replacement is established first for the scores in $\mathcal H$ and then for their squared norms. 
\begin{enumerate}
    \item We show that the first term of $T_n$ converges in distribution to $\|\mathcal{G}_\infty\|_{\Hcal}^2$, i.e., \begin{equation*}\left\|
\frac1{\sqrt n}\sum_{i=1}^n \widehat{e}_i\Phi(\mathbf W_i)
\right\|_{\Hcal}^2\rightarrow^D\|\mathcal{G}_\infty\|_{\mathcal{H}}^2.\end{equation*}
In order to establish this result, we follow the steps below:
\begin{enumerate}
\item The first term of $T_n$ can be rewritten as \begin{equation}\label{eq:part_a}\left\|
\frac1{\sqrt n}\sum_{i=1}^n \widehat{e}_i\Phi(\mathbf W_i)
\right\|_{\Hcal}^2=\left\|
\frac1{\sqrt n}\sum_{i=1}^n e_i t_i
\right\|_{\Hcal}^2+o_p(1)\end{equation} 
\item The leading term can be reduced to\begin{equation}\label{eq:part_b}\left\|
\frac1{\sqrt n}\sum_{i=1}^n e_it_i
\right\|_{\Hcal}^2=\left\|
\frac1{\sqrt n}\sum_{i=1}^n e_i h_p(\mathbf  W_i)
\right\|_{\Hcal}^2+o_p(1).\end{equation}
\item We further simplify the leading term as \begin{equation}\label{eq:part_c}\left\|
\frac1{\sqrt n}\sum_{i=1}^n e_i h_p(\mathbf  W_i)
\right\|_{\Hcal}^2=\left\|
\frac1{\sqrt n}\sum_{i=1}^n e_i h_\infty(\mathbf  W_i)
\right\|_{\Hcal}^2+o_p(1).\end{equation}
\item Lastly, we show that \[\frac1{\sqrt n}\sum_{i=1}^n e_i h_\infty(\mathbf  W_i)
\rightarrow^D
\mathcal{G}_\infty,
\] which implies that $\left\|\frac1{\sqrt n}\sum_{i=1}^n e_i h_\infty(\mathbf W_i)\right\|_{\mathcal{H}}^2 \rightarrow^D \left\|\mathcal{G}_\infty \right\|_{\mathcal{H}}^2$.
\end{enumerate}
\item We show that the second term of $T_n$ converges in probability to a constant $\sigma^2\mu_D$, i.e., 
\begin{equation}\label{eq:diag_null}\frac{1}{n-3}\sum_i d_i \widehat{e}_i^2\rightarrow^p \sigma^2\mu_D.
\end{equation}
\end{enumerate}

For simplicity of notation, we denote  \[\Phi_i = \Phi(\mathbf W_i),  \quad h_{p, i} = h_p(\mathbf W_i).
\]
To show \eqref{eq:part_a}, we use the fact that $\mathbf \Pi_n$ is symmetric and obtain
\begin{equation}\label{eq:Phi_t_equality}
\begin{aligned}
\frac{1}{\sqrt{n}}\sum_{i=1}^n\widehat{e}_i\Phi_i
&=\frac{1}{\sqrt{n}}\sum_{i=1}^n(\mathbf \Pi_ne + \mathbf \Pi_n R_{p})_i\Phi_i
=
\frac{1}{\sqrt{n}}\sum_{i=1}^n\sum_{j=1}^n
(\mathbf \Pi_n)_{ij} (e_j+R_{p,j})\Phi_i\\
&=\frac{1}{\sqrt{n}}
\sum_{j=1}^n(e_j+R_{p,j})
\left\{
\sum_{i=1}^n(\mathbf \Pi_n)_{ji}\Phi_i
\right\}=\frac{1}{\sqrt{n}}
\sum_{j=1}^n(e_j+R_{p,j})t_j,
\end{aligned}
\end{equation}
where $t_i =(\mathbf{\Pi}_n\boldsymbol\Phi)_i$ and the first equation is obtained by \eqref{eq:e_hat}. 

Next, we shall show that 
\[
\left\|\frac{1}{\sqrt{n}}\sum_i R_{p,i}t_i\right\|_{\Hcal}=o_p(1),
\]
which implies \eqref{eq:part_a}. 
By Cauchy--Schwarz inequality,
\begin{align*}
\left\|\frac1{\sqrt n}\sum_i R_{p,i}t_i\right\|_{\Hcal}
&\le\frac1{\sqrt n}
\left(\sum_i R_{p,i}^2\right)^{1/2}
\left(\sum_i\|t_i\|_{\Hcal}^2\right)^{1/2}\\
&\leq \sqrt{n}a_p \left(\frac{1}{n}\sum_i\|t_i\|_{\Hcal}^2\right)^{1/2},
\end{align*}%
due to the fact that $|R_{p,i}^2|\le a_p^2$ in (B3). Since $\mathbf \Pi_n$ is an orthogonal
projection and $\|\Phi(\mathbf W_i)\|^2=2\|\mathbf W_i\|\le2M$ by (B1), we also have
\[
\sum_i\|t_i\|_\mathcal{H}^2=\sum_i \|(\mathbf \Pi_n \boldsymbol\Phi)_i\|_\mathcal{H}^2\le\sum_i\|\Phi(\mathbf W_i)\|_\mathcal{H}^2\le 2nM.
\]
Hence, we obtain
\[
\left\|\frac1{\sqrt n}\sum_i R_{p,i}t_i\right\|_\Hcal
\le \sqrt{2M}\sqrt n\,a_p=o(1),
\]
which implies that 
\[
\frac1{\sqrt n}\sum_i\widehat e_i\Phi_i
=
\frac1{\sqrt n}\sum_ie_it_i+o_p(1).
\]
For squared norms, conditional on $\mathbf{W}_{1:n}=\{\mathbf W_1, \cdots, \mathbf W_n\}$, we have
\[
\E\left(\left\|\frac{1}{\sqrt{n}}\sum_ie_it_i\right\|_\mathcal{H}^2\mid \mathbf{W}_{1:n}\right)
=\frac{\sigma^2}{n}\sum_i\|t_i\|_\Hcal^2\le\sigma^2(2M),
\]
so $\left\|\frac{1}{\sqrt{n}}\sum_ie_it_i\right\|_\mathcal{H}^2=O_p(1)$, while $\left\|\frac1{\sqrt n}\sum_i R_{p,i}t_i\right\|_\Hcal=o_p(1)$. Hence
\begin{align*}
&\left\|\frac{1}{\sqrt{n}}\sum_ie_it_i+\frac1{\sqrt n}\sum_i R_{p,i}t_i\right\|_\mathcal{H}^2-\left\|\frac{1}{\sqrt{n}}\sum_ie_it_i\right\|_\mathcal{H}^2\\
&\quad=2\left\langle \frac{1}{\sqrt{n}}\sum_ie_it_i,\frac1{\sqrt n}\sum_i R_{p,i}t_i\right\rangle+\left\|\frac1{\sqrt n}\sum_i R_{p,i}t_i\right\|_\mathcal{H}^2\\
&\quad =o_p(1).
\end{align*}
This proves \eqref{eq:part_a}.
For \eqref{eq:part_b}, let $\mathbf t=(t_1,\ldots,t_n)^\top$ and recall $\boldsymbol\Phi$ and $\mathbf h_p$ defined above. Since $\Phi_i=m_\Phi+\mathcal B_p\mathbf g_{p,i}+h_{p,i}$, and \eqref{eq:zero_design} gives
\[
\mathbf\Pi_n(\mathbf 1m_\Phi)=\mathbf 0
\quad\text{and}\quad
\mathbf\Pi_n(\mathcal B_p\mathbf g_{p,1},\ldots,\mathcal B_p\mathbf g_{p,n})^\top=\mathbf\Pi_n\mathbf G(\beta_1^\Phi,\ldots,\beta_p^\Phi)^\top=\mathbf 0,
\]
we have
\[
\mathbf{t}=\mathbf \Pi_n\mathbf \Phi=\mathbf \Pi_n\mathbf{h}_p.
\]
Thus, we have
\begin{equation}\label{eq:t-h_p}
\mathbf{t}-\mathbf{h}_p=-\mathbf P_n\mathbf{h}_p.
\end{equation}
We shall first show that 
\begin{equation}\label{eq:t-h_p_square_norm}
\frac{1}{n} \sum_{i=1}^n\|t_i-h_{p,i}\|_{\Hcal}^2=O_p\left(\frac{\xi_p^2}{n}\right)=o_p(1).
\end{equation}
By using the explicit projection,
\[
\mathbf P_n\mathbf h_p=\mathbf1\overline{ h}_p+\mathbf B_n \widehat{\mathbf{Q}}_b^{-1}\mathbf C_n,
\]
where
\[
\overline h_p=\frac{1}{n}\sum_{i=1}^n h_{p,i}\in \mathcal{H},\qquad
\mathbf C_n=n^{-1}\mathbf B_n^\top \mathbf h_p=\frac{1}{n}\sum_{i=1}^n \mathbf{b}_{p,i} h_{p,i}\in \mathcal{H}^p.
\]
Based on this fact, we have
\[
\frac{1}{n}\|\mathbf P_n\mathbf h_p\|_{\mathcal{H}}^2=\frac{1}{n}\|\mathbf1\bar h_p\|_{\mathcal{H}}^2+\frac{1}{n}\|\mathbf B_n\widehat{\mathbf{Q}}_b^{-1}\mathbf C_n\|_{\mathcal{H}}^2 = \|\bar h_p\|_{\mathcal{H}}^2+\frac{1}{n}\|\mathbf B_n\widehat{\mathbf{Q}}_b^{-1} \mathbf C_n\|_{\mathcal{H}}^2,
\]
since $\langle \mathbf1\bar h_p, \mathbf B_n\widehat{\mathbf{Q}}_b^{-1} \mathbf C_n\rangle =0$ due to the fact that $\mathbf1^\top \mathbf B_n=0$.
Also, we have 
\begin{align*}
\E\|\bar h_p\|_{\mathcal{H}}^2 &= \E\left\|\frac{1}{n}\sum_i h_{p,i} \right\|_{\mathcal{H}}^2 = \frac{1}{n^2} \sum_i \E\|h_{p,i}\|_{\mathcal{H}}^2\\
&=\frac{1}{n}\E\|h_{p}(\mathbf W)\|_{\mathcal{H}}^2\leq \frac{\mu_D}{n}\rightarrow 0,
\end{align*}%
due to the fact that $\E\|h_{p}(\mathbf W)\|_{\mathcal{H}}^2 + \sum_{j=1}^p \|\beta_j^\Phi\|_\mathcal{H}^2= \E[\|\Phi(\mathbf W)-m_\Phi\|^2]=\mu_D$ because $\E\langle h_p(\mathbf W),\mathcal{B}_p \mathbf g_p(\mathbf Z)\rangle_{\mathcal{H}}=0$.
Also, 
\begin{align*}
\frac{1}{n}\|\mathbf B_n\widehat{\mathbf{Q}}_b^{-1} \mathbf C_n\|_{\mathcal{H}}^2&=
\langle \widehat{\mathbf{Q}}_b^{-1} \mathbf C_n, \frac{1}{n}\mathbf B_n^\top \mathbf B_n\widehat{\mathbf{Q}}_b^{-1} \mathbf C_n\rangle_{\mathcal{H}}=\langle \widehat{\mathbf{Q}}_b^{-1} \mathbf C_n, \widehat{\mathbf{Q}}_b\widehat{\mathbf{Q}}_b^{-1} \mathbf C_n\rangle_{\mathcal{H}} \\
&= \langle \widehat{\mathbf{Q}}_b^{-1}\mathbf C_n, \mathbf C_n\rangle_{\mathcal{H}}\leq \|\widehat{\mathbf{Q}}_b^{-1}\|_{2} \|\mathbf C_n\|_{\mathcal{H}^p}^2,
 \end{align*}
 where $\|\widehat{\mathbf{Q}}_b^{-1}\|_{2} =O_p(1)$ and note that
 \[
 \mathbf C_n =\frac{1}{n}\sum_{i=1}^n (\mathbf g_{p,i} -\bar{\mathbf g}_p) h_{p,i} = \frac{1}{n}\sum_{i=1}^n \mathbf g_{p,i}h_{p,i} - \bar{\mathbf g}_p\overline{h}_p,
 \]
 which gives us
 \begin{equation}\label{eq:C_n}
     \E\|\mathbf C_n\|_{\mathcal{H}^p}^2 \leq C\left( \E\left\|\frac{1}{n}\sum_{i=1}^n \mathbf g_{p,i}h_{p,i}\right\|_{\mathcal{H}^p}^2+\E\|\bar{\mathbf g}_p\overline{h}_p\|_{\mathcal{H}^p}^2\right).
 \end{equation}
 Note that 
 $\E[\mathbf g_{p,i}h_{p,i}]=\mathbf 0$ by the second property of $h_p$ listed above, so the cross terms vanish and
 \[
 \E\left\|\frac{1}{n}\sum_i \mathbf g_{p,i}h_{p,i}\right\|_{\mathcal{H}^p}^2=\frac{1}{n^2}\sum_{i=1}^n\E\|\mathbf g_{p,i}h_{p,i}\|_{\mathcal{H}^p}^2=\frac{1}{n}\E[\| \mathbf g_p\|^2\|h_{p}\|_{\mathcal{H}}^2]\leq\frac{\xi_p^2\mu_D}{n}\rightarrow 0,
 \]%
by (B2) and due to the fact that $\E\|h_{p}\|_{\mathcal{H}}^2\leq \mu_D$. Further, 
since $\|\bar{\mathbf g}_p\|\le\xi_p$ and $\E\|\bar h_p\|_{\mathcal H}^2\le\mu_D/n$ (shown above),
\[
\E\|\bar{\mathbf g}_p\bar{h}_p\|_{\mathcal{H}^p}^2\leq \E[\|\bar{\mathbf g}_p\|^2 \|\bar{h}_p\|_{\mathcal{H}}^2 ]\leq \xi_p^2 \E[\|\bar{h}_p\|_{\mathcal{H}}^2]\leq \frac{\xi_p^2\mu_D}{n}\rightarrow 0.
\] 
 Consequently, by \eqref{eq:C_n} with $C=2$, $\E\|\mathbf C_n\|_{\mathcal{H}^p}^2\le4\xi_p^2\mu_D/n$, so that $\|\mathbf C_n\|_{\mathcal{H}^p}^2=O_p(\xi_p^2/n)$. Together with \eqref{eq:t-h_p}, $\E\|\bar h_p\|_{\mathcal H}^2\le\mu_D/n$, $\|\widehat{\mathbf Q}_b^{-1}\|_2=O_p(1)$, and $\xi_p^2\ge1$, this implies \eqref{eq:t-h_p_square_norm}.

\noindent Notice that 
\[
\frac1{\sqrt n}\sum_{i=1}^ne_i t_i= \frac1{\sqrt n}\sum_{i=1}^ne_i h_{p,i}+ \frac1{\sqrt n}\sum_{i=1}^ne_i\{t_i-h_{p,i}\}. 
\]
Conditional on $\mathbf{W}_{1:n}$, we have
\[
\begin{aligned}
&\E\left[\left\|\frac1{\sqrt n}\sum_{i=1}^ne_i\{t_i-h_{p,i}\}\right\|_{\mathcal{H}}^2\mid \mathbf{W}_{1:n}\right]\\
&\quad=\frac1n\sum_{i,j=1}^n
\E[e_ie_j\mid \mathbf{W}_{1:n}]
\langle t_i-h_{p,i},t_j-h_{p,j}\rangle\\
&\quad=\frac{\sigma^2}{n}\sum_{i=1}^n\|t_i-h_{p,i}\|_{\mathcal{H}}^2=o_p(1),
\end{aligned}
\]%
by \eqref{eq:t-h_p_square_norm}. To pass from this conditional bound to convergence in probability, write $A_n=n^{-1/2}\sum_i e_i(t_i-h_{p,i})$ and $V_n=\E(\|A_n\|_{\mathcal H}^2\mid\mathbf W_{1:n})$. For any $\epsilon,v>0$, conditional Markov's inequality gives
\[
P(\|A_n\|_{\mathcal H}>\epsilon)
\le P(V_n>v)+\frac{v}{\epsilon^2}.
\]
Since $V_n=o_p(1)$, letting $n\to\infty$ and then $v\downarrow0$ yields $A_n=o_p(1)$. Define $\mathcal S_t=n^{-1/2}\sum_{i=1}^ne_it_i$ and $\mathcal S_h=n^{-1/2}\sum_{i=1}^ne_ih_{p,i}$. Since $\mathcal S_t-\mathcal S_h=A_n=o_p(1)$ and $\E\|\mathcal S_h\|_{\mathcal H}^2=\sigma^2\E\|h_p(\mathbf W)\|_{\mathcal H}^2\le\sigma^2\mu_D$, both scores are $O_p(1)$. Consequently,
\begin{equation}\label{eq:quad_diff}
\big|\|\mathcal S_t\|_{\mathcal H}^2-\|\mathcal S_h\|_{\mathcal H}^2\big|
\le\|\mathcal S_t-\mathcal S_h\|_{\mathcal H}
(\|\mathcal S_t\|_{\mathcal H}+\|\mathcal S_h\|_{\mathcal H})=o_p(1).
\end{equation}
This proves \eqref{eq:part_b}.

For \eqref{eq:part_c},
\begin{align*}
\E\left[\left\|\frac1{\sqrt n}\sum_{i=1}^n
e_i\{h_p(\mathbf W_i)-h_\infty(\mathbf W_i)\}\right\|_\Hcal^2\right]
&=
\E\left[e^2\|h_p(\mathbf W)-h_\infty(\mathbf W)\|_\Hcal^2\right]\\
&=\sigma^2\E\left[\|h_p(\mathbf W)-h_\infty(\mathbf W)\|_\Hcal^2\right]\to0.
\end{align*}%
Hence $\frac1{\sqrt n}\sum_{i=1}^n
e_i\{h_p(\mathbf W_i)-h_\infty(\mathbf W_i)\}=o_p(1)$ and by using the same arguments in \eqref{eq:quad_diff}, 
\begin{equation}\label{eq:h_0-h_infty}
\left\|\frac1{\sqrt n}\sum_{i=1}^ne_i h_p(\mathbf W_i)\right\|_{\mathcal{H}}^2
=
\left\|\frac1{\sqrt n}\sum_{i=1}^ne_i h_\infty(\mathbf W_i)\right\|_{\mathcal{H}}^2+o_p(1).
\end{equation}
Combining the approximations of the scores established in the proofs of \eqref{eq:part_a}, \eqref{eq:part_b}, and \eqref{eq:part_c}, namely $\|n^{-1/2}\sum_iR_{p,i}t_i\|_{\Hcal}=o(1)$, $n^{-1/2}\sum_ie_i(t_i-h_{p,i})=o_p(1)$, and $n^{-1/2}\sum_ie_i\{h_p(\mathbf W_i)-h_\infty(\mathbf W_i)\}=o_p(1)$, we have
\begin{equation}\label{eq:first_term_sim_final}
\frac1{\sqrt n}\sum_{i=1}^n\widehat e_i\Phi(\mathbf W_i)
=
\frac1{\sqrt n}\sum_{i=1}^ne_i h_\infty(\mathbf W_i)+o_p(1).
\end{equation}
Notice that $\{e_i h_\infty(\mathbf W_i)\}_{i=1}^n$ are i.i.d. Hilbert-valued random elements with
\[\E[e h_\infty(\mathbf W)]=0, \quad
\E\|eh_\infty(\mathbf W)\|^2\mathrel{=}\sigma^2\E\|h_\infty(\mathbf W)\|^2
=
\sigma^2\operatorname{tr}(\boldsymbol{\Sigma}_\infty)\mathrel{\le}\sigma^2\mu_D<\infty.
\]
Hence, by the central limit theorem for i.i.d.\ random elements in a separable Hilbert space \citep{araujo1980central} ($\mathcal H$ can be taken separable; see the proof of Theorem~\ref{thm:null_fixed_p}), we have
\begin{equation}\label{eq:CLT}
\frac1{\sqrt n}\sum_{i=1}^ne_i h_\infty(\mathbf W_i)\rightarrow^D \mathcal{G}_\infty,
\qquad
\operatorname{Cov}(\mathcal{G}_\infty)=\sigma^2\boldsymbol{\Sigma}_\infty.
\end{equation}
By the continuous mapping theorem and \eqref{eq:first_term_sim_final} and \eqref{eq:CLT}, we further obtain 
\begin{equation}\label{eq:part1}
\left\|\frac1{\sqrt n}\sum_{i=1}^n\widehat e_i\Phi(\mathbf W_i)\right\|_{\mathcal{H}}^2
\rightarrow^D
\|\mathcal G_\infty\|_{\mathcal{H}}^2.
\end{equation}

We are left to show \eqref{eq:diag_null}. We begin by establishing the following result, which will be useful throughout the proof.
\begin{equation}\label{eq:d_i_bound}
0\le d_i\le6M,\qquad d_i
= \frac{2\sum_jA_{ij}}{n-2}
-
\frac{\sum_{j,k}A_{jk}}{(n-1)(n-2)}.
\end{equation}
For the lower bound, fix $i$. Since $A_{ii}=0$ and, by the triangle inequality, $A_{jk}\le A_{ji}+A_{ik}$ for $j,k\neq i$, we have
\[
\begin{aligned}
\sum_{j,k}A_{jk}
&=2\sum_{j}A_{ij}+\sum_{\substack{j\neq k\\ j,k \neq i}}A_{jk}\\
&\le2\sum_{j}A_{ij}+2(n-2)\sum_{j}A_{ij}
=2(n-1)\sum_{j}A_{ij},
\end{aligned}
\]
so that $d_i\ge0$. For the upper bound, $A_{ij}\le\|\mathbf W_i\|+\|\mathbf W_j\|\le2M$ by Condition (B1), so that $d_i\le2\sum_jA_{ij}/(n-2)\le4(n-1)M/(n-2)\le6M$ for $n\ge4$. This proves \eqref{eq:d_i_bound}. Moreover, since $A_{ii}=0$ for all $i$, we have
\begin{align*}
\frac1n\sum_{i=1}^n d_i
&= \frac{2\sum_{i,j}A_{ij}}{n(n-2)}-\frac{n\sum_{j,k}A_{jk}}{n(n-1)(n-2)}\\
&=\sum_{i\neq j}A_{ij} \left(\frac{2}{n(n-2)}-\frac{1}{(n-1)(n-2)}\right)\\
&=\frac{1}{n(n-1)}
\sum_{i\ne j}A_{ij}.
\end{align*}
The right-hand side is a U-statistic with the bounded kernel $\|\mathbf W_i-\mathbf W_j\|$; therefore, by the law of large numbers for U-statistics,
\begin{equation}\label{eq:mu_D}
\frac1n\sum_{i=1}^n d_i
\rightarrow^{p}
\E\|\mathbf W-\mathbf W'\|
=
\mu_D.
\end{equation}

By \eqref{eq:e_hat}, we can write
\begin{equation}\label{eq:delta}
\widehat e=e+\delta_n,\qquad
\delta_n=-\mathbf P_ne+\mathbf \Pi_nR_p,
\end{equation}
so that $\widehat e_i^2-e_i^2=2e_i\delta_{n,i}+\delta_{n,i}^2$. Under the condition (B1) and the fact in \eqref{eq:d_i_bound}, we have
\[
\begin{aligned}
\left|\frac1n\sum_i d_i(\widehat e_i^2-e_i^2)\right|
&\le
12M
\left(\frac1n\sum_ie_i^2\right)^{1/2}
\left(\frac1n\sum_i\delta_{n,i}^2\right)^{1/2}\\
&\quad+
6M\frac1n\sum_i\delta_{n,i}^2.
\end{aligned}
\]
Since $n^{-1}\sum_ie_i^2\rightarrow^p\sigma^2$, it suffices to show that $n^{-1}\|\mathbf P_ne\|^2$ and $n^{-1}\|\mathbf\Pi_nR_p\|^2$ are $o_p(1)$. Because $\mathbf P_n$ is an orthogonal projection of rank at most $p+1$ that depends only on $\mathbf W_{1:n}$, and $\E(ee^\top\mid\mathbf W_{1:n})=\sigma^2\mathbf I_n$,
\[
\E\left(\|\mathbf P_ne\|^2\,\middle|\,\mathbf W_{1:n}\right)
=\E\left(e^\top\mathbf P_ne\,\middle|\,\mathbf W_{1:n}\right)
=\sigma^2\operatorname{tr}(\mathbf P_n)
\le\sigma^2(p+1),
\]
so that, by the conditional Markov inequality and Condition (B2),
\begin{equation}\label{eq:Pi_epsilon}
\frac{1}{n}\|\mathbf P_n e\|^2
=
O_p\left(\frac{p}{n}\right)
=
o_p(1).
\end{equation}
Moreover, since $\mathbf\Pi_n$ is an orthogonal projection and $|R_{p,i}|\le a_p$ by Condition (B3), $n^{-1}\|\mathbf\Pi_nR_p\|^2\le n^{-1}\sum_{i=1}^nR_{p,i}^2\le a_p^2=o(1)$. Combining the two bounds, we obtain
\begin{equation}\label{eq:delta_square}
\frac{1}{n}\|\delta_n\|^2
\le\frac{2}{n}\|\mathbf P_ne\|^2+\frac{2}{n}\|\mathbf \Pi_nR_p\|^2
=O_p\left(\frac{p}{n}+a_p^2\right)=o_p(1).
\end{equation}%
Thus,
\[
\frac1n\sum_i d_i\widehat e_i^2
=
\frac1n\sum_i d_ie_i^2+o_p(1).
\]
Note that
\begin{equation}\label{eq:diag_sim_part3}
\frac1n\sum_i d_ie_i^2
=
\sigma^2
\frac1n\sum_i d_i
+
\frac1n
\sum_i
d_i(e_i^2-\sigma^2).
\end{equation}

Conditional on $\mathbf{W}_{1:n}$, the $d_i$'s are fixed, while the $e_i$'s are i.i.d.\ and independent of $\mathbf W_{1:n}$, with $\E(e_i^2)=\sigma^2$ and $\operatorname{Var}(e_i^2)<\infty$ by Condition (B1). Hence, by \eqref{eq:d_i_bound},
\[
\begin{aligned}
\E\Big[\frac1n\sum_id_i(e_i^2-\sigma^2)\,\Big|\,\mathbf W_{1:n}\Big]&=0,\\
\operatorname{Var}\Big(\frac1n\sum_id_i(e_i^2-\sigma^2)\,\Big|\,\mathbf W_{1:n}\Big)&=\frac{\operatorname{Var}(e^2)}{n^2}\sum_id_i^2\le\frac{36M^2\operatorname{Var}(e^2)}{n},
\end{aligned}
\]
and the conditional Chebyshev inequality gives
\begin{equation}\label{eq:diag_sim_part4}
\frac1n
\sum_i
d_i(e_i^2-\sigma^2)
=
o_p(1).
\end{equation}
By \eqref{eq:mu_D}, \eqref{eq:diag_sim_part3}, and \eqref{eq:diag_sim_part4},
\begin{equation}\label{eq:d_i epsilon_i^2 limit}
\frac1n
\sum_i
d_ie_i^2
\rightarrow^{p}
\sigma^2\mu_D,
\end{equation}
which, combined with $n^{-1}\sum_id_i\widehat e_i^2=n^{-1}\sum_id_ie_i^2+o_p(1)$, implies that
\begin{equation}\label{eq:part2}
\frac1n
\sum_i
d_i\widehat{e}_i^2
\rightarrow^{p}
\sigma^2\mu_D.
\end{equation}
Consequently, since $n/(n-3)\to1$, by the expression for $T_n$ above, \eqref{eq:part1}, \eqref{eq:part2}, and Slutsky's theorem, we obtain
\[
T_n\mathrel{\rightarrow^D}
\|\mathcal G_\infty\|_{\Hcal}^2-\sigma^2\mu_D
=
\mathcal L_{0,\infty}.
\]
Finally, let $\rho_{\infty,1}\ge\rho_{\infty,2}\ge\cdots\ge0$ denote the eigenvalues of $\boldsymbol{\Sigma}_\infty$; by Condition (B4), $\operatorname{tr}(\boldsymbol{\Sigma}_\infty)>0$, so that $\rho_{\infty,1}>0$. The spectral representation provides
\[
\|\mathcal G_\infty\|^2
\overset D=
\sigma^2\sum_{j\ge1}\rho_{\infty,j}\widetilde{G}_j^2,
\]
where $\{\widetilde{G}_j\}_{j\geq 1}$ are independent standard normal random variables. Since $\sigma^2>0$ by Condition (B1), the term $\sigma^2\rho_{\infty,1}\widetilde G_1^2$ has a continuous distribution and is independent of the remaining nonnegative terms, so $\|\mathcal G_\infty\|_{\mathcal H}^2$ has a continuous distribution. Hence the cdf of
$\mathcal L_{0,\infty}$ is continuous. Weak convergence to a continuous cdf implies
uniform cdf convergence:
\[
\sup_t
\left|
P(T_n\le t)-P(\mathcal L_{0,\infty}\le t)
\right|
\to0.
\]
This completes the proof.
\end{proof}

\begin{proof}[Proof of Proposition~\ref{prop:fixed_p_feature}]
We construct the Gaussian quantities in Theorem~\ref{thm:null_fixed_p} jointly. We then remove the basis projection from the oracle feature-score limit and express the null limit as a shifted squared norm.

Let $\mathbf g_p(\mathbf Z)
=\breve{\boldsymbol{\phi}}_p(\mathbf Z)$
denote the centered and orthonormalized basis vector,
so that $\mathbb E[\mathbf g_p(\mathbf Z)]=\mathbf 0$ and $\mathbb E[\mathbf g_p(\mathbf Z)\mathbf g_p(\mathbf Z)^\top]=\mathbf I_p$, and recall that $h_p(\mathbf W)=\Phi(\mathbf W)-m_\Phi-\mathcal B_p\mathbf g_p(\mathbf Z)$ and $\boldsymbol{\Sigma}_p=\mathbb E[h_p(\mathbf W)\otimes h_p(\mathbf W)]$.
We note that $\Phi:\mathbb R^{s+q}\to\mathcal H$ satisfies
\begin{equation}\label{eq:fixed_equiv_kernel}
\left\langle
\Phi(\mathbf w)-m_\Phi,
\Phi(\mathbf v)-m_\Phi
\right\rangle_{\mathcal H}
=
-U(\mathbf w,\mathbf v).
\end{equation}
Moreover, by \eqref{grow:feature_supp},
\begin{equation}\label{eq:fixed_equiv_muD}
\mathbb E\|\Phi(\mathbf W)-m_\Phi\|_{\mathcal H}^2
=\mathbb E\|\Phi(\mathbf W)\|_{\mathcal H}^2-\|m_\Phi\|_{\mathcal H}^2
=2\mathbb E\|\mathbf W\|-\{2\mathbb E\|\mathbf W\|-\mu_D\}
=\mu_D.
\end{equation}%
We first construct the Gaussian variables appearing
in Theorem~\ref{thm:null_fixed_p} on a common
probability space.
Define
\[
\mathcal G_{0,n}
=\frac{1}{\sqrt n}\sum_{i=1}^ne_i\{\Phi(\mathbf W_i)-m_\Phi\}
\quad\text{and}\quad
\bm{\mathcal N}_n
=\frac{1}{\sqrt n}\sum_{i=1}^ne_i\mathbf g_p(\mathbf Z_i).
\]%
The pairs of summands are i.i.d. and centered in the product Hilbert space $\mathcal H\times\mathbb R^p$, because $e$ has mean zero and is independent of $\mathbf W$. By \eqref{eq:fixed_equiv_muD} and the orthonormality of $\mathbf g_p$, their second moments satisfy
\[
\E\!\left[e^2\{\|\Phi(\mathbf W)-m_\Phi\|_{\mathcal H}^2+\|\mathbf g_p(\mathbf Z)\|^2\}\right]
=\sigma^2(\mu_D+p)<\infty.
\]
The central limit theorem in $\mathcal H\times\mathbb R^p$ therefore yields
\begin{equation}\label{eq:fixed_equiv_jointCLT}
(\mathcal G_{0,n},\bm{\mathcal N}_n)
\rightarrow^{D}
(\mathcal G_0,\bm{\mathcal N}),
\end{equation}
where $(\mathcal G_0,\bm{\mathcal N})$
is jointly centered Gaussian with
\begin{equation}\label{eq:fixed_equiv_cov0}
\operatorname{Cov}(\mathcal G_0)
=
\sigma^2\mathbb E\left[
\{\Phi(\mathbf W)-m_\Phi\}
\otimes
\{\Phi(\mathbf W)-m_\Phi\}
\right], \qquad \operatorname{Cov}(\bm{\mathcal N})
=
\sigma^2\mathbf I_p,
\end{equation}
and with the cross-covariance operator of
$\mathcal G_0$ and $\bm{\mathcal N}$ given by
\begin{equation}\label{eq:fixed_equiv_crosscov}
\operatorname{Cov}
(\mathcal G_0,\bm{\mathcal N})
=
\sigma^2\mathcal B_p,
\end{equation}
because, for every $\mathbf u\in\mathbb R^p$,
\begin{align*}
\mathbb E[
\mathcal G_0(\bm{\mathcal N}^\top\mathbf u)]
&=
\mathbb E\left[
e^2\{\Phi(\mathbf W)-m_\Phi\}
\mathbf g_p(\mathbf Z)^\top\mathbf u
\right]\\
&=
\sigma^2\sum_{j=1}^p
u_j\mathbb E[
\{\Phi(\mathbf W)-m_\Phi\}g_j(\mathbf Z)]=
\sigma^2\mathcal B_p\mathbf u.
\end{align*}
Consequently, since $h_p(\mathbf W_i)=\Phi(\mathbf W_i)-m_\Phi-\mathcal B_p\mathbf g_p(\mathbf Z_i)$, \eqref{eq:fixed_equiv_jointCLT} and the continuous mapping theorem give
\[
\frac{1}{\sqrt n}\sum_{i=1}^n e_i h_p(\mathbf W_i)
=\mathcal G_{0,n}-\mathcal B_p\bm{\mathcal N}_n
\rightarrow^{D}
\mathcal G_0-\mathcal B_p\bm{\mathcal N}.
\]%
We therefore define
\begin{equation}\label{eq:fixed_equiv_Gp}
\mathcal G_p
=
\mathcal G_0-\mathcal B_p\bm{\mathcal N},
\end{equation}
where $\mathcal G_p$ is a centered Gaussian element
since it is a linear
transformation of jointly Gaussian elements.
Furthermore, write
\[
\mathcal B_p^*:\mathcal H\to\mathbb R^p,\qquad
\mathcal B_p^*x=(\langle\beta_1^\Phi,x\rangle_{\mathcal H},\ldots,\langle\beta_p^\Phi,x\rangle_{\mathcal H})^\top,
\]
for the adjoint of $\mathcal B_p$. By
\eqref{eq:fixed_equiv_cov0} and
\eqref{eq:fixed_equiv_crosscov},
\begin{equation}\label{eq:fixed_equiv_covGp}
\begin{aligned}
\operatorname{Cov}(\mathcal G_p)
&=\operatorname{Cov}(\mathcal G_0)-2\sigma^2\mathcal B_p\mathcal B_p^*+\sigma^2\mathcal B_p\mathcal B_p^*\\
&=\sigma^2\big\{\mathbb E[(\Phi(\mathbf W)-m_\Phi)\otimes(\Phi(\mathbf W)-m_\Phi)]-\mathcal B_p\mathcal B_p^*\big\}
=\sigma^2\boldsymbol{\Sigma}_p,
\end{aligned}
\end{equation}%
where the last equality holds because, by the orthonormality of $\mathbf g_p$, $\boldsymbol{\Sigma}_p=\mathbb E[h_p(\mathbf W)\otimes h_p(\mathbf W)]=\mathbb E[(\Phi(\mathbf W)-m_\Phi)\otimes(\Phi(\mathbf W)-m_\Phi)]-\mathcal B_p\mathcal B_p^*$.

We next establish that
\begin{equation}\label{eq:fixed_equiv_H}
\breve{\mathbf H}
=
\mathcal B_p^*\mathcal B_p.
\end{equation}
By the definition of the MDD matrix,
\[
(\breve{\mathbf H})_{jk}
=
-\mathbb E[
g_j(\mathbf Z)g_k(\mathbf Z')
\|\mathbf W-\mathbf W'\|].
\]
Using the facts in \eqref{eq:fixed_equiv_kernel},
and $\mathbb E[g_j(\mathbf Z)]=0$, for any $(j,k)$, 
we have
\begin{align*}
(\breve{\mathbf H})_{jk}
&=
\mathbb E\left[
g_j(\mathbf Z)g_k(\mathbf Z')
\left\langle
\Phi(\mathbf W)-m_\Phi,
\Phi(\mathbf W')-m_\Phi
\right\rangle_{\mathcal H}
\right]\\
&=
\left\langle
\mathbb E[
\{\Phi(\mathbf W)-m_\Phi\}g_j(\mathbf Z)],
\mathbb E[
\{\Phi(\mathbf W')-m_\Phi\}g_k(\mathbf Z')]
\right\rangle_{\mathcal H}\\
&=
\langle\beta_j^\Phi,\beta_k^\Phi
\rangle_{\mathcal H}=
(\mathcal B_p^*\mathcal B_p)_{jk},
\end{align*}
and obtain \eqref{eq:fixed_equiv_H}. 

We now identify the Gaussian vector
$\bm{\mathcal T}$ in
Theorem~\ref{thm:null_fixed_p}. Recall that
\[
K_p(\mathbf V)
=
-\mathbb E\left[
eU(\mathbf W,\mathbf W')
\mathbf g_p(\mathbf Z')
\mid\mathbf V
\right].
\]
By \eqref{eq:fixed_equiv_kernel}, we can rewrite the kernel $K_p$ as 
\[
\begin{aligned}
K_p(\mathbf V)
&=e\,\mathbb E\big[\langle\Phi(\mathbf W)-m_\Phi,\Phi(\mathbf W')-m_\Phi\rangle_{\mathcal H}\,\mathbf g_p(\mathbf Z')\mid\mathbf V\big]\\
&=e\,\mathcal B_p^*\{\Phi(\mathbf W)-m_\Phi\},
\quad\text{so that}\quad
\frac{2}{\sqrt n}\sum_{i=1}^n K_p(\mathbf V_i)=2\mathcal B_p^*\mathcal G_{0,n}.
\end{aligned}
\]%
Therefore, under the same joint Gaussian
construction, we obtain
\begin{equation}\label{eq:fixed_equiv_T}
\bm{\mathcal T}
=
2\mathcal B_p^*\mathcal G_0,
\end{equation}
with $\operatorname{Cov}(\bm{\mathcal T})
=
4\operatorname{var}(K_p(\mathbf V))$
and $\operatorname{Cov}(\bm{\mathcal N},\bm{\mathcal T})=2\operatorname{Cov}(\bm{\mathcal N},\mathcal B_p^*\mathcal G_0)=2\sigma^2\mathcal B_p^*\mathcal B_p=2\sigma^2\breve{\mathbf H}$.
Next, by \eqref{eq:fixed_equiv_Gp},
we expand
\begin{align}
\|\mathcal G_p\|_{\mathcal H}^2
&=
\|\mathcal G_0-\mathcal B_p
\bm{\mathcal N}\|_{\mathcal H}^2
\nonumber\\
&=
\|\mathcal G_0\|_{\mathcal H}^2
+\|\mathcal B_p\bm{\mathcal N}\|_{\mathcal H}^2
-2\langle\mathcal G_0,
\mathcal B_p\bm{\mathcal N}\rangle_{\mathcal H}
\nonumber\\
&=
\|\mathcal G_0\|_{\mathcal H}^2
+\bm{\mathcal N}^\top
\mathcal B_p^*\mathcal B_p
\bm{\mathcal N}
-2\bm{\mathcal N}^\top
\mathcal B_p^*\mathcal G_0
\nonumber\\
&=
\|\mathcal G_0\|_{\mathcal H}^2
+\bm{\mathcal N}^\top
\breve{\mathbf H}\bm{\mathcal N}
-\bm{\mathcal T}^\top\bm{\mathcal N},
\label{eq:fixed_equiv_expansion}
\end{align}
where the last equality follows from
\eqref{eq:fixed_equiv_H} and
\eqref{eq:fixed_equiv_T}.

It remains to show that 
\[\|\mathcal G_0\|_{\mathcal H}^2
-\sigma^2\mu_D = \sum_{k=1}^\infty\lambda_k (G_k^2-1),
\]
where the $G_k$ are constructed jointly with $\bm{\mathcal N}$ and $\bm{\mathcal T}$; this is needed because the law of $\mathcal L_{0,p}$ in Theorem~\ref{thm:null_fixed_p} depends on the joint covariance $\widetilde{\boldsymbol\Sigma}$ in \eqref{eq:MCLT}. By \eqref{eq:fixed_equiv_cov0}, $\operatorname{Cov}(\mathcal G_0)$ is the covariance operator $\boldsymbol\Sigma_{\boldsymbol\chi}$ of $\boldsymbol\chi(\mathbf V)=e\{\Phi(\mathbf W)-m_\Phi\}$ in the proof of Theorem~\ref{thm:null_fixed_p}, whose positive eigenvalues are the $\lambda_k$, with orthonormal eigenvectors $u_k$ and $\psi_k(\mathbf V)=\lambda_k^{-1/2}\langle\boldsymbol\chi(\mathbf V),u_k\rangle_{\mathcal H}$. Let $G_k=\lambda_k^{-1/2}\langle\mathcal G_0,u_k\rangle_{\mathcal H}$. Since $4h_{1,1}=2K_p$, we have
\[
\frac{1}{\sqrt n}\sum_{i=1}^n\psi_k(\mathbf V_i)=\frac{\langle\mathcal G_{0,n},u_k\rangle_{\mathcal H}}{\lambda_k^{1/2}},\quad
\boldsymbol\xi^\circ(e)=\bm{\mathcal N}_n,\quad
\frac{1}{\sqrt n}\sum_{i=1}^n4h_{1,1}(\mathbf V_i)=2\mathcal B_p^*\mathcal G_{0,n}.
\]
By \eqref{eq:fixed_equiv_jointCLT} and the continuous mapping theorem, for every $K$, the vector
\[
(G_1,\ldots,G_K,\bm{\mathcal N}^\top,\bm{\mathcal T}^\top)^\top
\]
has the limiting law $N(\mathbf 0,\widetilde{\boldsymbol\Sigma})$ in \eqref{eq:MCLT}; in particular, the $G_k$ are independent standard normal random variables. Moreover, $\E\langle\mathcal G_0,u\rangle_{\mathcal H}^2=\langle\operatorname{Cov}(\mathcal G_0)u,u\rangle_{\mathcal H}=0$ for every $u$ in the null space of $\operatorname{Cov}(\mathcal G_0)$, so, as in the proof of Theorem~\ref{thm:null_fixed_p}, $\mathcal G_0=\sum_k\lambda_k^{1/2}G_ku_k$ almost surely, and
\[
\|\mathcal G_0\|_{\mathcal H}^2
=
\sum_{k=1}^{\infty}\lambda_kG_k^2\qquad\text{almost surely.}
\]
Furthermore, by
\eqref{eq:fixed_equiv_cov0} and
\eqref{eq:fixed_equiv_muD}, $\sum_{k=1}^{\infty}\lambda_k=\operatorname{tr}\{\operatorname{Cov}(\mathcal G_0)\}=\sigma^2\mathbb E\|\Phi(\mathbf W)-m_\Phi\|_{\mathcal H}^2=\sigma^2\mu_D$. Combining this with \eqref{eq:fixed_equiv_expansion}, we obtain
\begin{equation}\label{eq:fixed_equiv_oracle}
\begin{aligned}
\|\mathcal G_p\|_{\mathcal H}^2-\sigma^2\mu_D
&=\|\mathcal G_0\|_{\mathcal H}^2-\sum_{k=1}^{\infty}\lambda_k+\bm{\mathcal N}^\top\breve{\mathbf H}\bm{\mathcal N}-\bm{\mathcal T}^\top\bm{\mathcal N}\\
&=\sum_{k=1}^{\infty}\lambda_k(G_k^2-1)+\bm{\mathcal N}^\top\breve{\mathbf H}\bm{\mathcal N}-\bm{\mathcal T}^\top\bm{\mathcal N}.
\end{aligned}
\end{equation}%
Hence $\mathcal L_{0,p}\overset{D}{=}\|\mathcal G_p\|_{\mathcal H}^2-\sigma^2\mu_D$, and the convergence in the proposition follows from Theorem~\ref{thm:null_fixed_p}. Finally, let $\rho_{p,1}\ge\rho_{p,2}\ge\cdots\ge0$ be the eigenvalues of $\boldsymbol\Sigma_p$. By \eqref{eq:fixed_equiv_covGp}, $\|\mathcal G_p\|_{\mathcal H}^2\overset{D}{=}\sigma^2\sum_{j\ge1}\rho_{p,j}\widetilde G_j^2$ with independent standard normal $\widetilde G_j$. If $\sigma^2\operatorname{tr}(\boldsymbol\Sigma_p)>0$, then $\sigma^2\rho_{p,1}>0$, and $\sigma^2\rho_{p,1}\widetilde G_1^2$ has a continuous distribution and is independent of the remaining nonnegative terms; hence $\mathcal L_{0,p}$ has a continuous distribution function and, in particular, is non-degenerate. This completes the proof.
\end{proof}

\begin{proof}[Proof of Theorem~\ref{thm:null_fixed_p_bootstrap}]
Write $T_n^*=n\widehat{\rm MDD}^{2*}(\widehat e\mid\mathbf W)$. In the notation introduced at the beginning of this section, the expansion \eqref{eq:nMDD_boot_lemma} of Lemma~\ref{lemma:fixed_p_bootstrap_remainder} reads
\begin{equation}\label{eq:nMDD_boot}
T_n^*=\frac{1}{n-3}\sum_{i\neq j}\widetilde A_{ij}\widetilde B_{ij}^{**}+\boldsymbol\xi(e\eta^*)^\top\mathbf S_n\boldsymbol\xi(e\eta^*)-\mathbf w_{\mathbf R^*}^\top\boldsymbol\xi(e\eta^*)+\mathcal R_{n,1}^*,
\end{equation}
where $\widetilde B_{ij}^{**}$ is the $\mathcal U$-centered version of $B_{ij}^{**}=\frac12(e_i\eta_i^*-e_j\eta_j^*)^2$ and $\mathcal R_{n,1}^*=o_p^*(1)$. By the second assertion of Lemma~\ref{lemma:sim1}, replacing $\boldsymbol\xi(e\eta^*)$ by $\boldsymbol\xi^\circ(e\eta^*)$ in the second and third terms changes them by $o_p^*(1)$. We analyze the leading terms and then establish their joint conditional limit.

We now treat the first term in \eqref{eq:nMDD_boot}. Since $\widetilde{\mathbf A}$ has zero diagonal and, by \eqref{eq:row_col_sum}, zero row and column sums, we have $\sum_{i\neq j}\widetilde A_{ij}\widetilde B_{ij}^{**}=\sum_{i\neq j}\widetilde A_{ij}B_{ij}^{**}$, and the terms of $B_{ij}^{**}=\frac12e_i^2\eta_i^{*2}+\frac12e_j^2\eta_j^{*2}-e_ie_j\eta_i^*\eta_j^*$ that depend on a single index contribute nothing to the right-hand side. Hence
\[
\begin{aligned}
\frac{1}{n-3}\sum_{i\neq j}\widetilde A_{ij}\widetilde B_{ij}^{**}
&=-\frac{1}{n-3}\sum_{i\neq j}\widetilde A_{ij}e_ie_j\eta_i^*\eta_j^*\\
&=\frac{1}{n-3}\sum_{i\neq j}U(\mathbf W_i,\mathbf W_j)(-e_ie_j)\eta_i^*\eta_j^*+\mathcal E_n^*,\\
\mathcal E_n^*&=-\frac{1}{n-3}\sum_{i\neq j}\{\widetilde A_{ij}-U(\mathbf W_i,\mathbf W_j)\}e_ie_j\eta_i^*\eta_j^*.
\end{aligned}
\]
The coefficients of $\eta_i^*\eta_j^*$ in $\mathcal E_n^*$ are symmetric in $(i,j)$, so (F1) in the proof of Lemma~\ref{lemma:fixed_p_bootstrap_remainder} holds with equality. Indeed, for a symmetric array $(c_{ij})$, the middle expression in (F1) satisfies $\sum_{i\neq j}c_{ij}(c_{ij}+c_{ji})$ then equals $2\sum_{i\neq j}c_{ij}^2$, gives
\[
\E^*[(\mathcal E_n^*)^2]=\frac{2}{(n-3)^2}\sum_{i\neq j}\{\widetilde A_{ij}-U(\mathbf W_i,\mathbf W_j)\}^2e_i^2e_j^2.
\]
Since $e_1,\ldots,e_n$ are i.i.d.\ and independent of $\mathbf W_1,\ldots,\mathbf W_n$ by Condition (A3) in Assumption~\ref{assum:0}, the expectation of the right-hand side is
\[
\frac{2\sigma^4}{(n-3)^2}\sum_{i\neq j}\E\{\widetilde A_{ij}-U(\mathbf W_i,\mathbf W_j)\}^2=\E[\mathcal E_n^2]=O(n^{-1}),
\]
where $\mathcal E_n$ is the term bounded in the proof of Theorem~\ref{thm:null_fixed_p}. Hence Markov's inequality gives $\E^*[(\mathcal E_n^*)^2]=O_p(n^{-1})$, so that $\mathcal E_n^*=O_p^*(n^{-1/2})$. Consequently,
\[
\frac{1}{n-3}\sum_{i\neq j}\widetilde A_{ij}\widetilde B_{ij}^{**}=\frac{1}{n-3}\sum_{i\neq j}6h_{2,2}(\mathbf V_i,\mathbf V_j)\eta_i^*\eta_j^*+o_p^*(1),
\]
where $6h_{2,2}(\mathbf V_i,\mathbf V_j)=U(\mathbf W_i,\mathbf W_j)(-e_ie_j)$.

The leading term is therefore
\[
\begin{aligned}
\mathcal U_{n,1}^*&:=\frac{1}{n-3}\sum_{i\neq j}6h_{2,2}(\mathbf V_i,\mathbf V_j)\eta_i^*\eta_j^*,\quad\text{where}\\
6h_{2,2}(\mathbf V_1,\mathbf V_2)&=\sum_{k=1}^\infty\lambda_k\psi_k(\mathbf V_1)\psi_k(\mathbf V_2),\qquad \E[\psi_k(\mathbf V)\psi_l(\mathbf V)]=\mathbf 1\{k=l\},
\end{aligned}
\]
is the eigen-expansion justified in the proof of Theorem~\ref{thm:null_fixed_p}. Since the sum defining $\mathcal U_{n,1}^*$ runs over $i\neq j$, the expansion is evaluated only off the diagonal.

Recall that $h_{1,1}(\mathbf V_i)=\frac12\E[U(\mathbf W,\mathbf W')(-e)\boldsymbol\phi_p(\mathbf Z')\mid\mathbf W=\mathbf W_i,e=e_i]$, which is defined in the proof of Theorem~\ref{thm:null_fixed_p}. Let $\mathbf m_i=\E[U(\mathbf W_i,\mathbf W')\boldsymbol\phi_p(\mathbf Z')\mid\mathbf W_i]$, so that $4h_{1,1}(\mathbf V_i)=-2e_i\mathbf m_i$, and define
\[
\mathbf L_n^*=\frac{4}{\sqrt n}\sum_{i=1}^n h_{1,1}(\mathbf V_i)\eta_i^*,\qquad
\mathbf N_n^*=\frac{1}{\sqrt n}\sum_{k=1}^n\boldsymbol\phi_p(\mathbf Z_k)e_k\eta_k^*,\qquad
\mathcal U_{n,3}^*=\mathbf L_n^{*\top}\mathbf N_n^*.
\]
Since $\sum_{i\neq j}\widetilde A_{ij}\widetilde{\mathbf R}_{ij}^*=\sum_{i\neq j}\widetilde A_{ij}\mathbf R_{ij}^*$ and the rows of $\widetilde{\mathbf A}$ sum to zero by \eqref{eq:row_col_sum}, the terms of $\mathbf R_{ij}^*$ that depend on a single index vanish, and
\[
\begin{aligned}
&\frac{1}{\sqrt n(n-3)}\sum_{i\neq j}\widetilde A_{ij}\widetilde{\mathbf R}_{ij}^*-\mathbf L_n^*
=-\frac{2}{\sqrt n}\sum_{i=1}^n e_i\eta_i^*\boldsymbol\omega_{n,i},\\
&\boldsymbol\omega_{n,i}=\frac{1}{n-3}\sum_{j\neq i}\widetilde A_{ij}\boldsymbol\phi_p(\mathbf Z_j)-\mathbf m_i.
\end{aligned}
\]
By (F2) in the proof of Lemma~\ref{lemma:fixed_p_bootstrap_remainder}, the conditional second moment of this difference is $4n^{-1}\sum_{i=1}^n e_i^2\|\boldsymbol\omega_{n,i}\|^2$. Each $\boldsymbol\omega_{n,i}$ depends only on $\mathbf W_1,\ldots,\mathbf W_n$, which are independent of $e_i$, so that $\E[n^{-1}\sum_{i=1}^n e_i^2\|\boldsymbol\omega_{n,i}\|^2]=\sigma^2\E\|\boldsymbol\omega_{n,1}\|^2$. To bound $\E\|\boldsymbol\omega_{n,1}\|^2$, let $a(\mathbf w)=\E\|\mathbf w-\mathbf W'\|$, $a_0=\E\|\mathbf W-\mathbf W'\|$, $\vartheta_i=R_i/(n-2)-a(\mathbf W_i)$, and $\vartheta_0=R_{++}/\{(n-1)(n-2)\}-a_0$. Since $U(\mathbf W_i,\mathbf W_j)=A_{ij}-a(\mathbf W_i)-a(\mathbf W_j)+a_0$, \eqref{eq:ucenter} gives $\widetilde A_{1j}-U(\mathbf W_1,\mathbf W_j)=\vartheta_0-\vartheta_1-\vartheta_j$ for $j\neq1$. Accordingly, $\boldsymbol\omega_{n,1}=\boldsymbol\omega_{n,1}^{(1)}+\boldsymbol\omega_{n,1}^{(2)}$, where
\[
\begin{aligned}
\boldsymbol\omega_{n,1}^{(1)}&=\frac{1}{n-3}\sum_{j\neq1}U(\mathbf W_1,\mathbf W_j)\boldsymbol\phi_p(\mathbf Z_j)-\mathbf m_1,\\
\boldsymbol\omega_{n,1}^{(2)}&=\frac{1}{n-3}\sum_{j\neq1}(\vartheta_0-\vartheta_1-\vartheta_j)\boldsymbol\phi_p(\mathbf Z_j).
\end{aligned}
\]
Given $\mathbf W_1$, the summands $U(\mathbf W_1,\mathbf W_j)\boldsymbol\phi_p(\mathbf Z_j)$, $j\neq1$, are i.i.d.\ with mean $\mathbf m_1$; since $\sup_{\mathbf z}\|\boldsymbol\phi_p(\mathbf z)\|\le C$ and $\E[U(\mathbf W,\mathbf W')^2]<\infty$, we have $\E\|\boldsymbol\omega_{n,1}^{(1)}\|^2=O(n^{-1})$. By the Cauchy--Schwarz inequality and the same bound on $\boldsymbol\phi_p$, $\|\boldsymbol\omega_{n,1}^{(2)}\|^2$ is at most a constant multiple of $\vartheta_0^2+\vartheta_1^2+(n-1)^{-1}\sum_{j\neq1}\vartheta_j^2$, whose expectation is $O(n^{-1})$ because $\E\|\mathbf W\|^2<\infty$: given $\mathbf W_i$, $R_i/(n-1)$ is an average of $n-1$ i.i.d.\ terms with mean $a(\mathbf W_i)$, and $R_{++}/\{n(n-1)\}$ is a $U$-statistic with mean $a_0$. Hence $\E\|\boldsymbol\omega_{n,1}\|^2=O(n^{-1})$, and Markov's inequality gives
\[
\E^*\Big\|\frac{1}{\sqrt n(n-3)}\sum_{i\neq j}\widetilde A_{ij}\widetilde{\mathbf R}_{ij}^*-\mathbf L_n^*\Big\|^2
=\frac{4}{n}\sum_{i=1}^n e_i^2\|\boldsymbol\omega_{n,i}\|^2=O_p(n^{-1}).
\]
Consequently, $\{\sqrt n(n-3)\}^{-1}\sum_{i\neq j}\widetilde A_{ij}\widetilde{\mathbf R}_{ij}^*=\mathbf L_n^*+o_p^*(1)$. Since $\mathbf N_n^*=O_p^*(1)$ by (F2), the third term in \eqref{eq:nMDD_boot} equals $-\mathcal U_{n,3}^*+o_p^*(1)$.

Therefore, we have 
\[n\widehat{\text{MDD}}^{2*}(\widehat{e} \mid \mathbf{W}) =\mathcal{U}_{n,1}^{*}+\mathcal{U}_{n,2}^{*}-\mathcal{U}_{n,3}^{*}+o_p^{*}(1),
\]
with 
$\mathcal U_{n,2}^*=\mathbf N_n^{*\top}\mathbf S_n\mathbf N_n^*$, where $\mathbf N_n^*=\boldsymbol\xi^\circ(e\eta^*)$ is defined above.

For simplicity of notation, we denote $\widetilde{\mathbf b}(\mathbf{V}_k):=\boldsymbol{\phi}_p(\mathbf{Z}_k)e_k$.
Next, we shall show that for any fixed $K$, 
the sum $\mathcal U_{n,1}^{*K}+\mathcal U_{n,2}^*-\mathcal U_{n,3}^*$ satisfies \eqref{eq:K_dist_boot} below, where \[
\begin{aligned}
\mathcal U_{n,1}^{*K}&=\frac{1}{n-3}\sum_{i\neq j}\sum_{k=1}^K\lambda_k\psi_k(\mathbf V_i)\psi_k(\mathbf V_j)\eta_i^*\eta_j^*\\
&=\frac{n}{n-3}\sum_{k=1}^K\lambda_k\Big\{\Big(\frac{1}{\sqrt n}\sum_{i=1}^n\psi_k(\mathbf V_i)\eta_i^*\Big)^2-\frac1n\sum_{i=1}^n\psi_k(\mathbf V_i)^2\eta_i^{*2}\Big\}.
\end{aligned}
\] 

Fix $K$ and let
\[
\begin{aligned}
\mathbf F_K(\mathbf V)&=\big(\psi_1(\mathbf V),\ldots,\psi_K(\mathbf V),\widetilde{\mathbf b}(\mathbf V)^\top,4h_{1,1}(\mathbf V)^\top\big)^\top,\\
\widehat{\boldsymbol\Sigma}_K&=\frac1n\sum_{i=1}^n\mathbf F_K(\mathbf V_i)\mathbf F_K(\mathbf V_i)^\top.
\end{aligned}
\]
Conditionally on the sample, $n^{-1/2}\sum_i\eta_i^*\mathbf F_K(\mathbf V_i)$ is exactly Gaussian with mean zero and covariance matrix $\widehat{\boldsymbol\Sigma}_K$, because the $\eta_i^*$ are i.i.d.\ $N(0,1)$ and independent of the sample. Since $\E\|\mathbf F_K(\mathbf V)\|^2<\infty$, the strong law of large numbers shows that $\widehat{\boldsymbol\Sigma}_K$ converges almost surely to $\widetilde{\boldsymbol\Sigma}=\E[\mathbf F_K(\mathbf V)\mathbf F_K(\mathbf V)^\top]$. Hence, for almost every sample, the conditional law of $n^{-1/2}\sum_i\eta_i^*\mathbf F_K(\mathbf V_i)$ converges to $N(\mathbf 0,\widetilde{\boldsymbol\Sigma})$, whether or not $\widetilde{\boldsymbol\Sigma}$ is singular. Moreover, for each $k\le K$,
\[
\begin{aligned}
\E^*\Big[\frac1n\sum_{i=1}^n\psi_k(\mathbf V_i)^2\eta_i^{*2}\Big]&=\frac1n\sum_{i=1}^n\psi_k(\mathbf V_i)^2\rightarrow1,\\
\operatorname{var}^*\Big[\frac1n\sum_{i=1}^n\psi_k(\mathbf V_i)^2\eta_i^{*2}\Big]&=\frac{2}{n^2}\sum_{i=1}^n\psi_k(\mathbf V_i)^4\rightarrow0
\end{aligned}
\]
almost surely. The second limit holds because
\[
\frac{1}{n^2}\sum_{i=1}^n\psi_k(\mathbf V_i)^4\le\Big\{\frac1n\max_{i\le n}\psi_k(\mathbf V_i)^2\Big\}\Big\{\frac1n\sum_{i=1}^n\psi_k(\mathbf V_i)^2\Big\},
\]
and $n^{-1}\max_{i\le n}\psi_k(\mathbf V_i)^2\to0$ almost surely when $\E[\psi_k(\mathbf V)^2]<\infty$. Hence, for almost every sample, $n^{-1}\sum_i\psi_k(\mathbf V_i)^2\eta_i^{*2}\rightarrow^{p^*}1$. Only second moments of $\mathbf F_K(\mathbf V)$ are used in this step.

 Therefore, with (\ref{eq:converges_p_refined}), we obtain
	\begin{eqnarray}
	\label{eq:K_dist_boot}
	&&\mathcal{U}_{n,1}^{*K}+\mathcal{U}_{n,2}^{*}-\mathcal{U}_{n,3}^{*}\rightarrow^{D^{*}} \sum_{k=1}^{K}\lambda_k (G_k^2 -1) +\bm{\mathcal{N}}^\top \breve{\bm{H}} \bm{\mathcal{N}}-\bm{\mathcal{T}}^\top\bm{\mathcal{N}}.
	\end{eqnarray}

Note that $\E[\{6h_{2,2}(\mathbf{V},\mathbf{V}')\}^2]=\sum_{k}\lambda_k^2<\infty$ which implies that
\begin{eqnarray*}
    \E^*\left[\left| \mathcal{U}_{n,1}^{*}-\mathcal{U}_{n,1}^{*K}\right|^2\right] &=&\frac{2}{(n-3)^2} \sum_{i \neq j} \left\{\sum_{k=K+1}^{\infty}\lambda_k \psi_k(\bm{V}_i)\psi_k(\bm{V}_j)\right\}^2\\
    &\rightarrow& 2\sum_{k=K+1}^{\infty}\lambda_k^2\quad\text{almost surely}
\end{eqnarray*}	
as $n \rightarrow \infty$, where the equality follows from (F1) in the proof of Lemma~\ref{lemma:fixed_p_bootstrap_remainder}, which holds with equality for a symmetric array, and the limit follows from the strong law of large numbers for $U$-statistics and the orthonormality of $\{\psi_k\}$. Hence, by Markov's inequality, for any $\delta>0$ and almost every sample, we have 
\begin{eqnarray}\label{eq:limsup_boot}
\lim_{K\rightarrow \infty}\limsup_{n\rightarrow \infty} P^*\left(\left| \mathcal{U}_{n,1}^{*}+\mathcal{U}_{n,2}^{*}-\mathcal{U}_{n,3}^{*}-\left(\mathcal{U}_{n,1}^{*K}+\mathcal{U}_{n,2}^{*}-\mathcal{U}_{n,3}^{*} \right)\right| >\delta\right) =0.
\end{eqnarray}
	Finally, since (\ref{eq:K_dist_boot}) and (\ref{eq:limsup_boot}) are both satisfied, we apply Theorem 2 in \citet{dehling2009new} conditionally on the sample, along the subsequences in the definition of $\rightarrow^{D^*}$, to conclude that
	\[ n \widehat{\rm MDD}^{2*}(\widehat{e} \mid \mathbf{W}) \rightarrow^{D*} \sum_{k=1}^{\infty}\lambda_k (G_k^2 -1) +\bm{\mathcal{N}}^\top\breve{\bm{H}}\bm{\mathcal{N}}-\bm{\mathcal{T}}^\top\bm{\mathcal{N}}.
	\]
\end{proof}

\begin{proof}[Proof of Theorem~\ref{thm:bootstrap}]
Notice that the bootstrap residual vector is
\begin{equation}\label{eq:e^*}
\widehat e^*
=
\mathbf \Pi_n(\widehat e\circ\eta^*),
\end{equation}
where $\widehat e\circ\eta^*
=
(\widehat e_1\eta_1^*,\ldots,
 \widehat e_n\eta_n^*)^\top$.
We note that Lemma~\ref{lemma:test_statistic} remains valid for the bootstrap residual $\widehat{e}^*$. Recall that our bootstrap statistic is defined as
\begin{align*}
T_n^*
:=
n\widehat{\operatorname{MDD}}^2
(\widehat e^*\mid \mathbf W) &=
\frac{n}{n-3}
\left\|
\frac1{\sqrt n}
\sum_{i=1}^n
\widehat e_i^*\Phi_i
\right\|_{\mathcal H}^2
-
\frac1{n-3}
\sum_{i=1}^n
d_i(\widehat e_i^*)^2\\
&=\frac{n}{n-3}
\left\|
\frac1{\sqrt n}
\sum_i
\widehat e_i\eta_i^*t_i\right\|_{\mathcal H}^2
-
\frac1{n-3}
\sum_{i=1}^n
d_i(\widehat e_i^*)^2,
\end{align*}
by the fact that
\begin{equation}\label{eq:t_i_Phi_i}
\begin{aligned}
\frac1{\sqrt n}
\sum_i\widehat e_i^*\Phi_i
&=
\frac1{\sqrt n}
\left\langle
\mathbf \Pi_n(\widehat e\circ\eta^*),
\boldsymbol{\Phi}
\right\rangle=
\frac1{\sqrt n}
\left\langle
\widehat e\circ\eta^*,
\mathbf \Pi_n\boldsymbol{\Phi}
\right\rangle=
\frac1{\sqrt n}
\sum_i
\widehat e_i\eta_i^*t_i,
\end{aligned}
\end{equation}
where $\boldsymbol{\Phi}=(\Phi_1,\cdots, \Phi_n )^\top$.

Similar to the proof of Theorem~\ref{thm:null}, the proof has two main parts:
\begin{enumerate}
\item We show that the first term of $T_n^*$ converges to $\|\mathcal{G}_{\infty}\|_{\mathcal H}^2$ in distribution in probability, i.e.,  
\[\left\|
\frac1{\sqrt n}
\sum_{i=1}^n
\widehat e_i^*\Phi_i
\right\|_{\mathcal H}^2 \rightarrow^{D^*} \|\mathcal{G}_{\infty}\|_{\mathcal H}^2\]
To show the result above, we follow the steps below:
\begin{enumerate}
\item First we show that $$\frac1{\sqrt n}
\sum_{i=1}^n
\widehat e_i^*\Phi_i=
\frac{1}{\sqrt n}
\sum_{i=1}^n
e_i\eta_i^*h_\infty(\mathbf W_i)+o_p^*(1)$$
\item Next we show that \begin{equation}\label{eq:boot-part1-b}\frac{1}{\sqrt n}
\sum_{i=1}^n
e_i\eta_i^*h_\infty(\mathbf W_i)
\rightarrow^{D^*}
\mathcal G_\infty,\end{equation}
\end{enumerate}
which implies that $\left\|\frac{1}{\sqrt n}
\sum_{i=1}^n
e_i\eta_i^*h_\infty(\mathbf W_i)\right\|_{\mathcal H}^2
\rightarrow^{D^*}
\|\mathcal G_\infty\|_{\mathcal H}^2$ by the continuous mapping theorem. 

\item We show that \begin{equation}\label{eq:2nd_part}
\frac1{n-3}
\sum_{i=1}^n
d_i(\widehat e_i^*)^2 \rightarrow^{p^*} \sigma^2 \mu_D.
\end{equation}
\end{enumerate}

Recall the definition of $\delta_n$ in \eqref{eq:delta} and the rate in \eqref{eq:delta_square}
\[
\delta_n
=
\widehat e-e
=
-\mathbf P_ne+\mathbf\Pi_n R_p, ~\qquad \frac{1}{n}\|\delta_n\|^2
=
O_p\left(\frac{p}{n}+a_p^2\right)
=
o_p(1).
\]
Notice that
\begin{equation}\label{eq:Z_n-Z_n,0}
\begin{aligned}
&\frac1{\sqrt n}
\sum_i
\widehat e_i\eta_i^*t_i-\frac{1}{\sqrt n}
\sum_{i=1}^n
e_i\eta_i^*h_\infty(\mathbf W_i)
=
\frac{1}{\sqrt n}
\sum_i
\left\{
\widehat e_i t_i
-
e_i h_\infty(\mathbf W_i)
\right\}\eta_i^*\\
&=
\frac{1}{\sqrt n}
\sum_i
(\widehat e_i-e_i)\eta_i^*t_i+
\frac{1}{\sqrt n}
\sum_i
e_i\eta_i^*
\{t_i-h_\infty(\mathbf W_i)\}\\
&=\frac{1}{\sqrt n}
\sum_i
\delta_{n,i}\eta_i^*t_i+\frac{1}{\sqrt n}
\sum_i
e_i\eta_i^*
\{t_i-h_\infty(\mathbf W_i)\}.
\end{aligned}
\end{equation}
We shall show that each term in \eqref{eq:Z_n-Z_n,0} is $o_p^*(1)$ which implies that \[
\frac1{\sqrt n}
\sum_i
\widehat e_i\eta_i^*t_i=\frac{1}{\sqrt n}
\sum_{i=1}^n
e_i\eta_i^*h_\infty(\mathbf W_i)+o_p^*(1).
\] 
For the first term, we have
\[
\begin{aligned}
\E^*\left\|\frac{1}{\sqrt n}
\sum_i
\delta_{n,i}\eta_i^*t_i\right\|_{\mathcal H}^2
&=
\frac{1}{n}
\sum_i
\delta_{n,i}^2
\|t_i\|_{\mathcal H}^2\\
&\leq \left(
\max_i\|t_i\|_{\mathcal H}^2
\right)
\frac{1}{n}\|\delta_n\|^2\\
&=
O_p(\xi_p^2)
O_p\left(
\frac{p}{n}+a_p^2
\right)=o_p(1),
\end{aligned}
\]by \eqref{eq:delta_square} and \eqref{eq:t_rate} in Lemma~\ref{lemma:t_rate} under the conditions (B2) and (B3) in Assumption~\ref{assum:1}. Here $\xi_p^2p/n\to0$ by (B2), while $\xi_p^2a_p^2=(\xi_p^2/n)(na_p^2)\to0$ by (B2) and (B3).
By Markov's inequality (see the beginning of this supplement), this gives
\begin{equation}\label{eq:Z_n-Z_n,0-part1}
\frac{1}{\sqrt n}
\sum_i
\delta_{n,i}\eta_i^*t_i=o_p^*(1).
\end{equation}

\noindent Next we consider the second term in \eqref{eq:Z_n-Z_n,0}. Let
\[
K_n:=\frac1n\sum_ie_i^2\|t_i-h_\infty(\mathbf W_i)\|_{\mathcal H}^2,
\qquad
D_n:=\frac1n\sum_i\|t_i-h_\infty(\mathbf W_i)\|_{\mathcal H}^2.
\]
Then $\E^*\|n^{-1/2}\sum_ie_i\eta_i^*\{t_i-h_\infty(\mathbf W_i)\}\|_{\mathcal H}^2=K_n$, and $\E(K_n\mid\mathbf W_{1:n})=\sigma^2D_n$ because $t_i$ and $h_\infty(\mathbf W_i)$ are functions of $\mathbf W_{1:n}$ and $e$ is independent of $\mathbf W$. The first expectation averages over the bootstrap weights with the full sample fixed; the second averages over the original errors with only the covariates fixed. Moreover,
\[
D_n\le\frac2n\sum_i\|t_i-h_p(\mathbf W_i)\|_{\mathcal H}^2+\frac2n\sum_i\|h_p(\mathbf W_i)-h_\infty(\mathbf W_i)\|_{\mathcal H}^2=o_p(1)
\]
by \eqref{eq:t-h_p_square_norm} and $\E\|h_p(\mathbf W)-h_\infty(\mathbf W)\|_{\mathcal H}^2\to0$. Since $K_n\ge0$, the conditional Markov inequality gives $P(K_n>u)\le P(D_n>v)+\sigma^2v/u$ for all $u,v>0$; letting $n\to\infty$ and then $v\downarrow0$ shows that $K_n=o_p(1)$. Hence, by Markov's inequality,
\begin{equation}\label{eq:Z_n-Z_n,0-part2}
\frac{1}{\sqrt n}
\sum_i
e_i\eta_i^*
\{t_i-h_\infty(\mathbf W_i)\}=o_p^*(1).
\end{equation}
By combining \eqref{eq:Z_n-Z_n,0}, \eqref{eq:Z_n-Z_n,0-part1}, and \eqref{eq:Z_n-Z_n,0-part2}, we obtain
\begin{equation}\label{eq:Z_n_Z_n0_op1}
\frac1{\sqrt n}
\sum_i
\widehat e_i\eta_i^*t_i=\frac{1}{\sqrt n}
\sum_{i=1}^n
e_i\eta_i^*h_\infty(\mathbf W_i)+o_p^*(1).
\end{equation}

\noindent We now show \eqref{eq:boot-part1-b}, using the truncation argument from the proof of Theorem~\ref{thm:null_fixed_p_bootstrap}. Write $\mathbb G_n^*=n^{-1/2}\sum_{i=1}^ne_i\eta_i^*h_\infty(\mathbf W_i)$. Let $\{u_j\}_{j\ge1}$ be an orthonormal basis of the separable Hilbert space $\mathcal H$, let $\mathbf Q_J:\mathcal H\to\mathcal H$ denote the orthogonal projection onto $\operatorname{span}\{u_1,\ldots,u_J\}$, and let $\mathrm{Id}$ denote the identity operator on $\mathcal H$. Conditional on the sample, $\mathbf Q_J\mathbb G_n^*$ is a centered $J$-dimensional Gaussian vector (in the coordinates $u_1,\ldots,u_J$), and, for $j,k\ge1$,
\[
\begin{aligned}
\E^*\big[\langle\mathbb G_n^*,u_j\rangle_{\mathcal H}\langle\mathbb G_n^*,u_k\rangle_{\mathcal H}\big]
&=\frac1n\sum_{i=1}^ne_i^2\langle h_\infty(\mathbf W_i),u_j\rangle_{\mathcal H}\langle h_\infty(\mathbf W_i),u_k\rangle_{\mathcal H},\\
\E^*\|(\mathrm{Id}-\mathbf Q_J)\mathbb G_n^*\|_{\mathcal H}^2
&=\frac1n\sum_{i=1}^ne_i^2\|(\mathrm{Id}-\mathbf Q_J)h_\infty(\mathbf W_i)\|_{\mathcal H}^2.
\end{aligned}
\]
Since $e$ is independent of $\mathbf W$ and $\E\|h_\infty(\mathbf W)\|_{\mathcal H}^2<\infty$, the strong law of large numbers shows that the right-hand sides converge almost surely to
\[
\sigma^2\E\big[\langle h_\infty(\mathbf W),u_j\rangle_{\mathcal H}\langle h_\infty(\mathbf W),u_k\rangle_{\mathcal H}\big]
\quad\text{and}\quad
\sigma^2\E\|(\mathrm{Id}-\mathbf Q_J)h_\infty(\mathbf W)\|_{\mathcal H}^2,
\]
respectively, where, for $j,k\le J$, the first limit is the $(j,k)$-th entry of the covariance matrix of $\mathbf Q_J\mathcal G_\infty$. As there are countably many indices $j,k,J\ge1$, all these limits hold simultaneously on an event $\Omega_0$ with $P(\Omega_0)=1$. Hence, on $\Omega_0$, $\mathbf Q_J\mathbb G_n^*\rightarrow^{D^*}\mathbf Q_J\mathcal G_\infty$ for every fixed $J$, and
\[
\lim_{J\to\infty}\limsup_{n\to\infty}\E^*\|(\mathrm{Id}-\mathbf Q_J)\mathbb G_n^*\|_{\mathcal H}^2
=\lim_{J\to\infty}\sigma^2\E\|(\mathrm{Id}-\mathbf Q_J)h_\infty(\mathbf W)\|_{\mathcal H}^2=0
\]
by dominated convergence, since $\mathbf Q_Jh_\infty(\mathbf w)\to h_\infty(\mathbf w)$ for every $\mathbf w$. By Markov's inequality, $\lim_{J\to\infty}\limsup_{n\to\infty}P^*(\|(\mathrm{Id}-\mathbf Q_J)\mathbb G_n^*\|_{\mathcal H}>\delta)=0$ on $\Omega_0$ for every $\delta>0$. Moreover, $\E\|(\mathrm{Id}-\mathbf Q_J)\mathcal G_\infty\|_{\mathcal H}^2=\sigma^2\E\|(\mathrm{Id}-\mathbf Q_J)h_\infty(\mathbf W)\|_{\mathcal H}^2\to0$, so that $\mathbf Q_J\mathcal G_\infty\rightarrow\mathcal G_\infty$ in distribution as $J\to\infty$. Applying Theorem 2 in \citet{dehling2009new}, we obtain \eqref{eq:boot-part1-b}, that is, $\mathbb G_n^*\rightarrow^{D^*}\mathcal G_\infty$.

\noindent Combining the result above and \eqref{eq:Z_n_Z_n0_op1} yields
\[
\frac1{\sqrt n}
\sum_i
\widehat e_i\eta_i^*t_i
\rightarrow^{D^*}
\mathcal{G}_\infty.
\]
Then by the continuous mapping theorem,
we further obtain
\begin{equation} \label{eq:Z_n_norm_convergence}
\left\|\frac1{\sqrt n}
\sum_i
\widehat e_i\eta_i^*t_i\right\|_{\mathcal H}^2
\rightarrow^{D^*}
\|\mathcal{G}_\infty\|_{\mathcal H}^2.
\end{equation}

It remains to show \eqref{eq:2nd_part}. We define
\[
\mathbf D_{\widehat e}
=
\operatorname{diag}
(\widehat e_1,\ldots,\widehat e_n),
\qquad
\mathbf D_d
=
\operatorname{diag}(d_1,\ldots,d_n).
\]
Then we have
\[
\widehat e^*
=
\mathbf \Pi_n\mathbf D_{\widehat e}\bm \eta^*,
\]
where $\bm \eta^*=(\eta_1^*, \cdots, \eta_n^*)^\top$
and therefore
\[
\begin{aligned}
\frac1{n-3}
\sum_i d_i(\widehat e_i^*)^2
&=
\frac{1}{n-3}
(\bm \eta^*)^\top
\mathbf D_{\widehat e}\mathbf \Pi_n\mathbf D_d\mathbf \Pi_n\mathbf D_{\widehat e}\bm \eta^*\\
&=\frac{1}{n-3}
(\bm \eta^*)^\top
\mathbf D_{\widehat e}\mathbf M_{d,n}\mathbf D_{\widehat e}\bm \eta^*
,
\end{aligned}
\]%
where $\mathbf M_{d,n} = \mathbf \Pi_n\mathbf D_d\mathbf \Pi_n$. 
We note that 
\begin{equation}\label{eq:tau}
\begin{aligned}
\zeta_n
:=
\E^*\left[ \frac1{n-3}
\sum_i d_i(\widehat e_i^*)^2\right]
&=
\frac{1}{n-3}
\operatorname{tr}
\left(
\mathbf D_{\widehat e}\mathbf M_{d,n}\mathbf D_{\widehat e}
\right)=
\frac{1}{n-3}
\sum_{i=1}^n
\widehat e_i^2(\mathbf M_{d,n})_{ii}.
\end{aligned}
\end{equation}
We next compare $(\mathbf M_{d,n})_{ii}$ with $d_i$. Since $\mathbf\Pi_n=\mathbf I_n-\mathbf P_n$, we have $\mathbf M_{d,n}=\mathbf D_d-\mathbf P_n\mathbf D_d-\mathbf D_d\mathbf P_n+\mathbf P_n\mathbf D_d\mathbf P_n$, and, since $\mathbf P_n$ is an orthogonal projection, $\sum_j(\mathbf P_n)_{ij}^2=(\mathbf P_n^2)_{ii}=(\mathbf P_n)_{ii}=\ell_i$. Hence, by \eqref{eq:d_i_bound},
\[
\begin{aligned}
|(\mathbf M_{d,n})_{ii}-d_i|
&=\Big|-2d_i\ell_i+\sum_{j=1}^nd_j(\mathbf P_n)_{ij}^2\Big|\\
&\le2(\max_jd_j)\ell_i+(\max_jd_j)\sum_{j=1}^n(\mathbf P_n)_{ij}^2=3(\max_jd_j)\ell_i\le18M\ell_i.
\end{aligned}
\]%
By (\ref{eq:tau}) and the fact above, it follows that 
\[
\begin{aligned}
\left|
\zeta_n
-
\frac{1}{n-3}
\sum_i d_i\widehat e_i^2
\right|
&\le
\frac{1}{n-3}
\sum_i
\widehat e_i^2
|(\mathbf M_{d,n})_{ii}-d_i|\\
&\le
\frac{
18M
}{n-3}
\left(
\max_i\ell_i
\right)
\sum_i\widehat e_i^2=o_p(1),
\end{aligned}
\]
since, by \eqref{eq:ell_i},
$\max_i\ell_i
=
O_p\left(\xi_p^2/n\right)
=o_p(1)$ because $\xi_p^2/n\to0$,
and $
\frac1n\sum_i\widehat e_i^2
=
O_p(1)
$.
Consequently,
\begin{equation}\label{eq:diag_sim_part1}
\zeta_n
-
\frac{1}{n-3}
\sum_i d_i\widehat e_i^2
=
o_p(1).
\end{equation}
By \eqref{eq:part2} in the proof of Theorem~\ref{thm:null}, $n^{-1}\sum_id_i\widehat e_i^2\rightarrow^p\sigma^2\mu_D$. Since $n/(n-3)\to1$, combining this with \eqref{eq:diag_sim_part1}, we have
\begin{equation}\label{eq:tau_mean_convergence}
\zeta_n
=
\E^*\left[ \frac1{n-3}
\sum_i d_i(\widehat e_i^*)^2\right]
\rightarrow^{p}
\sigma^2\mu_D.
\end{equation}
Next, we shall show that $\frac1{n-3}
\sum_i d_i(\widehat e_i^*)^2$ concentrates around its conditional mean
$\zeta_n$.
We notice that 
\begin{equation}\label{eq:var_Qn}
\begin{aligned}
\operatorname{Var}^*\left(\frac1{n-3}
\sum_i d_i(\widehat e_i^*)^2\right)
&=\operatorname{Var}^*\left(\frac1{n-3}
(\bm \eta^*)^\top \mathbf D_{\widehat e}\mathbf M_{d,n}\mathbf D_{\widehat e}\bm \eta^*\right)\\
&=
\frac{2}{(n-3)^2}
\left\|
\mathbf D_{\widehat e}\mathbf M_{d,n}\mathbf D_{\widehat e}
\right\|_F^2,
\end{aligned}
\end{equation}
because $\bm\eta^*$ is a standard Gaussian vector and $\mathbf D_{\widehat e}\mathbf M_{d,n}\mathbf D_{\widehat e}$ is symmetric. Indeed, orthogonally diagonalizing this matrix reduces the quadratic form to a weighted sum of independent squared standard normals, with the eigenvalues as weights; each squared standard normal has variance $2$, and the squared eigenvalues sum to the squared Frobenius norm.
Also, 
\[
\begin{aligned}
\|\mathbf D_{\widehat e}\mathbf M_{d,n}\mathbf D_{\widehat e}\|_F^2
&=
\sum_{i,j}
{\widehat e}_i^2{\widehat e}_j^2(\mathbf M_{d,n})_{ij}^2\le
\sum_i
{\widehat e}_i^4
\sum_j(\mathbf M_{d,n})_{ij}^2\\
&=\sum_i
{\widehat e}_i^4 (\mathbf M_{d,n}^2)_{ii}\leq \|\mathbf M_{d,n}\|_{2}^2\sum_i
{\widehat e}_i^4\\
&\leq (\|\mathbf \Pi_n\|_{2}^2
\|\mathbf D_d\|_{2})^2\sum_i
{\widehat e}_i^4 \leq 36M^2 \sum_i
{\widehat e}_i^4,
\end{aligned}
\]
where the first inequality uses $\widehat e_i^2\widehat e_j^2\le(\widehat e_i^4+\widehat e_j^4)/2$ and the symmetry of $\mathbf M_{d,n}$, $\|\mathbf \Pi_n\|_{2}\le1$, and $\|\mathbf D_d\|_{2}
=
\max_i|d_i|
\le6M$. 
By this fact, \eqref{eq:var_Qn}, and $\frac1n
\sum_{i=1}^n
\widehat e_i^4
=
O_p(1)$ which we show later, we obtain
\begin{equation}\label{eq:var_diag_boot}
\begin{aligned}
\operatorname{Var}^*\left(\frac1{n-3}
\sum_i d_i(\widehat e_i^*)^2\right)
&\le
\frac{2\cdot 36M^2}{(n-3)^2}
\sum_i\widehat e_i^4=
O_p(n^{-1})=o_p(1).
\end{aligned}
\end{equation}
By Chebyshev's inequality, \eqref{eq:var_diag_boot} gives $(n-3)^{-1}\sum_id_i(\widehat e_i^*)^2-\zeta_n=o_p^*(1)$, which, together with \eqref{eq:tau_mean_convergence}, implies
\begin{equation}\label{eq:Q_n_convergence}
\frac1{n-3}
\sum_i d_i(\widehat e_i^*)^2
\rightarrow^{p^*}
\sigma^2\mu_D.
\end{equation}

\noindent Now we are left to verify that $n^{-1}\sum_{i=1}^n\widehat e_i^4=O_p(1)$. Under the null hypothesis, $\widehat e=\mathbf\Pi_nY=\mathbf\Pi_ne+\mathbf\Pi_nR_p=e-\mathbf P_ne+\mathbf\Pi_nR_p$. Let $C_4=\max\{\E(e^4),3\sigma^4\}$. Conditional on $\mathbf W_{1:n}$, the matrix $\mathbf P_n$ is fixed and $(\mathbf P_ne)_i=\sum_{j=1}^n(\mathbf P_n)_{ij}e_j$. Since $\mathbf P_n^2=\mathbf P_n$, we have $\sum_j(\mathbf P_n)_{ij}^2=(\mathbf P_n)_{ii}\in[0,1]$ and $\sum_j(\mathbf P_n)_{ij}^4\le(\mathbf P_n)_{ii}^2$. On expanding the fourth power, only terms with four equal indices or two pairs of equal indices have nonzero expectations. Consequently,
\[
\begin{aligned}
\E\left[(\mathbf P_ne)_i^4\mid\mathbf W_{1:n}\right]
&=3\sigma^4\Big\{\sum_{j=1}^n(\mathbf P_n)_{ij}^2\Big\}^2+\{\E(e^4)-3\sigma^4\}\sum_{j=1}^n(\mathbf P_n)_{ij}^4\\
&\le C_4(\mathbf P_n)_{ii}^2\le C_4.
\end{aligned}
\]
The upper bound holds for either sign of $\E(e^4)-3\sigma^4$: if this coefficient is negative, its contribution can be dropped; otherwise, use $\sum_j(\mathbf P_n)_{ij}^4\le(\mathbf P_n)_{ii}^2$. Averaging over $i$ and applying Markov's inequality gives $n^{-1}\sum_i(\mathbf P_ne)_i^4=O_p(1)$. Moreover, $\sum_ix_i^4\le(\sum_ix_i^2)^2$ and $\|\mathbf\Pi_nR_p\|\le\|R_p\|$, so that, by (B3) in Assumption~\ref{assum:1},
\[
\frac1n\sum_{i=1}^n(\mathbf\Pi_nR_p)_i^4\le\frac1n\|R_p\|^4\le na_p^4=o(1).
\]
Since $n^{-1}\sum_ie_i^4\rightarrow^p\E e^4<\infty$ by (B1), the inequality $(a+b+c)^4\le27(a^4+b^4+c^4)$ yields
\begin{equation}
\frac1n\sum_{i=1}^n\widehat e_i^4
\le
\frac{27}{n}\sum_{i=1}^n\big\{e_i^4+(\mathbf P_ne)_i^4+(\mathbf\Pi_nR_p)_i^4\big\}
=O_p(1).
\label{eq:fitted-residual-fourth}
\end{equation}
Lastly, by (\ref{eq:Z_n_norm_convergence}), (\ref{eq:Q_n_convergence}), and Slutsky's theorem,
we obtain
\begin{equation}\label{eq:bootstrap_null_dist}
T_n^*
\rightarrow^{D^*}
\|\mathcal G_\infty\|_{\mathcal H}^2
-
\sigma^2\mu_D=\mathcal L_{0,\infty}.
\end{equation}

\noindent To strengthen \eqref{eq:bootstrap_null_dist} to uniform convergence, write $F_{0,\infty}(x)=P(\mathcal L_{0,\infty}\le x)$. Its continuity was established at the end of the proof of Theorem~\ref{thm:null}.

Let $F_n^*(x)=P^*(T_n^*\le x)$. By \eqref{eq:bootstrap_null_dist} and the continuity of $F_{0,\infty}$, $F_n^*(x)\rightarrow^pF_{0,\infty}(x)$ for every $x\in\mathbb R$. Fix $\delta>0$ and choose $-\infty<x_0<x_1<\cdots<x_K<\infty$ such that $F_{0,\infty}(x_0)\le\delta$, $1-F_{0,\infty}(x_K)\le\delta$, and $F_{0,\infty}(x_k)-F_{0,\infty}(x_{k-1})\le\delta$ for $k=1,\ldots,K$. For $x\in[x_{k-1},x_k]$, the monotonicity of $F_n^*$ and $F_{0,\infty}$ gives
\[
F_n^*(x_{k-1})-F_{0,\infty}(x_k)\le F_n^*(x)-F_{0,\infty}(x)\le F_n^*(x_k)-F_{0,\infty}(x_{k-1}),
\]
and, by the choice of the grid, the lower and upper bounds differ from $F_n^*(x_{k-1})-F_{0,\infty}(x_{k-1})$ and $F_n^*(x_k)-F_{0,\infty}(x_k)$, respectively, by at most $\delta$. For $x<x_0$, $|F_n^*(x)-F_{0,\infty}(x)|\le\max\{F_n^*(x_0),F_{0,\infty}(x_0)\}\le|F_n^*(x_0)-F_{0,\infty}(x_0)|+\delta$, and, in the same way, $|F_n^*(x)-F_{0,\infty}(x)|\le|F_n^*(x_K)-F_{0,\infty}(x_K)|+\delta$ for $x>x_K$. Therefore,
\[
\sup_{x\in\mathbb R}|F_n^*(x)-F_{0,\infty}(x)|\le\max_{0\le k\le K}|F_n^*(x_k)-F_{0,\infty}(x_k)|+\delta=o_p(1)+\delta.
\]
Since $\delta>0$ is arbitrary, we conclude that
\begin{equation}
\sup_{x\in\mathbb R}
\left|
P^*(T_n^*\le x)
-
P(\mathcal L_{0,\infty}\le x)
\right|
\rightarrow^{p}0.
\label{eq:uniform-bootstrap-cdf}
\end{equation}

This completes the proof.

\end{proof}

\subsection{Proofs for limiting distributions under the alternatives}
After a lemma for the fixed-$p$ bootstrap remainder, we prove the alternative limits for fixed and growing $p$, followed by the corresponding bootstrap power results.

\begin{lemma}\label{lemma:boostrap_remainder_alt}
Consider the fixed-$p$ setting of Theorem~\ref{thm:power_local_boot}. Let $\mathbf P_n$ be the projection onto the column space of the $n\times(p+1)$ matrix with rows $(1,\boldsymbol\phi_{p,i}^\top)$, let $\mathbf\Pi_n=\mathbf I_n-\mathbf P_n$, and let $\mathbf D_\eta=\operatorname{diag}(\eta_1^*,\ldots,\eta_n^*)$. For $\mathbf u,\mathbf u'\in\mathbb R^n$, define $\widehat{\mathbf e}^*(\mathbf u)=\mathbf\Pi_n\mathbf D_\eta\mathbf\Pi_n\mathbf u$ and
\[
B_n^*(\mathbf u,\mathbf u')=-\frac{\widehat{\mathbf e}^*(\mathbf u)^\top\widetilde{\mathbf A}\,\widehat{\mathbf e}^*(\mathbf u')}{n-3},
\qquad
T_n^*(\mathbf u)=B_n^*(\mathbf u,\mathbf u),
\]
and write $\boldsymbol\varepsilon=(\varepsilon_1,\ldots,\varepsilon_n)^\top$ and $\boldsymbol\Delta=(\Delta_1,\ldots,\Delta_n)^\top$ with $\Delta_i=\Delta(\mathbf W_i)$. Suppose that the conditions of Theorem~\ref{thm:power_local_boot} hold; in particular, Assumptions~\ref{assum:0} and \ref{assum:2} hold with $\varepsilon$ in place of $e$, and $\E[\Delta(\mathbf W)^8]<\infty$. Then $B_n^*(\mathbf u,\mathbf u')=O_p^*(1)$ for all $\mathbf u,\mathbf u'\in\{\boldsymbol\varepsilon,\boldsymbol\Delta\}$; in particular, $B_n^*(\boldsymbol\varepsilon,\boldsymbol\Delta)$, $T_n^*(\boldsymbol\varepsilon)$, and $T_n^*(\boldsymbol\Delta)$ are $O_p^*(1)$.
\end{lemma}

\begin{proof}[Proof of Lemma~\ref{lemma:boostrap_remainder_alt}]
Fix $\mathbf u,\mathbf u'\in\{\boldsymbol\varepsilon,\boldsymbol\Delta\}$, and let $\mathbf w=\mathbf\Pi_n\mathbf u$ and $\mathbf w'=\mathbf\Pi_n\mathbf u'$.

\textit{Step 1: an exact expansion.} Let $\mathbf F$ and $\mathbf F_c$ be the $n\times p$ matrices with rows $\boldsymbol\phi_{p,k}^\top$ and $(\boldsymbol\phi_{p,k}-\overline{\boldsymbol\phi}_p)^\top$, and let
\[
\boldsymbol\beta^*(\mathbf w)=\widehat{\mathbf Q}^{-1}\frac1n\sum_{k=1}^n(\boldsymbol\phi_{p,k}-\overline{\boldsymbol\phi}_p)\eta_k^*w_k
\]
be the slope of the least-squares fit of $\mathbf D_\eta\mathbf w$ on the intercept and the basis. Then $\widehat{\mathbf e}^*(\mathbf u)=\mathbf D_\eta\mathbf w-a\mathbf 1-\mathbf F_c\boldsymbol\beta^*(\mathbf w)$ for a scalar $a$. Since $\widetilde{\mathbf A}\mathbf 1=\mathbf 0$ by \eqref{eq:row_col_sum}, the term $a\mathbf 1$ does not contribute to $B_n^*$, and $\widetilde{\mathbf A}\mathbf F_c=\widetilde{\mathbf A}\mathbf F$ has rows $\mathbf v_i^\top$, where $\mathbf v_i=\sum_{j\neq i}\widetilde A_{ij}\boldsymbol\phi_{p,j}$. Moreover, $\mathbf F^\top\widetilde{\mathbf A}\mathbf F=n(n-3)\mathbf M_n$ with $\mathbf M_n$ as in (F3) of the proof of Lemma~\ref{lemma:fixed_p_bootstrap_remainder}. Using also $\widetilde A_{ii}=0$, we obtain
\[
\begin{aligned}
(n-3)B_n^*(\mathbf u,\mathbf u')
={}&-\sum_{i\neq j}\widetilde A_{ij}w_iw_j'\eta_i^*\eta_j^*
+\sum_{i=1}^n\eta_i^*\big\{w_i\mathbf v_i^\top\boldsymbol\beta^*(\mathbf w')+w_i'\mathbf v_i^\top\boldsymbol\beta^*(\mathbf w)\big\}\\
&-n(n-3)\,\boldsymbol\beta^*(\mathbf w)^\top\mathbf M_n\boldsymbol\beta^*(\mathbf w').
\end{aligned}
\]

\textit{Step 2: the residuals.} Write $\mathbf w=\mathbf u-\bar u\mathbf 1-\mathbf F_c\widehat{\boldsymbol\beta}_{\mathbf u}$, where $\bar u=n^{-1}\sum_ku_k$ and $\widehat{\boldsymbol\beta}_{\mathbf u}=\widehat{\mathbf Q}^{-1}n^{-1}\sum_k(\boldsymbol\phi_{p,k}-\overline{\boldsymbol\phi}_p)u_k$. For both $\mathbf u=\boldsymbol\varepsilon$ and $\mathbf u=\boldsymbol\Delta$, we have $\E[u_k\mid\mathbf Z_k]=0$. Thus the i.i.d.\ variables $u_k$ and vectors $\boldsymbol\phi_{p,k}u_k$ have mean zero and finite second moments, so the central limit theorem gives $\bar u=O_p(n^{-1/2})$ and $n^{-1}\sum_k\boldsymbol\phi_{p,k}u_k=O_p(n^{-1/2})$. Since
\[
\widehat{\boldsymbol\beta}_{\mathbf u}
=\widehat{\mathbf Q}^{-1}
\left(\frac1n\sum_k\boldsymbol\phi_{p,k}u_k-\overline{\boldsymbol\phi}_p\bar u\right),
\]
and $\|\widehat{\mathbf Q}^{-1}\|_2=O_p(1)$ and $\overline{\boldsymbol\phi}_p=O_p(n^{-1/2})$, we also obtain $\widehat{\boldsymbol\beta}_{\mathbf u}=O_p(n^{-1/2})$. Since $\sup_{\mathbf z}\|\boldsymbol\phi_p(\mathbf z)\|\le C$, it follows that $|w_k|\le|u_k|+\rho_n$ for all $k$, with $\rho_n=|\bar u|+2C\|\widehat{\boldsymbol\beta}_{\mathbf u}\|=O_p(n^{-1/2})$, and the same holds for $\mathbf w'$. Markov's inequality, the moment bounds in Assumption~\ref{assum:2} with $\varepsilon$ in place of $e$, and $\E[\Delta(\mathbf W)^8]<\infty$ give
\[
\sum_{i\neq j}\widetilde A_{ij}^2(1+\varepsilon_i^2+\Delta_i^2)(1+\varepsilon_j^2+\Delta_j^2)=O_p(n^2).
\]
To see this, the triangle inequality and \eqref{eq:ucenter} give, for a constant $C_0$ independent of $n\ge4$,
\[
\widetilde A_{ij}^2
\le C_0\left(\|\mathbf W_i\|^2+\|\mathbf W_j\|^2
+\frac1n\sum_{k=1}^n\|\mathbf W_k\|^2\right).
\]
Multiplying by $(1+\varepsilon_i^2+\Delta_i^2)(1+\varepsilon_j^2+\Delta_j^2)$, independence across observations and the Cauchy--Schwarz inequality show that each summand has bounded expectation, uniformly in $i\ne j$ and $n$. In particular, $\E[\|\mathbf W\|^2\Delta^2]\le(\E\|\mathbf W\|^4\,\E\Delta^4)^{1/2}<\infty$. Summing these expectations and applying Markov's inequality gives the stated $O_p(n^2)$ bound.
Consequently,
\[
\begin{aligned}
&\sum_{i\neq j}\widetilde A_{ij}^2w_i^2w_j'^2=O_p(n^2),\qquad
\frac1n\sum_{k=1}^nw_k^2=O_p(1),\\
&\sum_{i=1}^nw_i^2\|\mathbf v_i\|^2\le C^2n\sum_{i\neq j}\widetilde A_{ij}^2w_i^2=O_p(n^3),
\end{aligned}
\]
where the last bound uses $\|\mathbf v_i\|^2\le C^2n\sum_j\widetilde A_{ij}^2$ from (F3); the same bounds hold for $\mathbf w'$.

\textit{Step 3: the bounds.} By (F1), the first sum in Step 1 has conditional second moment at most $2\sum_{i\neq j}\widetilde A_{ij}^2w_i^2w_j'^2=O_p(n^2)$, so it is $O_p^*(n)$. By (F2), $\E^*\|\sum_i\eta_i^*w_i\mathbf v_i\|^2=\sum_iw_i^2\|\mathbf v_i\|^2=O_p(n^3)$ and
\[
\E^*\|\boldsymbol\beta^*(\mathbf w)\|^2\le\|\widehat{\mathbf Q}^{-1}\|_2^2\,\frac{4C^2}{n^2}\sum_{k=1}^nw_k^2=O_p(n^{-1}),
\]
so that the second term is $O_p^*(n^{3/2}\cdot n^{-1/2})=O_p^*(n)$ by the Cauchy--Schwarz inequality. Since $\|\mathbf M_n\|_2=O_p(1)$ by (F3), the third term is $O_p^*(n^2\cdot n^{-1})=O_p^*(n)$. Dividing by $n-3$ gives $B_n^*(\mathbf u,\mathbf u')=O_p^*(1)$.

\end{proof}

\begin{proof}[Proof of Theorem~\ref{thm:power_local_fixed_p}]

We consider three regimes: (i) $\sqrt n|c_n|\to\infty$ and $\sup_n|c_n|\le M_c<\infty$, (ii) $\sqrt n c_n\to\tau\in\mathbb R$, and (iii) $\sqrt n c_n\to0$; in each of them, $c_n$ is bounded. As in Section~\ref{sec:power} of the main text, $\varepsilon$ plays the role of $e$ and $\mathbf V=(\mathbf W,\varepsilon)$; accordingly, we use the notation introduced before Lemma~\ref{lemma:sim1} with $e$ replaced by $\varepsilon$, so that $\mathbf R_{ij}=(\varepsilon_i-\varepsilon_j)(\boldsymbol\phi_{p,i}-\boldsymbol\phi_{p,j})$. Recall from the main text that
\[
\mu_\Delta=\mathrm{MDD}^2(\Delta\mid\mathbf W)>0,
\qquad
\boldsymbol\nu_\Delta=2\E[U(\mathbf W,\mathbf W')\Delta(\mathbf W)\boldsymbol\phi_p(\mathbf Z')].
\]
For ease of notation, write $\Delta_i=\Delta(\mathbf X_i,\mathbf Z_i)$ and $u_i=\varepsilon_i+c_n\Delta_i$.

Under the alternative $H_{1,n}$ and Condition (A1), we have
\[
Y_i=\gamma_0+\boldsymbol\phi_{p,i}^\top\boldsymbol\beta_0+u_i,
\qquad
\E[\Delta_i\mid\mathbf Z_i]=0,
\]
which implies that $\widehat{\boldsymbol\beta}-\boldsymbol\beta_0=n^{-1/2}\boldsymbol\xi(u)$. Since $\widehat e_i=u_i-(\widehat\gamma-\gamma_0)-\boldsymbol\phi_{p,i}^\top(\widehat{\boldsymbol\beta}-\boldsymbol\beta_0)$, the intercept cancels in differences of residuals; with $\mathbf q_n=\boldsymbol\xi(u)$,
\[
\widehat e_i-\widehat e_j
=u_i-u_j-\frac{1}{\sqrt n}(\boldsymbol\phi_{p,i}-\boldsymbol\phi_{p,j})^\top\mathbf q_n.
\]
Let $\widetilde B_{ij}$, $\widetilde C_{ij}$, $\widetilde\Delta_{ij}$, and $\widetilde{\mathbf D}_{ij}$ be the $\mathcal U$-centered versions of
\[
\begin{aligned}
B_{ij}&=\tfrac12(\varepsilon_i-\varepsilon_j)^2,
&C_{ij}&=(\varepsilon_i-\varepsilon_j)(\Delta_i-\Delta_j),\\
\Delta_{ij}&=\tfrac12(\Delta_i-\Delta_j)^2,
&\mathbf D_{ij}&=(\Delta_i-\Delta_j)(\boldsymbol\phi_{p,i}-\boldsymbol\phi_{p,j}),
\end{aligned}
\]
respectively, and let $J_F=(n-3)^{-1}\sum_{i\neq j}\widetilde A_{ij}\widetilde F_{ij}$ for $F\in\{B,C,\Delta\}$. Squaring the residual difference gives
\[
\tfrac12(\widehat e_i-\widehat e_j)^2
=B_{ij}+c_nC_{ij}+c_n^2\Delta_{ij}
-\tfrac{1}{\sqrt n}(\mathbf R_{ij}+c_n\mathbf D_{ij})^\top\mathbf q_n
+\tfrac1n\mathbf q_n^\top\mathbf S_{ij}\mathbf q_n.
\]
Since $\mathcal U$-centering is linear, the test statistic admits the exact decomposition
\begin{equation}\label{eq:nMDD_alt}
n\widehat{\rm MDD}^2(\widehat e\mid\mathbf W)
=J_B+c_nJ_C+c_n^2J_\Delta
-(\mathbf w_{\mathbf R}+c_n\mathbf w_{\mathbf D})^\top\mathbf q_n
+\mathbf q_n^\top\mathbf S_n\mathbf q_n.
\end{equation}

Let $\mathbf g_n=\boldsymbol\xi^\circ(u)=\boldsymbol\xi^\circ(\varepsilon)+c_n\boldsymbol\xi^\circ(\Delta)$ and $\mathbf r_n=\mathbf q_n-\mathbf g_n$. Substituting $\mathbf q_n=\mathbf g_n+\mathbf r_n$ into \eqref{eq:nMDD_alt}, expanding $\mathbf g_n$, and using the symmetry of $\mathbf S_n$, we obtain the exact identity
\begin{equation}\label{eq:nMDD_alt_split}
n\widehat{\rm MDD}^2(\widehat e\mid\mathbf W)=T_{n,0}+T_{n,\Delta}+E_n,
\end{equation}
where
\[
\begin{aligned}
T_{n,0}&=J_B-\mathbf w_{\mathbf R}^\top\boldsymbol\xi^\circ(\varepsilon)
+\boldsymbol\xi^\circ(\varepsilon)^\top\mathbf S_n\boldsymbol\xi^\circ(\varepsilon),\\
T_{n,\Delta}&=c_n^2J_\Delta+c_nJ_C
-c_n\mathbf w_{\mathbf R}^\top\boldsymbol\xi^\circ(\Delta)
-c_n\mathbf w_{\mathbf D}^\top\boldsymbol\xi^\circ(\varepsilon)
-c_n^2\mathbf w_{\mathbf D}^\top\boldsymbol\xi^\circ(\Delta)\\
&\quad+2c_n\boldsymbol\xi^\circ(\Delta)^\top\mathbf S_n\boldsymbol\xi^\circ(\varepsilon)
+c_n^2\boldsymbol\xi^\circ(\Delta)^\top\mathbf S_n\boldsymbol\xi^\circ(\Delta),\\
E_n&=-(\mathbf w_{\mathbf R}+c_n\mathbf w_{\mathbf D})^\top\mathbf r_n
+2\mathbf g_n^\top\mathbf S_n\mathbf r_n+\mathbf r_n^\top\mathbf S_n\mathbf r_n.
\end{aligned}
\]
Here $T_{n,0}$ consists of the three leading terms of \eqref{eq:nmdd_fixed_p_stat} with $e$ replaced by $\varepsilon$, and $T_{n,\Delta}$ collects the terms contributed by the alternative. We next collect the orders of the building blocks. Apart from $\mathbf g_n$, $\mathbf r_n$, and $\sum_ku_k$, they do not involve $c_n$: they are functions of the i.i.d.\ sample $\{(\mathbf W_i,\varepsilon_i)\}_{i=1}^n$, whose distribution does not change with $n$.
\begin{enumerate}
\item[(O1)] Since $\E[\boldsymbol\phi_p(\mathbf Z)\varepsilon]=\E[\boldsymbol\phi_p(\mathbf Z)\Delta]=\mathbf 0$ and $\boldsymbol\phi_p$ is bounded, the central limit theorem gives $\|\boldsymbol\xi^\circ(\varepsilon)\|+\|\boldsymbol\xi^\circ(\Delta)\|=O_p(1)$; hence $\|\mathbf g_n\|=O_p(1)$. Moreover, $\|\mathbf S_n\|_2=O_p(1)$ by \eqref{eq:sim1_II}.
\item[(O2)] $J_B=O_p(1)$ and $\|\mathbf w_{\mathbf R}\|=O_p(1)$ by the arguments in the proof of Theorem~\ref{thm:null_fixed_p}, which apply because $\varepsilon$ is independent of $\mathbf W$ with $\E[\varepsilon]=0$.
\item[(O3)] $J_\Delta=n\mu_\Delta\{1+o_p(1)\}$, because $J_\Delta/n=\widehat{\rm MDD}^2(\Delta\mid\mathbf W)\rightarrow^p\mu_\Delta>0$ by the strong law of large numbers for $\mathcal U$-statistics \citep{hoeffding1961strong, lee1990u}.
\item[(O4)] $J_C=O_p(\sqrt n)$ and $\mathbf w_{\mathbf D}=-\sqrt n\,\boldsymbol\nu_\Delta+O_p(1)$. Indeed, $J_C/n$ and $\mathbf w_{\mathbf D}/\sqrt n$ are fourth-order $\mathcal U$-statistics whose kernels have finite second moments by Conditions (A2) and (A4), $\E[\Delta^4]<\infty$, and the independence of $\varepsilon$ and $\mathbf W$, and their means are $0$ and $-\boldsymbol\nu_\Delta$, respectively. For the means, note that $\widehat{\rm MDD}^2$ is unbiased for ${\rm MDD}^2$, so by polarization the $\mathcal U$-statistic built from $(a_i-a_j)(b_i-b_j)$ has mean $-2\E[(a-\E a)(b'-\E b')\|\mathbf W-\mathbf W'\|]$. This mean is $0$ for $(a,b)=(\varepsilon,\Delta)$, because $\varepsilon$ is independent of $\mathbf W$ with $\E[\varepsilon]=0$, and it equals $-2\E[\Delta(\mathbf W)\boldsymbol\phi_p(\mathbf Z')\|\mathbf W-\mathbf W'\|]=-\boldsymbol\nu_\Delta$ for $(a,b)=(\Delta,\boldsymbol\phi_p)$, because the centering terms of $U$ do not contribute when $\E[\Delta(\mathbf W)]=0$ and $\E[\boldsymbol\phi_p(\mathbf Z)]=\mathbf 0$.
\item[(O5)] $\|\mathbf r_n\|=O_p(n^{-1/2})$. Indeed,
\[
\mathbf r_n=(\widehat{\mathbf Q}^{-1}-\mathbf I_p)\mathbf g_n
-\widehat{\mathbf Q}^{-1}\overline{\boldsymbol\phi}_p\,\frac1{\sqrt n}\sum_{k=1}^nu_k.
\]
Here $\|\widehat{\mathbf Q}^{-1}\|_2=O_p(1)$ by \eqref{eq:Q-I}, $\overline{\boldsymbol\phi}_p=O_p(n^{-1/2})$ by \eqref{eq:phi_bar}, and $n^{-1/2}\sum_ku_k=O_p(1)$ because $\E[\varepsilon]=\E[\Delta]=0$. Moreover, by \eqref{eq:Q},
\[
\begin{aligned}
\widehat{\mathbf Q}-\mathbf I_p
&=\frac1n\sum_{k=1}^n(\boldsymbol\phi_{p,k}\boldsymbol\phi_{p,k}^\top-\mathbf I_p)
-\overline{\boldsymbol\phi}_p\overline{\boldsymbol\phi}_p^\top
=O_p(n^{-1/2}),\\
\widehat{\mathbf Q}^{-1}-\mathbf I_p
&=\widehat{\mathbf Q}^{-1}(\mathbf I_p-\widehat{\mathbf Q})=O_p(n^{-1/2}),
\end{aligned}
\]
where the first bound follows from Chebyshev's inequality and Condition (A4).
\end{enumerate}
By the Cauchy--Schwarz inequality and (O1)--(O5),
\[
\begin{aligned}
|E_n|
&\le\bigl(\|\mathbf w_{\mathbf R}\|+|c_n|\|\mathbf w_{\mathbf D}\|\bigr)\|\mathbf r_n\|
+\bigl(2\|\mathbf g_n\|+\|\mathbf r_n\|\bigr)\|\mathbf S_n\|_2\|\mathbf r_n\|\\
&=O_p(1+\sqrt n|c_n|)\,O_p(n^{-1/2})+O_p(n^{-1/2})
=O_p(n^{-1/2}+|c_n|).
\end{aligned}
\]
Thus $E_n=o_p(1)$ when $c_n\to0$. When $c_n$ does not tend to zero, $E_n$ need not be negligible, because $\|\mathbf w_{\mathbf D}\|$ may grow at the rate $\sqrt n$; in regime (i) we use instead that $E_n=o_p(nc_n^2)$. Finally, (O1)--(O4) give the orders of the terms of $T_{n,\Delta}$:
\begin{equation}\label{eq:extra_terms_alternative}
\begin{gathered}
c_n^2J_\Delta=nc_n^2\mu_\Delta\{1+o_p(1)\},\qquad
|c_nJ_C|+|c_n\mathbf w_{\mathbf D}^\top\boldsymbol\xi^\circ(\varepsilon)|=O_p(\sqrt n|c_n|),\\
|c_n^2\mathbf w_{\mathbf D}^\top\boldsymbol\xi^\circ(\Delta)|=O_p(\sqrt n c_n^2),\qquad
|c_n^2\boldsymbol\xi^\circ(\Delta)^\top\mathbf S_n\boldsymbol\xi^\circ(\Delta)|=O_p(c_n^2),\\
|c_n\mathbf w_{\mathbf R}^\top\boldsymbol\xi^\circ(\Delta)|+|2c_n\boldsymbol\xi^\circ(\Delta)^\top\mathbf S_n\boldsymbol\xi^\circ(\varepsilon)|=O_p(|c_n|),
\end{gathered}
\end{equation}%
while $T_{n,0}=O_p(1)$ by (O1) and (O2).

\textit{Regime (i).} Since $\sqrt n|c_n|\to\infty$, both $1+\sqrt n|c_n|$ and $n^{-1/2}+|c_n|$ are $o(nc_n^2)$. Hence $T_{n,0}$ and $E_n$ are $o_p(nc_n^2)$. Since also $\sqrt n c_n^2\le M_c\sqrt n|c_n|$, \eqref{eq:extra_terms_alternative} shows that every term of $T_{n,\Delta}$ other than $c_n^2J_\Delta=nc_n^2\mu_\Delta\{1+o_p(1)\}$ is $O_p(\sqrt n|c_n|)=o_p(nc_n^2)$. Dividing \eqref{eq:nMDD_alt_split} by $nc_n^2$, we obtain
\[
\frac{n\widehat{\rm MDD}^2(\widehat e\mid\mathbf W)}{nc_n^2}\rightarrow^p\mu_\Delta>0,
\]
and hence $n\widehat{\rm MDD}^2(\widehat e\mid\mathbf W)\rightarrow^p\infty$.

\textit{Regimes (ii) and (iii).} Now $c_n\to0$, so that $E_n=o_p(1)$ and \eqref{eq:nMDD_alt_split} becomes
\[
n\widehat{\rm MDD}^2(\widehat e\mid\mathbf W)=T_{n,0}+T_{n,\Delta}+o_p(1).
\]
By the proof of Theorem~\ref{thm:null_fixed_p} with $e$ replaced by $\varepsilon$, $T_{n,0}=\mathcal U_{n,1}+\mathcal U_{n,2}-\mathcal U_{n,3}+o_p(1)$, where $\mathcal U_{n,1}$, $\mathcal U_{n,2}$, and $\mathcal U_{n,3}$ are defined there.

Consider first the local case $\sqrt n c_n\to\tau\in\mathbb R$. By \eqref{eq:extra_terms_alternative}, all terms of $T_{n,\Delta}$ other than $c_n^2J_\Delta$, $c_nJ_C$, and $-c_n\mathbf w_{\mathbf D}^\top\boldsymbol\xi^\circ(\varepsilon)$ are $o_p(1)$. For these three terms, the strong law of large numbers for $\mathcal U$-statistics \citep{hoeffding1961strong, lee1990u}, the first-order Hoeffding projection argument used in the proof of Theorem~\ref{thm:null_fixed_p}, and (O4) give
\begin{equation}\label{eq:U_Delta}
\begin{aligned}
\mathcal U_{n,1}^{\Delta}:=c_n^2J_\Delta
&=nc_n^2\binom n4^{-1}\sum_{i<j<q<r}\breve h_4(\mathbf V_i,\mathbf V_j,\mathbf V_q,\mathbf V_r)
\rightarrow^p\tau^2\mu_\Delta,\\
c_nJ_C
&=\sqrt n c_n\cdot\sqrt n\binom n4^{-1}\sum_{i<j<q<r}h_4(\mathbf V_i,\mathbf V_j,\mathbf V_q,\mathbf V_r)\\
&=\tau\,\mathcal U_{n,2}^{\Delta}+o_p(1),\\
-c_n\mathbf w_{\mathbf D}^\top\boldsymbol\xi^\circ(\varepsilon)
&=\sqrt n c_n\,\boldsymbol\nu_\Delta^\top\boldsymbol\xi^\circ(\varepsilon)+o_p(1)
=\tau\,\mathcal U_{n,3}^{\Delta}+o_p(1),
\end{aligned}
\end{equation}
where $\mathcal U_{n,2}^{\Delta}=4n^{-1/2}\sum_{i=1}^nh_{4,1}(\mathbf V_i)$ and $\mathcal U_{n,3}^{\Delta}=\boldsymbol\nu_\Delta^\top\boldsymbol\xi^\circ(\varepsilon)$. Here the kernels of $J_C/n$ and $J_\Delta/n$ are
\[
\begin{aligned}
h_4(\mathbf V_i,\mathbf V_j,\mathbf V_q,\mathbf V_r)
&=\frac1{4!}\sum_{(s,t,u,v)}^{(i,j,q,r)}C_{st}\bigl(A_{uv}+A_{st}-A_{su}-A_{tv}\bigr)\in\mathbb R,\\
\breve h_4(\mathbf V_i,\mathbf V_j,\mathbf V_q,\mathbf V_r)
&=\frac1{4!}\sum_{(s,t,u,v)}^{(i,j,q,r)}A_{st}\bigl(\Delta_{uv}+\Delta_{st}-\Delta_{su}-\Delta_{tv}\bigr)\in\mathbb R,
\end{aligned}
\]
with the summation over all permutations of $(i,j,q,r)$ as in the proof of Theorem~\ref{thm:null_fixed_p}, and
\[
h_{4,1}(\mathbf V_1)
=\frac12\E[(-\varepsilon\Delta')U(\mathbf W,\mathbf W')\mid\mathbf V=\mathbf V_1]
\]
is the projection kernel of $h_4$. Combining these results with the expansion of $T_{n,0}$ above, we obtain
\begin{equation}
\label{eq:nMDD_sim1_fixed_p_final_alternative}
\begin{aligned}
n\widehat{\mathrm{MDD}}^2(\widehat e\mid\mathbf W)
&=\mathcal U_{n,1}+\mathcal U_{n,2}-\mathcal U_{n,3}\\
&\quad+\tau^2\mu_\Delta+\tau\mathcal U_{n,2}^{\Delta}+\tau\mathcal U_{n,3}^{\Delta}+o_p(1).
\end{aligned}
\end{equation}

\noindent To derive the limiting distribution, we use the truncation argument from the proof of Theorem~\ref{thm:null_fixed_p}, with $\boldsymbol\psi^K(\mathbf V)=(\psi_1(\mathbf V),\ldots,\psi_K(\mathbf V))^\top$ as there. For each fixed positive integer $K$, the multivariate central limit theorem gives
\begin{equation}\label{eq:MCLT_alternative}
\frac1{\sqrt n}\sum_{i=1}^n\big(\boldsymbol\psi^K(\mathbf V_i)^\top,\varepsilon_i\boldsymbol\phi_{p,i}^\top,4h_{1,1}(\mathbf V_i)^\top,4h_{4,1}(\mathbf V_i)\big)^\top\rightarrow^{D}N(\mathbf0,\widetilde{\boldsymbol\Sigma}^{\Delta}),
\end{equation}
where $\widetilde{\boldsymbol\Sigma}^{\Delta}$ is the covariance matrix of each summand; the second block on the left-hand side is $\boldsymbol\xi^\circ(\varepsilon)$. Denote the four blocks of the Gaussian limit by $(G_1,\ldots,G_K)^\top$, $\boldsymbol{\mathcal N}$, $\boldsymbol{\mathcal T}$, and $\mathcal C_\Delta$, and let $\mathcal D_\Delta=\boldsymbol\nu_\Delta^\top\boldsymbol{\mathcal N}$. Because $4h_{4,1}(\mathbf V)=2C(\mathbf V)$ and $\operatorname{Var}\{\varepsilon\boldsymbol\phi_p(\mathbf Z)\}=\sigma^2\mathbf I_p$, we have $\mathcal C_\Delta\sim N\{0,4\operatorname{var}(C(\mathbf V))\}$ and $\mathcal D_\Delta\sim N(0,\sigma^2\|\boldsymbol\nu_\Delta\|^2)$, as in Theorem~\ref{thm:power_local_fixed_p}. The quantities $\mathcal U_{n,1}^K$ (defined in the proof of Theorem~\ref{thm:null_fixed_p}), $\mathcal U_{n,3}$, $\mathcal U_{n,2}^{\Delta}$, and $\mathcal U_{n,3}^{\Delta}$ are continuous functions of the left-hand side of \eqref{eq:MCLT_alternative} and of $n^{-1}\sum_i\psi_k(\mathbf V_i)^2\rightarrow^p1$, $k\le K$, while $\mathcal U_{n,2}=\boldsymbol\xi^\circ(\varepsilon)^\top\mathbf S_n\boldsymbol\xi^\circ(\varepsilon)$ with $\mathbf S_n\rightarrow^p\breve{\mathbf H}$ by \eqref{eq:converges_p_refined}. Hence, for fixed $K$,
\begin{equation}\label{eq:fixed_K_dist_alternative}
\begin{aligned}
&\mathcal U_{n,1}^K+\mathcal U_{n,2}-\mathcal U_{n,3}+\tau\mathcal U_{n,2}^{\Delta}+\tau\mathcal U_{n,3}^{\Delta}\\
&\quad\rightarrow^{D}\sum_{k=1}^K\lambda_k(G_k^2-1)+\boldsymbol{\mathcal N}^\top\breve{\mathbf H}\boldsymbol{\mathcal N}-\boldsymbol{\mathcal T}^\top\boldsymbol{\mathcal N}+\tau\mathcal C_\Delta+\tau\mathcal D_\Delta.
\end{aligned}
\end{equation}
As shown in the proof of Theorem~\ref{thm:null_fixed_p}, $\E[(\mathcal U_{n,1}-\mathcal U_{n,1}^K)^2]=2(n-1)n^{-1}\sum_{k>K}\lambda_k^2$, which tends to zero as $K\to\infty$. Therefore, for any $\delta>0$,
\begin{equation}\label{eq:limsup_alternative}
\lim_{K\to\infty}\limsup_{n\to\infty}P\big(|\mathcal U_{n,1}-\mathcal U_{n,1}^K|\ge\delta\big)=0.
\end{equation}
By combining Theorem 2 of \citet{dehling2009new} with \eqref{eq:U_Delta}, \eqref{eq:nMDD_sim1_fixed_p_final_alternative}, \eqref{eq:fixed_K_dist_alternative}, and \eqref{eq:limsup_alternative}, it follows that
\[
n\widehat{\rm MDD}^2(\widehat e\mid\mathbf W)\rightarrow^{D}\mathcal L_{0,p}+\tau\mathcal C_\Delta+\tau\mathcal D_\Delta+\tau^2\mu_\Delta=\mathcal L_{p,\tau},
\]
where $\mathcal L_{0,p}=\sum_{k=1}^{\infty}\lambda_k(G_k^2-1)+\boldsymbol{\mathcal N}^\top\breve{\mathbf H}\boldsymbol{\mathcal N}-\boldsymbol{\mathcal T}^\top\boldsymbol{\mathcal N}$ is the limit in \eqref{eq:null_fixed_p} of the main text.

Finally, the third regime $\sqrt n\,c_n\to0$ is the special case $\tau=0$ of the preceding local-alternative result: all alternative-specific terms in \eqref{eq:extra_terms_alternative} are $o_p(1)$, so that $n\widehat{\rm MDD}^2(\widehat e\mid\mathbf W)=T_{n,0}+o_p(1)$, and the arguments in the proof of Theorem~\ref{thm:null_fixed_p} give
\[
n\widehat{\rm MDD}^2(\widehat e\mid\mathbf W)
\rightarrow^{D}
\sum_{k=1}^{\infty}\lambda_k(G_k^2-1)+\boldsymbol{\mathcal N}^\top\breve{\mathbf H}\boldsymbol{\mathcal N}
-\boldsymbol{\mathcal T}^\top\boldsymbol{\mathcal N}.
\]

This completes the proof.
\end{proof}

\begin{proof}[Proof of Theorem~\ref{thm:power_local_growing_p}] The proof follows the steps of the proof of Theorem~\ref{thm:null}. We first show that $m_\Delta=\E[\Delta(\mathbf W)h_\infty(\mathbf W)]$. We then split the feature score into an innovation term, a signal term, and an approximation term in \eqref{eq:score-decomposition}, and finally treat the three regimes for $c_n$, handling the diagonal term of $T_n$ separately in each. Recall that under the alternative $H_{1,n}$, we have
\[
Y=f(\mathbf Z)
+c_n\Delta(\mathbf W)+\varepsilon=
\gamma_p+\boldsymbol{\beta}_p^\top \mathbf g_p(\mathbf Z)
+
R_p
+
c_n\Delta(\mathbf W)
+
\varepsilon,
\]

\noindent where $\E[\Delta(\mathbf W)\mid \mathbf Z]=0$, so that $\E[\Delta(\mathbf W)]=0$ and
\[
\E[\Delta(\mathbf W)\mathbf g_{p}(\mathbf Z)]
=\E\left[\mathbf g_{p}(\mathbf Z)\E[\Delta(\mathbf W)\mid \mathbf Z]\right]=\mathbf 0.
\]
Hence, for every $p$,
\begin{equation}\label{eq:signal-hp}
\begin{aligned}
\E[\Delta(\mathbf W)h_p(\mathbf W)]
&=\E[\Delta(\mathbf W)\Phi(\mathbf W)]-\E[\Delta(\mathbf W)]m_\Phi-\mathcal{B}_p\E[\Delta(\mathbf W)\mathbf g_p(\mathbf Z)]\\
&=\E[\Delta(\mathbf W)\Phi(\mathbf W)].
\end{aligned}
\end{equation}
Since $\E[\Delta(\mathbf W)^2]<\infty$ and $\E\|h_p(\mathbf W)-h_\infty(\mathbf W)\|_{\mathcal H}^2\to0$, the Cauchy--Schwarz inequality gives
\[
\begin{aligned}
&\big\|\E[\Delta(\mathbf W)\{h_p(\mathbf W)-h_\infty(\mathbf W)\}]\big\|_{\mathcal H}\\
&\quad\le\E\big[|\Delta(\mathbf W)|\,\|h_p(\mathbf W)-h_\infty(\mathbf W)\|_{\mathcal H}\big]\\
&\quad\le\{\E[\Delta(\mathbf W)^2]\}^{1/2}\big\{\E\|h_p(\mathbf W)-h_\infty(\mathbf W)\|_{\mathcal H}^2\big\}^{1/2}\rightarrow0,
\end{aligned}
\]
so that letting $p\to\infty$ in \eqref{eq:signal-hp} yields $m_\Delta=\E[\Delta(\mathbf W)h_\infty(\mathbf W)]=\E[\Delta(\mathbf W)\Phi(\mathbf W)]$.

For ease of notation, we still write $\Delta_i=\Delta(\mathbf{X}_i, \mathbf{Z}_i)$. Recall our test statistic using Lemma~\ref{lemma:test_statistic}, i.e., 
\[
T_n
=n\widehat{\MDD}^{\,2}(\widehat e\mid \mathbf W)=
\frac{n}{n-3}
\left\|
\frac1{\sqrt n}\sum_{i=1}^n \widehat{e}_i\Phi_i
\right\|_{\Hcal}^2
-
\frac{1}{n-3}\sum_i d_i \widehat{e}_i^2.
\]Under the alternative, we have
\begin{equation}
\widehat{e}=\mathbf \Pi_nY
=
\mathbf \Pi_n\varepsilon
+
c_n\mathbf \Pi_n\Delta
+
\mathbf \Pi_nR_p.
\label{eq:res_alternative}
\end{equation}
Using the exact projection identity established in (\ref{eq:Phi_t_equality}), we obtain
\begin{align}
\frac1{\sqrt n}
\sum_{i=1}^n\widehat{e}_i\Phi_i
={}&
\frac1{\sqrt n}\sum_{i=1}^n\varepsilon_i t_i
+
\sqrt n\,c_n
\left(
\frac1n\sum_{i=1}^n\Delta_i t_i
\right)
+
\frac1{\sqrt n}
\sum_{i=1}^nR_{p,i}t_i.
\label{eq:score-decomposition}
\end{align}
From the proof of Theorem~\ref{thm:null}, where the bounds preceding \eqref{eq:quad_diff} and \eqref{eq:h_0-h_infty} show that both $n^{-1/2}\sum_i\varepsilon_i(t_i-h_{p,i})$ and $n^{-1/2}\sum_i\varepsilon_i\{h_{p,i}-h_\infty(\mathbf W_i)\}$ are $o_p(1)$ in $\mathcal H$, and by \eqref{eq:CLT}, we already have
\begin{equation}
\frac1{\sqrt n}\sum_{i=1}^n\varepsilon_i t_i
\rightarrow^D \mathcal{G}_\infty.
\label{eq:innovation-clt}
\end{equation}

\noindent Moreover, since $|R_{p,i}|\le a_p$ by (B3) and $\sum_{i=1}^n\|t_i\|_{\mathcal H}^2\le\sum_{i=1}^n\|\Phi_i\|_{\mathcal H}^2\le2Mn$ by (B1), the Cauchy--Schwarz inequality gives
\begin{equation}\label{eq:approx-negligible}
\Big\|\frac1{\sqrt n}\sum_{i=1}^nR_{p,i}t_i\Big\|_{\mathcal H}
\le\frac1{\sqrt n}\Big(\sum_{i=1}^nR_{p,i}^2\Big)^{1/2}\Big(\sum_{i=1}^n\|t_i\|_{\mathcal H}^2\Big)^{1/2}
\le\sqrt{2M}\,\sqrt n\,a_p=o(1).
\end{equation}
Furthermore, we claim that
\begin{equation}\label{eq:signal-score}
\frac1n\sum_{i=1}^n\Delta_i t_i\rightarrow^{p}m_\Delta.
\end{equation}
With $h_{\infty,i}=h_\infty(\mathbf W_i)$, write $n^{-1}\sum_i\Delta_it_i-m_\Delta=\mathcal I_1+\mathcal I_2+\mathcal I_3$, where
\[
\mathcal I_1=\frac1n\sum_{i}\Delta_i(t_i-h_{p,i}),\quad
\mathcal I_2=\frac1n\sum_{i}\Delta_i(h_{p,i}-h_{\infty,i}),\quad
\mathcal I_3=\frac1n\sum_{i}\Delta_ih_{\infty,i}-m_\Delta.
\]
By the Cauchy--Schwarz inequality,
\[
\begin{aligned}
\|\mathcal I_1\|_{\mathcal H}
&\le\Big(\frac1n\sum_{i=1}^n\Delta_i^2\Big)^{1/2}\Big(\frac1n\sum_{i=1}^n\|t_i-h_{p,i}\|_{\mathcal H}^2\Big)^{1/2},\\
\E\|\mathcal I_2\|_{\mathcal H}
&\le\E\big[|\Delta(\mathbf W)|\,\|h_p(\mathbf W)-h_\infty(\mathbf W)\|_{\mathcal H}\big]\\
&\le\{\E[\Delta(\mathbf W)^2]\}^{1/2}\big\{\E\|h_p(\mathbf W)-h_\infty(\mathbf W)\|_{\mathcal H}^2\big\}^{1/2}\rightarrow0.
\end{aligned}
\]
Since $\E[\Delta(\mathbf W)^2]<\infty$, we have $n^{-1}\sum_i\Delta_i^2=O_p(1)$, so that $\mathcal I_1=o_p(1)$ by \eqref{eq:t-h_p_square_norm}; moreover, $\mathcal I_2=o_p(1)$ by Markov's inequality. Finally,
\[
\E\|\Delta(\mathbf W)h_\infty(\mathbf W)\|_{\mathcal H}\le\{\E[\Delta(\mathbf W)^2]\,\E\|h_\infty(\mathbf W)\|_{\mathcal H}^2\}^{1/2}<\infty,
\]
so the strong law of large numbers for i.i.d.\ random elements of the separable Hilbert space $\mathcal H$ with a finite first moment yields $\mathcal I_3\rightarrow^p0$, because $\E[\Delta(\mathbf W)h_\infty(\mathbf W)]=m_\Delta$. This proves \eqref{eq:signal-score}.

Together with \eqref{eq:signal-score},
\eqref{eq:score-decomposition}, and \eqref{eq:approx-negligible}, therefore it becomes
\begin{equation}
\frac1{\sqrt n}
\sum_{i=1}^n\widehat{e}_i\Phi_i
=
\frac1{\sqrt n}\sum_{i=1}^n\varepsilon_i t_i
+
\sqrt n\,c_n\{m_\Delta+o_p(1)\}
+
o_p(1).
\label{eq:master-score}
\end{equation}

\noindent Consider first bounded $c_n$ with $\sqrt n\,|c_n|\to\infty$. Dividing \eqref{eq:master-score} by $\sqrt n\,c_n$ and using $n^{-1/2}\sum_{i=1}^n\varepsilon_it_i=O_p(1)$ by \eqref{eq:innovation-clt}, we obtain $(\sqrt n\,c_n)^{-1}n^{-1/2}\sum_{i=1}^n\widehat e_i\Phi_i\rightarrow^p m_\Delta$ in $\mathcal H$, and the continuous mapping theorem yields
\begin{equation}\label{eq:large-score-square}
\frac{1}{nc_n^2}\Big\|\frac{1}{\sqrt n}\sum_{i=1}^n\widehat e_i\Phi_i\Big\|_{\mathcal H}^2\rightarrow^p\|m_\Delta\|_{\mathcal H}^2=\mu_\Delta.
\end{equation}
For the second term of $T_n$, \eqref{eq:d_i_bound} and $\|\mathbf\Pi_nY\|\le\|\varepsilon+R_p+c_n\Delta\|$ give
\[
\begin{aligned}
0\le\frac{1}{nc_n^2}\Big\{\frac1n\sum_{i=1}^nd_i\widehat e_i^2\Big\}
&\le\frac{6M}{n^2c_n^2}\|\mathbf\Pi_nY\|^2\\
&\le\frac{18M}{nc_n^2}\Big(\frac1n\|\varepsilon\|^2+\frac1n\|R_p\|^2\Big)+\frac{18M}{n}\cdot\frac1n\|\Delta\|^2.
\end{aligned}
\]
The factor $c_n^2$ from the squared signal cancels with the normalization in the last term. Here $n^{-1}\|\varepsilon\|^2\rightarrow^p\sigma^2$, $n^{-1}\|R_p\|^2\le a_p^2=o(1)$, and $n^{-1}\|\Delta\|^2=O_p(1)$ because $\E[\Delta(\mathbf W)^2]<\infty$. Since $nc_n^2=(\sqrt n\,|c_n|)^2\to\infty$, it follows that
\begin{equation}\label{eq:large-diagonal-negligible}
\frac{1}{nc_n^2}\Big\{\frac1n\sum_{i=1}^nd_i\widehat e_i^2\Big\}\rightarrow^p0.
\end{equation}

We therefore obtain
\begin{equation}\label{eq:large-Tn}
\begin{aligned}
\frac{T_n}{n c_n^2}
&=
\frac{n}{n-3}
\Bigg[
\frac{1}{n c_n^2}
\left\|
\frac{1}{\sqrt{n}}
\sum_{i=1}^n \widehat{e}_i\Phi_i
\right\|_{\mathcal H}^2
-
\frac{1}{n c_n^2}
\left\{
\frac{1}{n}
\sum_{i=1}^n d_i\widehat{e}_i^2
\right\}
\Bigg]\rightarrow^{p} \mu_\Delta>0,
\end{aligned}
\end{equation}
by \eqref{eq:large-score-square} and \eqref{eq:large-diagonal-negligible}. The positivity of $\mu_\Delta$ is established at the end of the proof; together with that fact, this proves part~(i). 

Next we consider the local alternative with 
$\sqrt n\,c_n\rightarrow\tau \in \R$.
By \eqref{eq:innovation-clt} and \eqref{eq:master-score}, we have
\[
\frac1{\sqrt n}
\sum_{i=1}^n\widehat{e}_i\Phi_i
\rightarrow^D
\mathcal G_\infty+\tau m_\Delta.
\]
Applying the continuous mapping theorem, we obtain
\begin{equation}
\left\|
\frac1{\sqrt n}
\sum_{i=1}^n\widehat{e}_i\Phi_i
\right\|_{\mathcal H}^2
\rightarrow^D
\|\mathcal G_\infty+\tau m_\Delta\|_{\mathcal H}^2.
\label{eq:squared-score-local}
\end{equation}

\noindent It remains to study the second term of $T_n$ when $c_n\to0$. By \eqref{eq:res_alternative}, $\widehat e=\varepsilon+k_n$ with $k_n=\delta_n+c_n\mathbf\Pi_n\Delta$, where $\delta_n=-\mathbf P_n\varepsilon+\mathbf\Pi_nR_p$ is as in \eqref{eq:delta} with $e$ replaced by $\varepsilon$. By \eqref{eq:delta_square} and $\|\mathbf\Pi_n\Delta\|\le\|\Delta\|$,
\[
\frac1n\|k_n\|^2\le\frac2n\|\delta_n\|^2+\frac{2c_n^2}{n}\|\Delta\|^2
=O_p\Big(\frac pn+a_p^2\Big)+c_n^2\,O_p(1)=o_p(1).
\]
Since $\widehat e_i^2-\varepsilon_i^2=2\varepsilon_ik_{n,i}+k_{n,i}^2$ and $0\le d_i\le6M$ by \eqref{eq:d_i_bound}, the Cauchy--Schwarz inequality gives
\[
\Big|\frac1n\sum_{i=1}^nd_i(\widehat e_i^2-\varepsilon_i^2)\Big|
\le12M\Big(\frac1n\|\varepsilon\|^2\Big)^{1/2}\Big(\frac1n\|k_n\|^2\Big)^{1/2}+\frac{6M}{n}\|k_n\|^2=o_p(1).
\]
Together with \eqref{eq:d_i epsilon_i^2 limit}, with $e$ replaced by $\varepsilon$, this yields
\begin{equation}\label{eq:diagonal-local}
\frac1n\sum_{i=1}^nd_i\widehat{e}_i^2=\frac1n\sum_{i=1}^nd_i\varepsilon_i^2+o_p(1)\rightarrow^{p}\sigma^2\mu_D.
\end{equation}

Combining
\eqref{eq:squared-score-local} and 
\eqref{eq:diagonal-local} provides
\begin{equation}\label{eq:dist_alternative_case2}
T_n
\rightarrow^D
\|\mathcal G_\infty+\tau m_\Delta\|_{\mathcal H}^2
-
\sigma^2\mu_D.
\end{equation}
This proves part~(ii).

If
$\sqrt n\,c_n\rightarrow0$,
we simply take $\tau=0$ in the preceding result. Hence
\begin{equation}\label{eq:dist_alternative_case3}
T_n
\rightarrow^{D}
\|\mathcal G_\infty\|_{\mathcal H}^2-\sigma^2\mu_D
=
\mathcal{L}_{0,\infty}.
\end{equation}
and part~(iii) follows.

\noindent We are left to show that $\mu_\Delta>0$. Let $(\mathbf W',\Delta')$ be an independent copy of $(\mathbf W,\Delta)$. Since $\E[\Delta]=0$, $\E[\|\mathbf W\|+|\Delta|]<\infty$, and $\E[\|\mathbf W\|\,|\Delta|]<\infty$, the terms $\E[\Delta\Delta'\|\mathbf W\|]$ and $\E[\Delta\Delta'\|\mathbf W'\|]$ vanish, and \eqref{grow:feature_supp} gives
\begin{equation}\label{eq:signal-inner-product}
\begin{aligned}
\mu_\Delta=\|\E[\Delta\Phi(\mathbf W)]\|_{\mathcal H}^2
&=\E\big[\Delta\Delta'\langle\Phi(\mathbf W),\Phi(\mathbf W')\rangle_{\mathcal H}\big]\\
&=-\E[\Delta\Delta'\|\mathbf W-\mathbf W'\|]=\operatorname{MDD}^2(\Delta\mid\mathbf W).
\end{aligned}
\end{equation}
If $\mu_\Delta=0$, the characterization property of the martingale difference divergence gives $\E(\Delta\mid\mathbf W)=\E(\Delta)=0$ almost surely. Since $\Delta=\Delta(\mathbf W)$ is $\mathbf W$-measurable, $\E(\Delta\mid\mathbf W)=\Delta$ almost surely, so that $\Delta=0$ almost surely, which contradicts the assumption that $\Delta$ is non-trivial, that is, $P(\Delta\neq0)>0$. Hence $\mu_\Delta=\operatorname{MDD}^2(\Delta\mid\mathbf W)>0$.

This completes the proof.
\end{proof}

\begin{proof}[Proof of Theorem~\ref{thm:power_local_boot}]
We use the notation of Lemma~\ref{lemma:boostrap_remainder_alt}.
Under $H_{1,n}$ and Condition (A1) of Assumption~\ref{assum:0} with $\varepsilon$ in place of $e$, the response vector is $\mathbf Y=\gamma_0\mathbf 1+\mathbf F\boldsymbol\beta_0+c_n\boldsymbol\Delta+\boldsymbol\varepsilon$, where $\mathbf F$ has rows $\boldsymbol\phi_{p,i}^\top$, $\varepsilon_i$ is independent of $\mathbf W_i$ with $\E[\varepsilon_i]=0$, and $\E[\Delta_i\mid\mathbf Z_i]=0$. Since $\mathbf\Pi_n$ annihilates $\mathbf 1$ and the columns of $\mathbf F$, the residual vector is $\widehat{\mathbf e}=\mathbf\Pi_n\mathbf Y=\mathbf\Pi_n(\boldsymbol\varepsilon+c_n\boldsymbol\Delta)$. The bootstrap responses are $\mathbf Y^*=\mathbf P_n\mathbf Y+\mathbf D_\eta\widehat{\mathbf e}$, so the bootstrap residual vector is $\widehat{\mathbf e}^*=\mathbf\Pi_n\mathbf Y^*=\widehat{\mathbf e}^*(\boldsymbol\varepsilon+c_n\boldsymbol\Delta)$. By \eqref{eq:base} in the proof of Lemma~\ref{lemma:test_statistic}, applied with $\mathbf v=\mathbf D_\eta\widehat{\mathbf e}$, the bootstrap statistic is $T_n^*=T_n^*(\boldsymbol\varepsilon+c_n\boldsymbol\Delta)$. Because $B_n^*$ is bilinear and symmetric,
\begin{equation}\label{eq:boot_quadratic}
T_n^*=T_n^*(\boldsymbol\varepsilon)+2c_nB_n^*(\boldsymbol\varepsilon,\boldsymbol\Delta)+c_n^2T_n^*(\boldsymbol\Delta).
\end{equation}
Here $T_n^*(\boldsymbol\varepsilon)$ is exactly the bootstrap statistic computed from the data $\{(\mathbf X_i,Y_i-c_n\Delta_i,\mathbf Z_i)\}_{i=1}^n$, which satisfy the null hypothesis with regression error $\varepsilon$. Since Assumptions~\ref{assum:0} and \ref{assum:2} hold with $\varepsilon$ in place of $e$, Theorem~\ref{thm:null_fixed_p_bootstrap} gives $T_n^*(\boldsymbol\varepsilon)\rightarrow^{D^*}\mathcal L_{0,p}$. By Lemma~\ref{lemma:boostrap_remainder_alt}, the other two terms in \eqref{eq:boot_quadratic} are $O_p^*(|c_n|+c_n^2)$, so that
\begin{equation}\label{eq:boot_alt_expansion}
T_n^*=T_n^*(\boldsymbol\varepsilon)+O_p^*(|c_n|+c_n^2).
\end{equation}

\textit{Parts (ii) and (iii).} If $\sqrt n\,c_n\to\tau\in\mathbb R$ or $\sqrt n\,c_n\to0$, then $c_n\to0$, and \eqref{eq:boot_alt_expansion} gives $T_n^*\rightarrow^{D^*}\mathcal L_{0,p}$. Next, we show that $\mathcal L_{0,p}$ is non-degenerate. Because $\E[\Delta(\mathbf W)\mid\mathbf Z]=0$, we have $m_\Delta=\E[\Delta(\mathbf W)h_p(\mathbf W)]$, so the Cauchy--Schwarz inequality gives $0<\mu_\Delta\le\E[\Delta(\mathbf W)^2]\operatorname{tr}(\boldsymbol\Sigma_p)$ and hence $\operatorname{tr}(\boldsymbol\Sigma_p)>0$. Since also $\sigma^2>0$, Proposition~\ref{prop:fixed_p_feature}, applied to the null data $\{(\mathbf X_i,Y_i-c_n\Delta_i,\mathbf Z_i)\}_{i=1}^n$, shows that $\mathcal L_{0,p}$ is distributed as a nondegenerate weighted sum of independent $\chi^2_1$ variables minus $\sigma^2\mu_D$; its distribution function is continuous and strictly increasing on $(-\sigma^2\mu_D,\infty)$, an interval that contains $Q_{(1-\alpha),\mathcal L_{0,p}}$ because the weighted sum is positive almost surely. Combined with $T_n^*\rightarrow^{D^*}\mathcal L_{0,p}$, these two properties give $Q_{(1-\alpha),n}^*\rightarrow^p Q_{(1-\alpha),\mathcal L_{0,p}}$. 

To verify that the local limit has no atoms, use the common Gaussian construction in the proof of Proposition~\ref{prop:fixed_p_feature}, with $\varepsilon$ in place of $e$. By \eqref{eq:fixed_equiv_kernel}, $\E[\Delta(\mathbf W)]=0$, and the definitions of $C$ and $\boldsymbol\nu_\Delta$ in the main text,
\[
\begin{aligned}
C(\mathbf V)&=\varepsilon\langle\Phi(\mathbf W)-m_\Phi,m_\Delta\rangle_{\mathcal H},\\
\boldsymbol\nu_\Delta&=-2\mathcal B_p^*m_\Delta.
\end{aligned}
\]
Since $4h_{4,1}=2C$, the joint convergence in \eqref{eq:MCLT_alternative} identifies $\mathcal C_\Delta=2\langle\mathcal G_0,m_\Delta\rangle_{\mathcal H}$ and $\mathcal D_\Delta=\boldsymbol\nu_\Delta^\top\boldsymbol{\mathcal N}$ on this common construction. Thus, using $\mathcal G_p=\mathcal G_0-\mathcal B_p\boldsymbol{\mathcal N}$ from \eqref{eq:fixed_equiv_Gp},
\[
\begin{aligned}
\mathcal C_\Delta+\mathcal D_\Delta
&=2\langle\mathcal G_0-\mathcal B_p\boldsymbol{\mathcal N},m_\Delta\rangle_{\mathcal H}\\
&=2\langle\mathcal G_p,m_\Delta\rangle_{\mathcal H}.
\end{aligned}
\]
Together with \eqref{eq:fixed_equiv_oracle} and $\mu_\Delta=\|m_\Delta\|_{\mathcal H}^2$, this gives
\[
\mathcal L_{p,\tau}\overset D=\|\mathcal G_p+\tau m_\Delta\|_{\mathcal H}^2-\sigma^2\mu_D.
\]
Choose a unit eigenvector $u$ of $\sigma^2\boldsymbol\Sigma_p$ with eigenvalue $\vartheta>0$, which exists by the non-degeneracy established above. Write $\mathcal G_p=\sqrt{\vartheta}\,Zu+\mathcal G_\perp$, where $Z\sim N(0,1)$ is independent of $\mathcal G_\perp$ and $\langle\mathcal G_\perp,u\rangle_{\mathcal H}=0$, and put $m_{\Delta,\perp}=m_\Delta-\langle m_\Delta,u\rangle_{\mathcal H}u$. Conditionally on $\mathcal G_\perp$, the variable in the last display is
\[
\big(\sqrt{\vartheta}\,Z+\tau\langle m_\Delta,u\rangle_{\mathcal H}\big)^2
+\|\mathcal G_\perp+\tau m_{\Delta,\perp}\|_{\mathcal H}^2-\sigma^2\mu_D.
\]
Its first term is a nonconstant quadratic in a continuously distributed scalar Gaussian variable, and the remaining terms are fixed under this conditioning. Therefore $\mathcal L_{p,\tau}$ has no atoms. The second and third assertions therefore follow from Theorem~\ref{thm:power_local_fixed_p}(ii)--(iii) and Slutsky's theorem.

\textit{Part (i).} Suppose that $n^{1/2}|c_n|\rightarrow\infty$ and $\sup_n|c_n|\le M_c$, which includes fixed alternatives. Since $T_n^*(\boldsymbol\varepsilon)=O_p^*(1)$ by Lemma~\ref{lemma:boostrap_remainder_alt}, \eqref{eq:boot_alt_expansion} gives $T_n^*=O_p^*(1)$; that is, for every $\delta>0$ there exist $K_\delta<\infty$ and $n_\delta$ such that $P\{P^*(|T_n^*|>K_\delta)>\delta\}<\delta$ for all $n\ge n_\delta$. Fix $\delta<\min(\alpha,1-\alpha)$. On the event $\{P^*(|T_n^*|>K_\delta)\le\delta\}$, we have $P^*(T_n^*>K_\delta)\le\delta<\alpha$ and $P^*(T_n^*<-K_\delta)\le\delta<1-\alpha$, and hence $|Q_{(1-\alpha),n}^*|\le K_\delta$; therefore $P(|Q_{(1-\alpha),n}^*|>K_\delta)<\delta$ for all $n\ge n_\delta$, that is, $Q_{(1-\alpha),n}^*=O_p(1)$. Since $n\widehat{\rm MDD}^{2}(\widehat e\mid\mathbf W)\xrightarrow{p}\infty$ by Theorem~\ref{thm:power_local_fixed_p}(i), we conclude that $P\{n\widehat{\rm MDD}^{2}(\widehat e\mid\mathbf W)>Q_{(1-\alpha),n}^*\}\to1$, which is the first assertion of Theorem~\ref{thm:power_local_boot}.

\end{proof}

\begin{proof}[Proof of Theorem~\ref{thm:power_local_boot_growing_p}]
The representation of $T_n^*$ at the beginning of the proof of Theorem~\ref{thm:bootstrap}, which follows from Lemma~\ref{lemma:test_statistic} and \eqref{eq:t_i_Phi_i}, does not use the null hypothesis; thus $T_n^*=\frac{n}{n-3}\|n^{-1/2}\sum_i\widehat e_i\eta_i^*t_i\|_{\mathcal H}^2-(n-3)^{-1}\sum_id_i(\widehat e_i^*)^2$ also under the alternative.

Under the alternative, we have \eqref{eq:res_alternative}, i.e.,
\[
\widehat e
=
\mathbf \Pi_nY
=
\mathbf \Pi_n\varepsilon
+
\mathbf \Pi_nR_p
+
c_n\mathbf \Pi_n\Delta,
\]
where we have an extra term $c_n\mathbf \Pi_n\Delta$ compared with the residual under the null in \eqref{eq:e_hat}. 
By this fact, the bootstrap feature-score is identical to
\begin{equation}\label{eq:Zn-alternative}
\frac1{\sqrt n}
\sum_i
\widehat e_i\eta_i^*t_i =  \frac{1}{\sqrt{n}} \sum_{i=1}^n (\mathbf \Pi_n\varepsilon)_i \eta_i^*t_i + \frac{1}{\sqrt{n}} \sum_{i=1}^n (\mathbf \Pi_nR_p)_i \eta_i^*t_i + \frac{c_n}{\sqrt{n}} \sum_{i=1}^n (\mathbf \Pi_n\Delta)_i \eta_i^*t_i,
\end{equation}
where 
\begin{align*}
\E^*\left\|\frac{c_n}{\sqrt{n}} \sum_{i=1}^n (\mathbf \Pi_n\Delta)_i \eta_i^*t_i\right\|_{\mathcal{H}}^2&\leq \frac{c_n^2}{n}
\sum_{i=1}^n
(\mathbf \Pi_n\Delta)_i^2
\|t_i\|_{\mathcal H}^2\\&\le
c_n^2
\left(
\max_i\|t_i\|_{\mathcal H}^2
\right)
\frac1n\|\Delta\|^2=
O_p(\xi_p^2c_n^2),
\end{align*}
by \eqref{eq:t_rate}. 

\textit{Parts (ii) and (iii).} We first consider sequences satisfying $\sqrt n\,c_n\to\tau\in\mathbb R$, which include the case $\sqrt n\,c_n\to0$ ($\tau=0$). Since $c_n=O(n^{-1/2})$, we have
$\xi_p^2c_n^2
=
O\left(\xi_p^2/n\right)$.
Assumption~(B2) implies $\xi_p^2/{n}\to0$,
and therefore
\begin{equation}
\frac{c_n^2}{n}
\sum_{i=1}^n
(\mathbf \Pi_n\Delta)_i^2
\|t_i\|_{\mathcal H}^2
=
o_p(1).
\label{eq:thm53-bootstrap-departure-feature}
\end{equation}

The remaining part of the bootstrap feature score in \eqref{eq:Zn-alternative} is handled by the
same argument in the proof of
Theorem~\ref{thm:bootstrap} obtaining \eqref{eq:Z_n_Z_n0_op1} and \eqref{eq:boot-part1-b}, applied with $\varepsilon$ in place of $e$ (note that $(\mathbf\Pi_n\varepsilon)_i+(\mathbf\Pi_nR_p)_i=\varepsilon_i+\delta_{n,i}$ with $\delta_n=-\mathbf P_n\varepsilon+\mathbf\Pi_nR_p$), and satisfies
\[
\frac{1}{\sqrt{n}} \sum_{i=1}^n (\mathbf \Pi_n\varepsilon)_i \eta_i^*t_i + \frac{1}{\sqrt{n}} \sum_{i=1}^n (\mathbf \Pi_nR_p)_i \eta_i^*t_i \rightarrow^{D^*}\mathcal G_\infty
\quad.
\]
Thus, by Slutsky's theorem and
\eqref{eq:thm53-bootstrap-departure-feature}, we obtain
\begin{equation}
\frac1{\sqrt n}
\sum_i
\widehat e_i\eta_i^*t_i
\rightarrow^{D^*}
\mathcal G_\infty.
\label{eq:thm53-bootstrap-score-limit}
\end{equation}
By the conditional continuous mapping theorem,
\begin{equation}
\left\|\frac1{\sqrt n}
\sum_i
\widehat e_i\eta_i^*t_i\right\|_{\mathcal H}^2
\rightarrow^{D^*}
\|\mathcal G_\infty\|_{\mathcal H}^2.
\label{eq:thm53-bootstrap-squared-score}
\end{equation}
Since $c_n\to0$, $n^{-1}\sum_i\widehat e_i^2\le3n^{-1}\sum_i(\varepsilon_i^2+R_{p,i}^2+c_n^2\Delta_i^2)=O_p(1)$, so \eqref{eq:diag_sim_part1}, whose proof uses only this bound and \eqref{eq:ell_i}, remains valid under the alternative; \eqref{eq:tau} is an identity. Together with \eqref{eq:diagonal-local} in the proof of Theorem~\ref{thm:power_local_growing_p}, which holds whenever $c_n\to0$, this gives
\[
\zeta_n=\E^*\left[\frac{1}{n-3}\sum_id_i(\widehat e_i^*)^2\right]=\frac1n\sum_{i=1}^nd_i\widehat e_i^2+o_p(1)\rightarrow^p\sigma^2\mu_D.
\]
Let $\widetilde e_i=(\mathbf\Pi_n\varepsilon)_i+(\mathbf\Pi_nR_p)_i$, so that $\widehat e_i=\widetilde e_i+c_n(\mathbf\Pi_n\Delta)_i$.
It remains to show that the bootstrap diagonal term concentrates around $\zeta_n$. The conditional variance bound \eqref{eq:var_diag_boot} in the proof of Theorem~\ref{thm:bootstrap} does not use the null hypothesis, so it suffices to show that $n^{-1}\sum_i\widehat e_i^4=O_p(1)$. The argument that establishes $n^{-1}\sum_i\widehat e_i^4=O_p(1)$ in the proof of Theorem~\ref{thm:bootstrap}, applied with $\varepsilon$ in place of $e$, gives $n^{-1}\sum_i\widetilde e_i^4=O_p(1)$. For the signal term, $\sum_i x_i^4\le(\sum_i x_i^2)^2$ and the contraction property of $\mathbf\Pi_n$ give
\[
\begin{aligned}
\frac1n\sum_i\{c_n(\mathbf\Pi_n\Delta)_i\}^4
&\le\frac{c_n^4}{n}\|\mathbf\Pi_n\Delta\|^4
\le\frac{c_n^4}{n}\Big(\sum_i\Delta_i^2\Big)^2\\
&=O_p(nc_n^4)=O_p(n^{-1}).
\end{aligned}
\]
Here $\sum_i\Delta_i^2=O_p(n)$ uses only $\E\Delta^2<\infty$, and $c_n=O(n^{-1/2})$ supplies the final rate; no fourth moment of $\Delta$ is needed. Hence
\[
\frac1n\sum_i\widehat e_i^4\le\frac8n\sum_i\widetilde e_i^4+\frac8n\sum_i\{c_n(\mathbf\Pi_n\Delta)_i\}^4=O_p(1),
\]
and \eqref{eq:var_diag_boot} together with Chebyshev's inequality gives $(n-3)^{-1}\sum_i d_i(\widehat e_i^*)^2-\zeta_n=o_p^*(1)$. Combining these results, we obtain
\begin{equation}
\frac1{n-3}
\sum_{i=1}^n
d_i(\widehat e_i^*)^2
\rightarrow^{p^*}
\sigma^2\mu_D.
\label{eq:thm53-bootstrap-diagonal}
\end{equation}
Combining
\eqref{eq:thm53-bootstrap-squared-score} and
\eqref{eq:thm53-bootstrap-diagonal}, together with Slutsky's theorem yields
\begin{equation}
T_n^*
\rightarrow^{D^*}
\mathcal{L}_{0,\infty}
=
\|\mathcal G_\infty\|_{\mathcal H}^2
-
\sigma^2\mu_D.
\label{eq:thm53-bootstrap-null-limit}
\end{equation}

\noindent Let $\vartheta_1\ge\vartheta_2\ge\cdots\ge0$ be the eigenvalues of $\sigma^2\boldsymbol\Sigma_\infty$, so that $\vartheta_1>0$ by (B1) and (B4), and $\|\mathcal G_\infty\|_{\mathcal H}^2\overset{D}{=}\sum_{j\ge1}\vartheta_j\widetilde G_j^2$ with $\widetilde G_1,\widetilde G_2,\ldots$ i.i.d.\ standard normal. As shown at the end of the proof of Theorem~\ref{thm:null}, the distribution function of $\mathcal L_{0,\infty}$ is continuous. It is also strictly increasing on $(-\sigma^2\mu_D,\infty)$, because $\vartheta_1\widetilde G_1^2$ has a positive density on $(0,\infty)$ and the independent remainder $\sum_{j\ge2}\vartheta_j\widetilde G_j^2\ge0$ is smaller than any $\eta>0$ with positive probability; this interval contains $Q_{(1-\alpha),\mathcal L_{0,\infty}}$ because $\|\mathcal G_\infty\|_{\mathcal H}^2>0$ almost surely. To justify the assertion about the remainder, note that $\sum_j\vartheta_j<\infty$. For any $\eta>0$, choose $J$ so that $\sum_{j>J}\vartheta_j<\eta/4$. Markov's inequality gives $P(\sum_{j>J}\vartheta_j\widetilde G_j^2<\eta/2)>1/2$, while the independent finite sum $\sum_{j=2}^J\vartheta_j\widetilde G_j^2$ is smaller than $\eta/2$ with positive probability. Combined with \eqref{eq:thm53-bootstrap-null-limit}, these properties give $Q_{(1-\alpha),n}^*\rightarrow^pQ_{(1-\alpha),\mathcal L_{0,\infty}}$. 

Moreover, let $u_1$ be a unit eigenvector for $\vartheta_1$. Then $\langle\mathcal G_\infty,u_1\rangle_{\mathcal H}\sim N(0,\vartheta_1)$ is independent of $\mathcal G_\infty-\langle\mathcal G_\infty,u_1\rangle_{\mathcal H}u_1$, and, conditionally on the latter, $\|\mathcal G_\infty+\tau m_\Delta\|_{\mathcal H}^2$ equals $(\langle\mathcal G_\infty,u_1\rangle_{\mathcal H}+\tau\langle m_\Delta,u_1\rangle_{\mathcal H})^2$ plus a constant; hence $\mathcal L_{\infty,\tau}$ has no atoms. 

By Theorem~\ref{thm:power_local_growing_p}(ii)--(iii), $T_n\rightarrow^D\mathcal L_{\infty,\tau}$, where $\mathcal L_{\infty,0}=\mathcal L_{0,\infty}$. Therefore, by Slutsky's theorem, under $\sqrt n\,c_n\to\tau\in\mathbb R$,
\[
P\left(T_n>Q_{(1-\alpha),n}^*\mid H_{1,n}\right)\rightarrow P\left(\mathcal L_{\infty,\tau}>Q_{(1-\alpha),\mathcal L_{0,\infty}}\right),
\]
and the limit equals $\alpha$ when $\sqrt n\,c_n\to0$.

\textit{Part (i).} Now suppose that $\sqrt n\,|c_n|\to\infty$ and $\sup_n|c_n|<\infty$. By \eqref{eq:large-Tn}, it suffices to show that the bootstrap critical value is negligible relative to $nc_n^2$.

\noindent We shall show that 
\begin{equation}
\E^*|T_n^*|
=
O_p(1+\xi_p^2c_n^2).
\label{eq:thm53-bootstrap-moment-large}
\end{equation}
Notice that by \eqref{eq:Zn-alternative}, we have
\begin{align*}
&\E^*\left\|\frac{1}{\sqrt{n}}\sum_i \widehat{e}_i\eta_i^* t_i\right\|_{\mathcal H}^2
=
\frac1n
\sum_{i=1}^n
\widehat e_i^2
\|t_i\|_{\mathcal H}^2\\
&\quad\leq \frac3n
\sum_{i=1}^n
(\mathbf\Pi_n\varepsilon)_i^2
\|t_i\|_{\mathcal H}^2
+
\frac3n
\sum_{i=1}^n
(\mathbf\Pi_nR_p)_i^2
\|t_i\|_{\mathcal H}^2
\nonumber\\
&\quad+
\frac{3c_n^2}{n}
\sum_{i=1}^n
(\mathbf \Pi_n\Delta)_i^2
\|t_i\|_{\mathcal H}^2,
\label{eq:thm53-bootstrap-score-second-moment}
\end{align*}
where $\frac{c_n^2}{n}
\sum_{i=1}^n
(\mathbf \Pi_n\Delta)_i^2
\|t_i\|_{\mathcal H}^2=
O_p(\xi_p^2c_n^2)$.
The first two terms are $O_p(1)$. Indeed, since $(\mathbf\Pi_n\varepsilon)_i^2\le2\varepsilon_i^2+2(\mathbf P_n\varepsilon)_i^2$, they are bounded by a constant multiple of
\[
\frac1n\sum_{i=1}^n\varepsilon_i^2\|t_i\|_{\mathcal H}^2+\max_i\|t_i\|_{\mathcal H}^2\,\frac1n\big(\|\mathbf P_n\varepsilon\|^2+\|\mathbf\Pi_nR_p\|^2\big).
\]
Given $\mathbf W_{1:n}$, the first part has mean $\sigma^2n^{-1}\sum_i\|t_i\|_{\mathcal H}^2\le\sigma^2n^{-1}\sum_i\|\Phi(\mathbf W_i)\|_{\mathcal H}^2\le2M\sigma^2$, and the second part is $O_p(\xi_p^2p/n+\xi_p^2a_p^2)=o_p(1)$ by \eqref{eq:t_rate}, \eqref{eq:Pi_epsilon}, and $n^{-1}\|\mathbf\Pi_nR_p\|^2\le a_p^2$. This implies that  
\begin{equation}
\E^*\left\|\frac{1}{\sqrt{n}}\sum_i \widehat{e}_i\eta_i^* t_i\right\|_{\mathcal H}^2
=
O_p(1+\xi_p^2c_n^2).
\label{eq:thm53-bootstrap-score-moment-large}
\end{equation}

\noindent For the second term in $T_n^*$, recall that $\zeta_n=\E^*[(n-3)^{-1}\sum_id_i(\widehat e_i^*)^2]$. Since $d_i\ge0$ and $\mathbf\Pi_n$ is an orthogonal projection, \eqref{eq:d_i_bound} and \eqref{eq:res_alternative} give
\[
\begin{aligned}
0\le\zeta_n
&\le\frac{6M}{n-3}\E^*\|\mathbf\Pi_n(\widehat e\circ\eta^*)\|^2
\le\frac{6M}{n-3}\|\widehat e\|^2\\
&\le\frac{18M}{n-3}\big(\|\varepsilon\|^2+\|R_p\|^2+c_n^2\|\Delta\|^2\big),
\end{aligned}
\]
so that, because $\{c_n\}$ is bounded,
\begin{equation}
\zeta_n=O_p(1+c_n^2)=O_p(1).
\label{eq:thm53-bootstrap-diagonal-moment-large}
\end{equation}
Since $|T_n^*|\le\frac{n}{n-3}\|n^{-1/2}\sum_i\widehat e_i\eta_i^*t_i\|_{\mathcal H}^2+(n-3)^{-1}\sum_id_i(\widehat e_i^*)^2$, \eqref{eq:thm53-bootstrap-score-moment-large} and \eqref{eq:thm53-bootstrap-diagonal-moment-large} give \eqref{eq:thm53-bootstrap-moment-large}.

\noindent Recall that $Q_{1-\alpha,n}^*$ is the conditional $(1-\alpha)$
quantile of $T_n^*$. By conditional Markov's inequality, for every $a>0$, we have
\[
P^*(|T_n^*|>a)
\le
\frac{\E^*|T_n^*|}{a}.
\]
Suppose first that $\E^*|T_n^*|>0$, and let $a
=
2\E^*|T_n^*|/ \min(\alpha,1-\alpha)$. Then we have
\[
P^*(T_n^*>a)
\le
P^*(|T_n^*|>a)
\le
\frac{\min(\alpha,1-\alpha)}{2}
<
\alpha,
\]
and likewise $P^*(T_n^*<-a)\le P^*(|T_n^*|>a)<1-\alpha$. The first bound gives $P^*(T_n^*\le a)>1-\alpha$, so $Q_{1-\alpha,n}^*\le a$; the second gives $P^*(T_n^*\le x)<1-\alpha$ for every $x<-a$, so $Q_{1-\alpha,n}^*\ge-a$. If $\E^*|T_n^*|=0$, then $P^*(T_n^*=0)=1$ and $Q_{1-\alpha,n}^*=0$. Hence, in either case,
\[
|Q_{1-\alpha,n}^*|
\le
\frac{2}{\min(\alpha,1-\alpha)}
\E^*|T_n^*|.
\]
Consequently, by
\eqref{eq:thm53-bootstrap-moment-large},
\begin{equation}
Q_{1-\alpha,n}^*
=
O_p(1+\xi_p^2c_n^2),
\label{eq:thm53-critical-value-bound}
\end{equation}
which provides
\[
\frac{Q_{1-\alpha,n}^*}{nc_n^2}
=
O_p\left(
\frac1{nc_n^2}
+
\frac{\xi_p^2}{n}
\right)=o_p(1),
\]
since $1/(nc_n^2)\to 0$ and $\xi_p^2/n\to 0$ under $\sqrt n\,|c_n|\to\infty$ and Assumption~(B2). 
By the fact in \eqref{eq:large-Tn}, it follows that
\begin{equation}
P\left(
T_n>Q_{1-\alpha,n}^*
\right)
\longrightarrow1,
\label{eq:thm53-part-i}
\end{equation}
which completes the proof. 

\end{proof}
\section{Additional Results for WHO Life Expectancy Data}\label{sec:who}
This section presents the optimal predictors selected by the CMDD method and the corresponding prediction accuracies across a range of tuning parameter values. Specifically, we examined integer values of $p \in [5, 15]$, where, as in Section~\ref{sec:real data} of the main text, $p$ is the number of B-spline basis functions for each selected predictor. The results are qualitatively similar to those reported in Section~\ref{sec:real data} of the main text.

For the majority of $p$ values tested, the set of optimal predictors is identical to the one identified for $p=10$ in the main text: Income composition of resources, Adult mortality, BMI, and Hepatitis B. Deviations were observed in only three cases:
\begin{itemize}
    \item For $p=14$, the optimal set consisted of three predictors, excluding BMI from the set selected with $p=10$.
    \item For $p \in \{9, 15\}$, Schooling replaced BMI, yielding a slightly different four-predictor set.
\end{itemize}

Table~\ref{table:variable_selection_appendix} summarizes the predictors selected in these specific instances, while Table~\ref{table:prediction_appendix} reports the corresponding Prediction Mean Squared Errors (PMSE). Notably, the CMDD method consistently achieved a lower PMSE than all alternative methods across all tested values of $p$.

\begin{table}[H]
\centering
\caption{Selected predictors for the CMDD method across different values of $p$.}
\label{table:variable_selection_appendix}
\begin{tabular}{@{}lp{0.6\textwidth}@{}}
\toprule
\textbf{$p$} & \textbf{Selected Predictors} \\
\midrule
\textbf{$p=14$} & Income composition of resources, Adult mortality, Hepatitis B\\
\textbf{$p\in \{9, 15\}$} & Income composition of resources, Adult mortality, Hepatitis B, Schooling \\
\textbf{$p\in[5, 15] \setminus \{9, 14, 15\}$} & Income composition of resources, Adult mortality, BMI, Hepatitis B \\
\bottomrule
\end{tabular}
\end{table}

\begin{table}[H]
\centering
\caption{Prediction accuracy (PMSE) for the CMDD method across different values of $p$.}
\label{table:prediction_appendix}
\begin{tabular}{@{}lccc@{}}
\toprule
& \textbf{$p=14$} & \textbf{$p \in \{9, 15\}$} & \textbf{$p \in [5, 15] \setminus \{9, 14, 15\}$}\\
\midrule
PMSE & 0.0867 & 0.0908 & 0.0861 \\
\bottomrule
\end{tabular}
\end{table}

\end{document}